\documentclass[%
 reprint,
superscriptaddress, 
nofootinbib,
 amsmath,amssymb,
 aps,
]{revtex4-2}

\usepackage{todonotes}
\usepackage{enumitem}   
\usepackage{subcaption} 

\usepackage{mathtools}%
\usepackage{tabularx}%
 \usepackage{multirow}
\usepackage{braket}
\usepackage{graphicx}
\usepackage{dcolumn}
\usepackage{bm}
\usepackage[version=4]{mhchem} 
\usepackage{adjustbox}
\usepackage{algorithm}
\usepackage{algpseudocode}

\usepackage{hyperref}

\usepackage{footnotehyper}
\usepackage{CJK}
\usepackage{nameref}
\usepackage{zref-xr}
\usepackage{zref-user}
\zxrsetup{toltxlabel}

\begin{document}

\preprint{APS/123-QED}

\title{Unitary Partitioning and the Contextual Subspace \\ Variational Quantum Eigensolver}

\author{Alexis Ralli}
\email{alexis.ralli.18@ucl.ac.uk}
\affiliation{
Centre for Computational Science, Department of Chemistry, University College London, London, WC1H 0AJ, United Kingdom
}
\author{Tim Weaving}
\email{timothy.weaving.20@ucl.ac.uk}
\affiliation{
Centre for Computational Science, Department of Chemistry, University College London, London, WC1H 0AJ, United Kingdom
}
\author{Andrew Tranter}
\email{tufts@atranter.net}
\affiliation{
 Department of Physics and Astronomy, Tufts University, Medford, Massachusetts 02155, USA
}
\affiliation{Cambridge Quantum Computing, 9a Bridge Street Cambridge, CB2 1UB, United Kingdom
}
\author{William M. Kirby}
\email{william.kirby@tufts.edu}
\affiliation{
 Department of Physics and Astronomy, Tufts University, Medford, Massachusetts 02155, USA
}
\author{Peter J. Love}
\email{peter.love@tufts.edu}
\affiliation{
 Department of Physics and Astronomy, Tufts University, Medford, MA 02155, USA
}
\affiliation{Computational Science Initiative, Brookhaven National Laboratory, Upton, New York 11973, USA}
\author{Peter V. Coveney}
\email{p.v.coveney@ucl.ac.uk}
\affiliation{
Centre for Computational Science, Department of Chemistry, University College London, London, WC1H 0AJ, United Kingdom
}
\affiliation{
Informatics Institute, University of Amsterdam, Amsterdam, 1098 XH, Netherlands
}


\date{\today}

\begin{abstract}
The contextual subspace variational quantum eigensolver (CS-VQE) is a hybrid quantum-classical algorithm that approximates the ground-state energy of a given qubit Hamiltonian. It achieves this by separating the Hamiltonian into contextual and noncontextual parts. The ground-state energy is approximated by classically solving the noncontextual problem, followed by solving the contextual problem using VQE, constrained by the noncontextual solution. In general, computation of the contextual correction needs fewer qubits and measurements compared with solving the full Hamiltonian via traditional VQE. We simulate CS-VQE on different tapered molecular Hamiltonians and apply the unitary partitioning measurement reduction strategy to further reduce the number of measurements required to obtain the contextual correction. Our results indicate that CS-VQE combined with measurement reduction is a promising approach to allow feasible eigenvalue computations on noisy intermediate-scale quantum devices. We also provide a modification to the CS-VQE algorithm; the CS-VQE algorithm previously could cause an exponential increase in Hamiltonian terms, but with this modification now at worst will scale quadratically. 
\end{abstract}

\maketitle


\section{Introduction}
\zlabel{sec:Intro}

One of the fundamental goals of quantum chemistry is to solve the time-independent non-relativistic Schr{\"o}dinger equation. The eigenvalues and eigenvectors obtained allow different molecular properties to be studied from first principles. Standard methods project the problem onto a Fock space ($\eta$ electrons distributed in $M$ orbitals) and solve. Under this approximation, the problem scales exponentially with system size, where the number of Slater determinants (configurations) scales as $\binom{M}{\eta} $ making the problem classically intractable \cite{szabo2012modern}. Quantum computers can efficiently represent the full configuration interaction (FCI) Hilbert space and offer a potential way to efficiently solve such molecular problems \cite{aspuru2005simulated, babbush2018encoding}. This use case is often the canonical example of where the first quantum computers will be advantageous over conventional computers \cite{elfving2020will, mccaskey2019quantum}.

In the fault-tolerant regime, quantum phase estimation (QPE) \cite{kitaev1995quantum} provides a practical way to perform quantum chemistry simulations in polynomial time \cite{aspuru2005simulated}. However, current noisy intermediate-scale quantum (NISQ) devices cannot implement this algorithm due to the deep quantum circuits and long coherence times required \cite{o2016scalable, mohammadbagherpoor2019experimental}. 

The constraints on present-day devices have given rise to a family of quantum-classical algorithms that leverage as much classical processing as possible to reduce the quantum resources required to solve the problem at hand. Common examples of NISQ algorithms are the variational quantum eigensolver (VQE) \cite{peruzzo2014variational}, quantum approximate optimization algorithm (QAOA) \cite{farhi2014quantum} and variational quantum linear solver (VQLS) \cite{bravo2019variational}.   A good example is the recently proposed entanglement forging method \cite{eddins2022doubling}, where the electronic structure problem for \ce{H2O} was reduced from a $10$-qubit problem to multiple $5$ qubit problems that were each studied using conventional VQE and classically combined. Recently another novel approach known as the quantum-classical hybrid quantum Monte Carlo (QC-QMC) method was used to unbias the sign problem in the projector Monte Carlo (PMC) method, which implements imaginary time evolution \cite{huggins2022unbiasing}. At a high level, the accuracy of a constrained PMC calculation is determined by the quality of trial wave functions. Quantum computers offer a way to efficiently store highly entangled trial wave functions and measure certain overlaps, which would require exponential resources classically. Huggins \textit{et al.} performed QC-QMC simulations of different chemical systems on Google's Sycamore processor and obtained results competitive with state-of-the-art classical methods \cite{huggins2022unbiasing}.

The contextual-subspace VQE algorithm is another hybrid quantum-classical approach \cite{kirby2021contextual}. It gives an approximate simulation method, where the quantum resources required can be varied for a trade off in accuracy. This allows problems to be studied where the full Hamiltonian would normally be too large to investigate on current NISQ hardware. This was shown in the original CS-VQE paper, where chemical accuracy for various molecular systems was reached using significantly fewer qubits compared with the number required for VQE on the full system \cite{kirby2021contextual}. As CS-VQE reduces the number of qubits required for simulation, the number of terms in a Hamiltonian requiring separate measurements is also reduced.

A natural question that arises from this is whether measurement reduction schemes can be utilized to reduce the overall measurement cost of these already reduced CS-VQE Hamiltonians \cite{zhao2019measurement, kandala2017hardware, verteletskyi2020measurement,izmaylov2019revising, cotler2020quantum, bonet2019nearly, gokhale2019n, jena2019pauli, huggins2019efficient, gokhale2019minimizing, crawford2021efficient, huang2020predicting, hadfield2020measurements, huang2021efficient}. The goal of this work was to investigate the possible reductions given by the unitary partitioning strategy \cite{zhao2019measurement, izmaylov2019unitary, ralli2021implementation} and whether chemical accuracy on larger molecules can be reached on currently available NISQ hardware.

This paper is structured as follows. Section \zref{sec:Theory} summarizes the CS-VQE algorithm. Here we provide a modification to the unitary partitioning step of the CS-VQE algorithm the CS-VQE algorithm previously could cause the number of terms in a Hamiltonian to exponentially increase, but with this modification now will at worst cause a quadratic increase. Section \zref{sec:num_results} is split into a description of the method in Section \zref{sec:Method} and two main parts, Sections \zref{sec:example_will} and \zref{sec:M_reduction}. Section \zref{sec:example_will} examines a model problem to exemplify each step of the CS-VQE algorithm. Section \zref{sec:M_reduction} gives the numerical results of applying unitary partitioning measurement reduction to a test bed of different molecular structure Hamiltonians, where the contextual subspace approximation has been employed.

\section{Background}
\zlabel{sec:Theory}
To keep our discussion self-contained and establish notation, we summarize the necessary background theory of the contextual-subspace VQE algorithm in this section.

\subsection{\zlabel{sec:contextuality}Contextuality}
 The foundation of quantum contextuality is the Bell-Kochen-Specker (BKS) theorem \cite{kochen1975problem}. In lay terms, every measurement provides a classical probability distribution (via the spectral theorem) and a joint distribution can be built as a product over all possible measurements \cite{budroni2019contextuality}. The BKS theorem proves that it is impossible to reproduce the probabilities of every possible measurement outcome for a quantum system as marginals of this joint probability distribution \cite{budroni2021quantum}. This is related to how quantum mechanics does not allow models that are locally causal in a classical sense \cite{de2017graph}. Contextuality is a generalization of nonlocality \cite{de2017graph, cabello2021converting}. This means that quantum measurement cannot be understood as simply revealing a pre-existing value of some underlying hidden variable \cite{howard2014contextuality, mermin1990simple}. Bell's theorem also reaches a similar conclusion against hidden variables \cite{bell1964einstein}, but in a different way. 



A good example of this phenomenon is the ``Peres-Mermin square" \cite{mermin1990simple, peres1990incompatible}, where no state preparation is involved and only observables are considered. We include an example in Appendix \zref{sec:Peres_square} and remark on the relation to VQE. Colloquially, for a noncontextual problem it is possible to assign deterministic outcomes to observables simultaneously without contradiction; however, for a contextual problem this is not possible \cite{kirby2021contextual}.

The following subsections set out the contextual subspace VQE algorithm and we provide an alternate way to construct $U_{\mathcal{W}}$ (defined below) compared to the original work \cite{kirby2021contextual}. This modification addresses the exponential scaling part of the method. Further background on the full CS-VQE algorithm is given in the Supplemental Material \cite{SuppMaterial}. 

\subsection{\zlabel{sec:CS_VQE_overview}Contextual subspace VQE}

Consider a Hamiltonian expressed as:

\begin{equation}
    \zlabel{eq:PauliHamilt}
    \begin{aligned}
    H_{\text{full}} &= \sum_{a} c_{a}P_{a} =  \sum_{a} c_{a} \bigg( \bigotimes_{j=0}^{n-1} \sigma_{j}^{(a)} \bigg) \\ 
    &= \sum_{a} c_{a} \big( \sigma_{0}^{(a)} \otimes \sigma_{1}^{(a)} \otimes ... \otimes \sigma_{n-1}^{(a)} \big),
    \end{aligned}
\end{equation}

\noindent where $c_{a}$ are real coefficients. Each Pauli operator $P_{a}$ is made up of an $n$-fold tensor product of single qubit Pauli matrices $\sigma_{j} \in \{\mathcal{I}, X, Y, Z\}$, where $j$ indexes the qubit the operator acts on. The CS-VQE algorithm is based on separating such a Hamiltonian into a contextual and noncontextual part \cite{kirby2021contextual}:

\begin{equation}
    \zlabel{eq:Hamilt}
    H_{\text{full}} = H_{\text{con}} + H_{\text{noncon}}.
\end{equation}
As it is possible to assign definite values to all terms in $H_{\text{noncon}}$ without contradiction, a classical hidden variable model (or quasiquantized model) can be used to represent this system \cite{Spekkens2016}.

In \cite{kirby2020classical}, such a model is constructed along with a classical algorithm to solve it. This was based on the work of Spekkens \cite{spekkens2007evidence, spekkens2016quasi}. Solving this model yields a noncontextual ground-state.

Once that solution is obtained, the remaining contextual part of the problem is solved. Solutions to $H_{\text{con}}$ must be consistent with the noncontextual ground-state, which defines a subspace of allowed states \cite{kirby2021contextual}. By projecting the problem into this subspace the overall energy is given by:

\begin{equation}
    \zlabel{eq:full_Hamilt_exp}
    \begin{aligned}
    E(\vec{\theta}; \vec{q}, \vec{r} \:) 
    =  &E_{\text{noncon}}(\vec{q}, \vec{r} \:) + E_{\text{con}}(\vec{\theta}; \vec{q}, \vec{r} \:)\\
    =  &E_{\text{noncon}}(\vec{q}, \vec{r} \:) +  \\
       & \frac{\bra{\psi_{\text{con}}(\vec{\theta})} \; Q_{\mathcal{W}}^{\dagger}U_{\mathcal{W}}^{\dagger} H_{\text{con}}U_{\mathcal{W}} Q_{\mathcal{W}} \; \ket{\psi_{\text{con}}(\vec{\theta})}}{\bra{\psi_{\text{con}}(\vec{\theta})} Q_{\mathcal{W}}^{\dagger} Q_{\mathcal{W}} \; \ket{\psi_{\text{con}}(\vec{\theta})}}\\
    = &\frac{\bra{\psi_{\text{con}}(\vec{\theta})} \; Q_{\mathcal{W}}^{\dagger} \; U_{\mathcal{W}}^{\dagger} H_{\text{full}}U_{\mathcal{W}}\; Q_{\mathcal{W}} \; \ket{\psi_{\text{con}}(\vec{\theta})}}{\bra{\psi_{\text{con}}(\vec{\theta})} Q_{\mathcal{W}}^{\dagger} Q_{\mathcal{W}} \; \ket{\psi_{\text{con}}(\vec{\theta})}}.
    \end{aligned}
\end{equation}
We have written $U_{\mathcal{W}}$ rather than $U_{\mathcal{W}}(\vec{q}, \vec{r})$ to simplify our notation. 

The vector $(\vec{q}, \vec{r})$ should be thought of as parameters that define a particular noncontextual state: normally, this will be a parameterization for the noncontextual ground-state \cite{kirby2020classical, kirby2021contextual}. This vector has a size of at most $2n +1$ for a Hamiltonian defined on $n$ qubits \cite{kirby2020classical}. From $(\vec{q}, \vec{r})$, we define a set of stabilizers $\mathcal{W}$ which stabilize that particular noncontextual state \cite{ kirby2021contextual}. The unitary $U_{\mathcal{W}}$ maps each of these stabilizers to a distinct single-qubit Pauli matrix; details of this are covered in \zref{sec:NonMappingConSub}. By enforcing the eigenvalue of these single-qubit Pauli operators we define a subspace of allowed quantum states that are consistent with the noncontextual state. To constrain the problem to this subspace, we use the projector $ Q_{\mathcal{W}}$. Note that this is not a unitary operation, hence the renormalization in equation \zref{eq:full_Hamilt_exp}. By projecting our contextual Hamiltonian into this subspace $H_{\text{con}} \mapsto H_{\text{con}}^{\mathcal{W}} = Q_{\mathcal{W}}^{\dagger} U_{\mathcal{W}}^{\dagger}H_{\text{con}}U_{\mathcal{W}} Q_{\mathcal{W}}$, we ensure that solutions to $H_{\text{con}}^{\mathcal{W}}$ remain in the subspace consistent with the noncontextual solution \cite{ kirby2021contextual}. In other words, this operation means that solutions to $H_{\text{con}}^{\mathcal{W}}$ will remain consistent with the noncontextual solution. Section \zref{sec:NonMappingConSub} goes into detail on this.

The contextual trial or ansatz state is prepared as $\ket{\psi_{\text{con}}(\vec{\theta})} = \; U_{\mathcal{W}}^{\dagger} V(\vec{\theta}) \ket{0}^{\otimes n}$, where $V(\vec{\theta})$ is the parameterized operator that prepares it. The projector $Q_{\mathcal{W}}$, in equation \zref{eq:full_Hamilt_exp}, then projects this state into the subspace of possible states consistent with the noncontextual ground-state. Again this depends on which stabilizer eigenvalues are fixed. Note that as $Q_{\mathcal{W}}$ is not a unitary operation the state must be renormalized. Further analysis of the contextual subspace VQE projection ansatz is provided in \cite{weaving2022stabilizer}.  In section \zref{sec:noncon_problem} we discuss how to solve the noncontextual problem.

\subsection{\zlabel{sec:noncon_problem}Noncontextual Hamiltonian}

For a given noncontextual Hamiltonian, we define $\mathcal{S}^{H_{\text{noncon}}}$ to be the set of $P_{i}$ present in  $H_{\text{noncon}}$. This set can be expanded as two subsets denoted as $\mathcal{Z}$ and $\mathcal{T}$, representing the set of fully commuting operators and its complement respectively \cite{kirby2020classical, kirby2021contextual}. The set $\mathcal{T}$ can be expanded into $N$ cliques, where operators within a clique must all commute with each other and operators between cliques must pairwise anticommute. This is because commutation forms an equivalence relation on $\mathcal{T}$ if and only if $\mathcal{S}^{H_{\text{noncon}}}$ is noncontextual \cite{kirby2020classical, kirby2021contextual}.

A hidden variable model for such a system can be built, where the set of observables $\mathcal{R}$ that define the phase-space points of the hidden variable model is \cite{kirby2020classical, kirby2021contextual, raussendorf2020phase}:

\begin{equation}
    \zlabel{eq:R_ind_set}
    \begin{aligned}
    \mathcal{R} &\equiv \{ P_{0}^{(j)} | j=0,1,\hdots, N-1 \} \cup \{ G_{0}, G_{1}, G_{2},... \} \\
    &\equiv \{ P_{0}^{(j)} | j=0,1,\hdots, N-1 \} \cup \mathcal{G}.
    \end{aligned}
\end{equation}
The set $\mathcal{G}$ represents an independent set of Pauli operators that generates the set of commuting observables $\mathcal{Z}$. Each $P_{0}^{(j)}$ corresponds to a chosen Pauli operator in the $j$-th clique of $\mathcal{T}$: by convention we say that this is the first operator in the set, but this can be any operator in the $j$th clique.

With respect to the phase-space model given in \cite{kirby2020classical}, a valid noncontextual state is defined by the parameters $(\vec{q}, \vec{r})$, which set the expectation value of the operators in $\mathcal{R}$ (Equation \zref{eq:R_ind_set}). Each operator in $\mathcal{G}$ is assigned the value $\langle G_{i} \rangle=q_{i}=\pm1$. The operators in $\mathcal{T}$ are assigned the values $\langle P_{0}^{(j)} \rangle=r_{j}$, where $\vec{r}$ is a unit vector ($|\vec{r}|=1$) \cite{kirby2020classical, kirby2021contextual}. The number of elements in $\vec{q}$ and $\vec{r}$ are $|\mathcal{G}|$ and $N$ respectively. For $n$ qubits the size of $|\mathcal{R}|$ is bounded by $2n+1$, which bounds the size of $(\vec{q}, \vec{r})$ \cite{kirby2020classical, raussendorf2020phase}.

The observables for the $N$ anticommuting $P_{0}^{(j)}$ operators in $\mathcal{R}$ can be combined into the observable \cite{kirby2020classical}:

\begin{equation}
    \zlabel{eq:anticomm}
    \begin{aligned}
    A(\vec{r}) = \sum_{j=0}^{N-1} r_{j} P_{0}^{(j)}.
    \end{aligned}
\end{equation}
We denote the set of Pauli operators making up this operator as $\mathcal{A} \equiv \{P_{0}^{(j)} | j=0,1\hdots,N-1 \}$.

The expectation value of $A(\vec{r})$ is assigned by the hidden variable model to always be $+1$, due to:

\begin{equation}
    \zlabel{eq:A_eigval}
    \begin{aligned}
    \langle A(\vec{r}) \rangle= \sum_{j=0}^{N-1} r_{j} \langle P_{0}^{(j)} \rangle = \sum_{j=0}^{N-1} r_{j} r_{j} = \sum_{j=0}^{N-1} |r_{j}|^{2} = +1,
    \end{aligned}
\end{equation}

\noindent using $\langle P_{0}^{(j)} \rangle=r_{j}$ and $|\vec{r}|=1$.

The expectation value for $H_{\text{noncon}}$ can be induced, by setting the expectation values of operators in $\mathcal{R}$ (Equation \zref{eq:R_ind_set}), as this set generates $\mathcal{S}^{H_{\text{noncon}}}$ \cite{kirby2020classical, kirby2021contextual}.  To find the ground-state of $H_{\text{noncon}}$, we perform a brute force search over this space. For each possible $\pm 1$ combination of expectation values for each $G_{j}$ ($2^{|\mathcal{G}|}$ possibilities), the energy is minimized with respect to the unit vector $\vec{r}$ - that sets the expectation values $ \langle P_{0}^{(j)} \rangle=r_{j}$. The Supplemental Material provides further algorithmic details \cite{SuppMaterial}. The vector $(\vec{q}, \vec{r})$  that was found to give the lowest energy defines the noncontextual ground-state. Each noncontextual state $(\vec{q}, \vec{r})$, corresponds to subspaces of quantum states, which we will describe in subsection \zref{sec:contextual_subs}.

\subsection{\zlabel{sec:contextual_subs}Contextual subspace}

In \cite{zhao2019measurement} and \cite{izmaylov2019unitary} it was shown that an operator constructed as a normalized linear combination of pairwise anticommuting Pauli operators, such as $A(\vec{r})$ (Equation \zref{eq:anticomm}),  is equivalent to a single Pauli operator up to a unitary rotation $R$. We can therefore write $A(\vec{r}) \mapsto P_{0}^{(k)} = R A(\vec{r})  R^{\dagger}$ for a selected  $P_{0}^{(k)} \in \mathcal{A}$. We write the set $\mathcal{R}_{\vec{r}}$ (Equation \zref{eq:R_ind_set}) under this transformation:

\begin{equation}
    \zlabel{eq:R_ind_set_scriptA}
    \begin{aligned}
    \mathcal{R}_{\vec{r}} &\equiv \{A(\vec{r}) \} \cup \mathcal{G} &\mapsto \mathcal{R}^{'} &\equiv \underbrace{\{ P_{0}^{(k)} \}}_{P_{0}^{(k)} = R A(\vec{r})  R^{\dagger}} \cup \; \mathcal{G}.
    \end{aligned}
\end{equation}
It will be shown later that the unitary $R$ is constructed from the operators in $\mathcal{A}$. This means that the terms in $\mathcal{G}$ are unaffected by this transformation, as operators in $\mathcal{G}$ and so must universally commute so must commute with $R$. 


A given noncontextual state $(\vec{q}, \vec{r})$ is equivalent to the joint expectation value assignment of $\langle G_{i} \rangle = q_{i} = \pm 1$ and $\langle A(\vec{r}) \rangle = +1$. This defines a set of stabilizers:
\begin{equation}
    \zlabel{eq:stabilizers_set_nonrot}
    \begin{aligned}
    \mathcal{W}_{all} &\equiv \{ q_{0}G_{0},q_{1}G_{1},..., q_{|\mathcal{G}|-1}G_{|\mathcal{G}|-1},  A(\vec{r}) \},
    \end{aligned}
\end{equation}
which by definition must stabilize that noncontextual state $(\vec{q}, \vec{r})$ or more precisely, the subspace of quantum states corresponding to it \footnote{Note that a stabilizer for a state leaves it unchanged. For example, if $O$ stabilizes $\ket{\psi}$ then $O\ket{\psi}=\ket{\psi}$}. Note that $A(\vec{r})$ is not a conventional stabilizer, but is unitarily equivalent to a single qubit operator $P_{0}^{(k)}$ \cite{zhao2019measurement, izmaylov2019unitary}.

We can consider this problem under the unitary transform defined in Equation \zref{eq:R_ind_set_scriptA}. The stabilizers in $\mathcal{W}_{all}$ become:

\begin{equation}
    \zlabel{eq:stabilizers_set_rot}
    \begin{aligned}
    \mathcal{W}_{all}^{'} &\equiv \{ q_{0}G_{0},q_{1}G_{1},..., q_{|\mathcal{G}|-1}G_{|\mathcal{G}|-1}, \xi \: P_{0}^{(k)} \},
    \end{aligned}
\end{equation}
which defines a regular set of stabilizers for the noncontextual state $(\vec{q}, \vec{r})$, which defines a subspace of quantum states. Here $\xi=\pm 1$, is determined by $R A(\vec{r})  R^{\dagger} = \xi P_{0}^{(k)}$ and can always be chosen to be $+1$, which we do throughout this paper.

Altogether, when certain noncontextual stabilizers are fixed (by the noncontextual state) they specify a subspace of allowed quantum states that will be consistent with that noncontextual state and thus define the constraints for the contextual part of the problem. We refer to this subspace as the contextual subspace \cite{kirby2021contextual}.


\subsection{\zlabel{sec:NonMappingConSub}Mapping a contextual subspace to a stabilizer subspace}

In CS-VQE, the expectation value of the full Hamiltonian is obtained according to Equation \zref{eq:full_Hamilt_exp}. First, the noncontextual problem is solved yielding the noncontextual state $(\vec{q}, \vec{r})$ - normally the ground-state $(\vec{q}_{0}, \vec{r}_{0})$.  The Supplemental Material shows how $(\vec{q}_{0}, \vec{r}_{0})$ can be obtained via a brute force approach \cite{SuppMaterial}. Next the contextual Hamiltonian is projected into the subspace of allowed quantum states consistent with the defined noncontextual state. This constraint is imposed via: $H_{\text{full}} \mapsto H_{\text{full}}^{\mathcal{W}} = Q_{\mathcal{W}}^{\dagger}U_{\mathcal{W}}^{\dagger}H_{\text{full}}U_{\mathcal{W}} Q_{\mathcal{W}}$, where the expectation value is then found on a quantum device.

The unitary operation $U_{\mathcal{W}}$ is defined by the set of contextual stabilizers $\mathcal{W} \subseteq \mathcal{W}_{all}$ (equation \zref{eq:stabilizers_set_nonrot}), whose eigenvalue we fix according to the noncontextual state. If $A(\vec{r}) \in \mathcal{W}$, meaning that $\langle A(\vec{r}) \rangle$ is fixed to be $+1$, then the steps summarized in Equation \zref{eq:R_ind_set_scriptA} must first be performed to reduce $A(\vec{r})$ to a single Pauli operator. Clifford operators $V_{i}(P)$ are then used to map each $P \in \mathcal{W}$ to a single-qubit $Z$ operator. Each $V_{i}$ is made up of at most two $\frac{\pi}{2}$ Clifford rotations, generated by Pauli operators, per element in  $\mathcal{W}$. In \cite{kirby2021contextual} it was shown that at most there will be $2n$ of these rotations, where $n$ is the number of qubits the problem is defined on \cite{nielsen2011quantum}. We can write this operator as:
\begin{equation}
    \zlabel{eq:rotation_U}
    \begin{aligned}
    U_{\mathcal{W}}^{\dagger}(\vec{q}, \vec{r}) &= \begin{cases}
       \prod_{P_{i} \in \mathcal{W}\subseteq \mathcal{W}_{all}} V_{i}(P_{i}) & \text{if}\ A(\vec{r}) \not\in \mathcal{W}  \\
       \Big(\prod_{P_{i} \in \mathcal{W}\subseteq \mathcal{W}_{all}} V_{i}(P_{i}) \Big)R  & \text{if } A(\vec{r}) \in \mathcal{W}.
    \end{cases}
    \end{aligned}
\end{equation}

Applying $U_{\mathcal{W}}^{\dagger}\mathcal{W}U_{\mathcal{W}} =\mathcal{W}^{Z}$ results in a set of single-qubit $Z$ Pauli operators. An implementation note is that each operator $V_{i}(P_{i})$ in $U_{\mathcal{W}}$ depends on the others. This can be seen by expanding $U_{\mathcal{W}}^{\dagger}\mathcal{W}U_{\mathcal{W}}$. Therefore each $V_{i}$ operator is dependent on the stabilizers in $\mathcal{W}$ and the order in which they occur. We recursively define each $V_{i}$ as follows:

\begin{enumerate}
  \item Set $\mathcal{W} = R\mathcal{W}R^{\dagger}$ if and only if $A(\vec{r}) \in \mathcal{W}$.
  \item Find the unitary $V_{0}$ mapping the first Pauli operator $P_{0} \in \mathcal{W}$ to a single qubit Pauli operator.
  \item Apply this operator to each operator in the set: $V_{0}\mathcal{W}V_{0}^{\dagger} = \mathcal{W}^{(0)}$.
  \item Find the unitary $V_{1}$ mapping $V_{0}P_{1}V_{0}^{\dagger} \in \mathcal{W}^{(0)}$ to a single-qubit $Z$ Pauli operator.
  \item Apply this operator to all operators in the set: $V_{1}\mathcal{W}^{(0)}V_{1}^{\dagger} = \mathcal{W}^{(1)}$
  \item Repeat this procedure from step (3) until all the operators are mapped to single qubit $Z$ Pauli operators: $\mathcal{W} \mapsto
 \mathcal{W}^{Z}$. 
\end{enumerate}

Finally, the eigenvalue of each single-qubit $Z$ Pauli stabilizer in $\mathcal{W}^{Z}$ is defined by the vector $\vec{q}$ of the noncontextual ground-state $(\vec{q}, \vec{r})$, note that $\langle A(\vec{r}) \rangle$ is fixed to $+1$ and thus $\vec{r}$ is not important here. $U_{\mathcal{W}}$ can flip the sign of these assignments, but it is efficient to classical determine by tracking how $U_{\mathcal{W}}$ affects the sign of the operators in $\mathcal{W}$.

To project the Hamiltonian into the subspace consistent with the noncontextual state, we first perform the following rotation $H_{\text{full}} \mapsto H_{\text{full}}^{'} = U_{\mathcal{W}}^{\dagger}H_{\text{full}}U_{\mathcal{W}}$. As this is a unitary transform, the resultant operator has the same spectrum as before. We then restrict the rotated Hamiltonian to the correct subspace by enforcing the eigenvalue of the operators in $\mathcal{W}^{Z}$, where the outcomes are defined by the noncontextual state. As each operator in $\mathcal{W}^{Z}$ only acts nontrivially on a unique qubit, each stabilizer fixes the state of that qubit to be either $\ket{0}$ or $\ket{1}$. We write this state as:

\begin{equation}
    \zlabel{eq:fixed_state}
    \begin{aligned}
    \ket{\psi_{\text{fixed}}} = \bigotimes_{P_{v} \in \mathcal{W}^{Z}} \ket{i}_{v} \begin{cases}
       i=0 & \text{if}\ \langle P_{v} \rangle = +1 \\
       i=1 & \text{if}\ \langle P_{v} \rangle = -1 ,
    \end{cases}
    \end{aligned}
\end{equation}
where $v$ indexes the qubit a given single-qubit stabilizer acts on and $ \langle P_{v} \rangle$ is defined by the noncontextual state. We can write the projector onto this state as:
\begin{equation}
    \zlabel{eq:Q_proj}
    \begin{aligned}
        Q_{\mathcal{W}} = \ket{\psi_{\text{fixed}}} \bra{\psi_{\text{fixed}}} \otimes \mathcal{I}_{(n-|\mathcal{W}^{Z}|)}
    \end{aligned}
\end{equation}
where $\mathcal{I}_{(n-|\mathcal{W}^{Z}|)}$ is the identity operator acting on the $(n-|\mathcal{W}^{Z}|)$ qubits not fixed by the single-qubit $P_{v}$ stabilizers. The action on a general state $\ket{\phi}$ is:
\begin{equation}
    \begin{aligned}
        Q_{\mathcal{W}}  \ket{\phi} = \ket{\psi_{\text{fixed}}} \bra{\psi_{\text{fixed}}}\phi\rangle \otimes \ket{\phi}_{(n-|\mathcal{W}^{Z}|)}
    \end{aligned}
\end{equation}
where $Q_{\mathcal{W}}$ has only fixed the state of qubits $v$ and thus each stabilizer $P_{v}$ removes $1$ qubit from the problem. As the states of these qubits are fixed, the expectation values of the single-qubit Pauli matrices indexed on qubits $v$ are known. Thus the Pauli operators in the rotated Hamiltonian $H_{\text{full}} \mapsto H_{\text{full}}^{'} = U_{\mathcal{W}}^{\dagger}H_{\text{full}}U_{\mathcal{W}}$ acting on these qubits can be updated accordingly and the Pauli matrices on qubits $v$ can be dropped. Any term in the rotated Hamiltonian that anticommutes with a fixed generator $P_{v}$ is forced to have an expectation value of zero and can be completely removed from the problem Hamiltonian. The resultant Hamiltonian acts on $|\mathcal{W}^{Z}|$ fewer qubits. We denote this operation as $H_{\text{full}} \mapsto H_{\text{full}}^{\mathcal{W}} = Q_{\mathcal{W}}^{\dagger} U_{\mathcal{W}}^{\dagger}H_{\text{full}}U_{\mathcal{W}} Q_{\mathcal{W}}$. The noncontextual approximation will be stored in the identity term of the problem and therefore does not need to be tracked separately.



The choice of which stabilizer eigenvalues to fix (i.e. what is included in $\mathcal{W}$) and which to allow to vary remains an open question of the CS-VQE algorithm. The number of possible stabilizer combinations will be $\sum_{i=1}^{|\mathcal{W}_{all}|} \binom{|\mathcal{W}_{all}|}{i} = 2^{|\mathcal{W}_{all}|}-1$. Rather than searching over all $2^{|\mathcal{W}_{all}|}-1$ combinations of stabilizers to fix, in this paper we use the heuristic given in \cite{kirby2021contextual}. This begins at the full noncontextual approximation, where $\mathcal{W}$ contains all possible stabilizers. We then add a qubit  to the quantum correction, by removing an operator from $\mathcal{W}$ and greedily choosing each pair that gives the lowest ground-state energy estimate \cite{kirby2021contextual}. Alternative strategies on how to do this remain an open question of CS-VQE. A possible way to approach this problem is to look at the priority of different terms in $H_{\text{con}}$ \cite{poulin2014trotter}. Note that the quality of the approximation is sensitive to which stabilizers are included in $\mathcal{W}$. When fewer stabilizers are considered (included in $\mathcal{W}$), the resultant rotated Hamiltonians will act on more qubits and approximate the true ground-state energy better.

In \cite{kirby2021contextual}, Kirby \textit{et al.} construct $R$ as a sequence of rotations (exponentiated Pauli operators) defined by $A(\vec{r})$ as in the unitary partitioning method \cite{zhao2019measurement, izmaylov2019unitary}. We denote this operation $R_{S}$. The Supplemental Material gives the full definition of this operator \cite{SuppMaterial}. If $R_{S}$ is considered as just an arbitrary sequence of exponentiated Pauli operator rotations, then the transformation $H_{\text{full}} \mapsto H_{\text{full}}^{R_{S}}= R_{S}H_{\text{full}}R_{S}^{\dagger}$ results in an operator whose terms have increased by a factor of $\mathcal{O}(2^{N})$, where $N$ is the number of cliques defined from $\mathcal{T}$ \cite{kirby2021contextual}. This presents a possible roadblock for the CS-VQE algorithm, as classically precomputing $U_{\mathcal{W}}^{\dagger} H_{\text{full}}U_{\mathcal{W}}$ could cause the number of terms to exponentially increase. We give a further analysis of this in the Supplemental Material \cite{SuppMaterial}. Additional structure between $R_{S}$ and $H_{\text{full}}$ can make the base of the exponent slightly lower; however, the scaling still remains exponential in the number of  qubits $n$, where $|\mathcal{A}| \leq 2n+1$ \cite{kirby2020classical}. The only case in which there is not an exponential increase in terms is for the trivial instance that $R_{S}$ commutes with $H_{\text{full}}$.  In the next section, we provide an alternative construction of $R$ via a linear combination of unitaries (LCU) that results in only a quadratic increase in the number of terms of the Hamiltonian when transformed. This avoids the need to apply the unitary partitioning operator $R$ (via a sequence of rotations) coherently in the quantum circuit after the ansatz circuit, which was proposed in \cite{kirby2021contextual}.

\subsection{\zlabel{sec:LinComb_P} Linear combination of unitaries construction of $R$}

In the unitary partitioning method \cite{zhao2019measurement, izmaylov2019unitary}, it was shown that $R$ could also be built as a linear combination of Pauli operators\cite{zhao2019measurement, ralli2021implementation}. We provide the full construction in the Supplemental Material \cite{SuppMaterial}. We denote the operator as $R_{LCU}$. Rotating a general Hamiltonian $H_{\text{full}}$ by this operation $R_{LCU}$ results in:

\begin{equation}
    \zlabel{eq:RHR_LCU}
\begin{aligned}
    R_{LCU} H_{\text{full}} R_{LCU}^{\dagger} = & \sum_{i}^{|H_{\text{full}}|} (\mu_{i}) P_{i} + \\
    &\sum_{j}^{|\mathcal{A}|-1}  \sum_{\substack{i \\  \forall \{P_{j}P_{k}, P_{i} \} =0}}^{|H_{\text{full}}|} \mu_{ij} P_{j}P_{k} P_{i} + \\
&\sum_{j}^{|\mathcal{A}|-1}\sum_{i}^{|H_{\text{full}}|} \sum_{\substack{l > j \\ \forall \{P_{i}, P_{j}P_{l} \}=0 }}^{|\mathcal{A}|-1}  \mu_{ijl} P_{i} P_{j} P_{l}.
\end{aligned}
\end{equation}
The Pauli operators $P_{j}$, $P_{k}$ and $P_{l}$ are operators in $\mathcal{A}$, further details are covered in the Supplemental Material \cite{SuppMaterial}. Overall, this unitary transformation causes the number of terms in the Hamiltonian to scale as $\mathcal{O}\big( \; |H_{\text{full}}|\cdot |\mathcal{A}|^{2} \; \big) $.  This scaling is quadratic in the size of $\mathcal{A}$ and as $|\mathcal{A}| \leq 2n+1$ \cite{kirby2020classical}, the number of terms in the rotated system will at worst scale quadratically with the number of qubits $n$. In a different context, this scaling result was also obtained for involutory linear combinations of entanglers \cite{lang2020unitary}. Overall, unlike the sequence of rotations approach, this non-Clifford operation doesn't cause the number of terms in a Hamiltonian to increase exponentially.

The transformation given in Equation  \zref{eq:RHR_LCU} $H_{\text{full}} \mapsto H_{\text{full}}^{LCU} = R_{LCU}^{\dagger} H_{\text{full}} R_{LCU}$ is performed classically in CS-VQE. This is efficient to do because it just involves Pauli operator multiplication, which can be done symbolically or via a symplectic approach \cite{dehaene2003clifford}. This operation could be applied within the quantum circuit. However, in contrast to the deterministic sequence of rotations approach, this implementation would be probabilistic as it requires post selection on an ancillary register \cite{ralli2021implementation, zhao2019measurement, wiebe2012hamiltonian,  Low2019hamiltonian}. Amplitude amplification techniques could improve this, but would require further coherent resources \cite{OblivAmp14, grover1997quantum,guerreschi2019repeat, boyer1998tight}. Performing this transformation in a classical pre-processing step therefore reduces the coherent resources required and at worst increases the number of terms needing measuring quadratically with respect to the number of qubits.

\subsection{\zlabel{sec:CS_summary}CS-VQE implementation}

In \cite{kirby2021contextual}, $ U_{\mathcal{W}}(\vec{q}, \vec{r})$ was fixed to include all the stabilizers of the noncontextual ground-state $\mathcal{W} \equiv \mathcal{W}_{all}$ (Equation \zref{eq:stabilizers_set_nonrot}), rather than possible subsets $\mathcal{W} \subseteq \mathcal{W}_{all}$. The whole Hamiltonian was mapped according to $H_{\text{full}} \mapsto H_{\text{full}}^{'} = U_{\mathcal{W}_{all}}^{\dagger}H_{\text{full}}U_{\mathcal{W}_{all}}$. In general, $A(\vec{r}) \in \mathcal{W}$ and $U_{\mathcal{W}_{all}}$ will therefore normally include the unitary partitioning operator $R$. The problem with this approach is that the unitary $R$ is not a Clifford operation and the transformation can cause the number of terms in the Hamiltonian to increase. This increase is exponential if $R_{S}$ is used and quadratic if $R_{LCU}$ is employed. As this step can generate more terms, $R$ should only be included in $U_{\mathcal{W}}$ if the eigenvalue of $A(\vec{r})$ is fixed to $+1$, otherwise it is a redundant operation as the spectrum of the operator rotated by $R$ is unchanged. We therefore modify the CS-VQE algorithm to  construct $U_{\mathcal{W}}$ from the CS-VQE noncontextual generator eigenvalues that are fixed. This means that $\mathcal{W} \subseteq \mathcal{W}_{all}$ and ensures that the number of terms can only increase if the eigenvalue of $A(\vec{r})$ is fixed.

\section{Numerical Results}
\zlabel{sec:num_results}
We describe the method in Section \zref{sec:Method} and then split our results into the two remaining sections. First, we explore a toy problem, showing the steps of the CS-VQE algorithm. We show how classically applying $R$ without fixing the eigenvalue of $A(\vec{r})$ to $+1$ can unnecessarily increase the number of terms in a Hamiltonian without changing its spectrum. Finally, in Section \zref{sec:M_reduction} we apply measurement reduction combined with CS-VQE to a set of electronic structure Hamiltonians and show that this can significantly reduce the number of terms requiring separate measurement. The raw data for these results are supplied in the Supplemental Material \cite{SuppMaterial}.

\subsection{\zlabel{sec:Method} Method}

We investigated the same electronic structure Hamiltonians considered in the original CS-VQE paper \cite{kirby2021contextual}. All molecules considered had a multiplicity of $1$ and thus a singlet ground-state. The same qubit tapering was performed to remove the $\mathbb{Z}_{2}$ symmetries \cite{bravyi2017tapering}. For each tapered Hamiltonian, we generate a set of reduced Hamiltonians  $\{Q_{\mathcal{W}}^{\dagger}U_{\mathcal{W}}^{\dagger}H_{\text{full}}U_{\mathcal{W}} Q_{\mathcal{W}} \}$ where the size of $\mathcal{W}$ varies from $1$ to $|\mathcal{W}_{all}|$, representing differing noncontextual approximations, as summarized in Section \zref{sec:CS_summary}. To generate the different CS-VQE Hamiltonians, we modify the original CS-VQE source code used in \cite{kirby2021contextual, ContextualSubspaceVQEgithub}. The code was modified to implement the unitary partitioning step of CS-VQE if and only if the eigenvalue of $A(\vec{r})$ was fixed. This ensured that the number of terms in the rotated Hamiltonian did not increase unnecessarily, as described in Section \zref{sec:CS_summary}. 

For each electronic structure Hamiltonian generated in this way, we then apply the unitary partitioning measurement reduction scheme to further reduce the number of terms requiring separate measurement \cite{zhao2019measurement, izmaylov2019unitary, ralli2021implementation}. Partitioning into anticommuting sets was performed using NETWORKX \cite{SciPyProceedings_11}. A graph of the qubit Hamiltonian is built, where nodes represent Pauli operators and edges are between nodes that commute. A graph coloring can be used to find the anticommuting cliques of the graph. This searches for the minimum number of colors required to color the graph, where no neighbors of a node can have the same color as the node itself. The “largest first” coloring strategy in NETWORKX was used in all cases \cite{SciPyProceedings_11, welsh1967upper}.

We calculate the ground-state energy of each Hamiltonian in this paper by directly evaluating the lowest eigenvalues. This was achieved by diagonalizing them on a conventional computer.

\subsection{\zlabel{sec:example_will} Toy example}

\begin{figure}[t]
\centering
\includegraphics[scale=0.55]{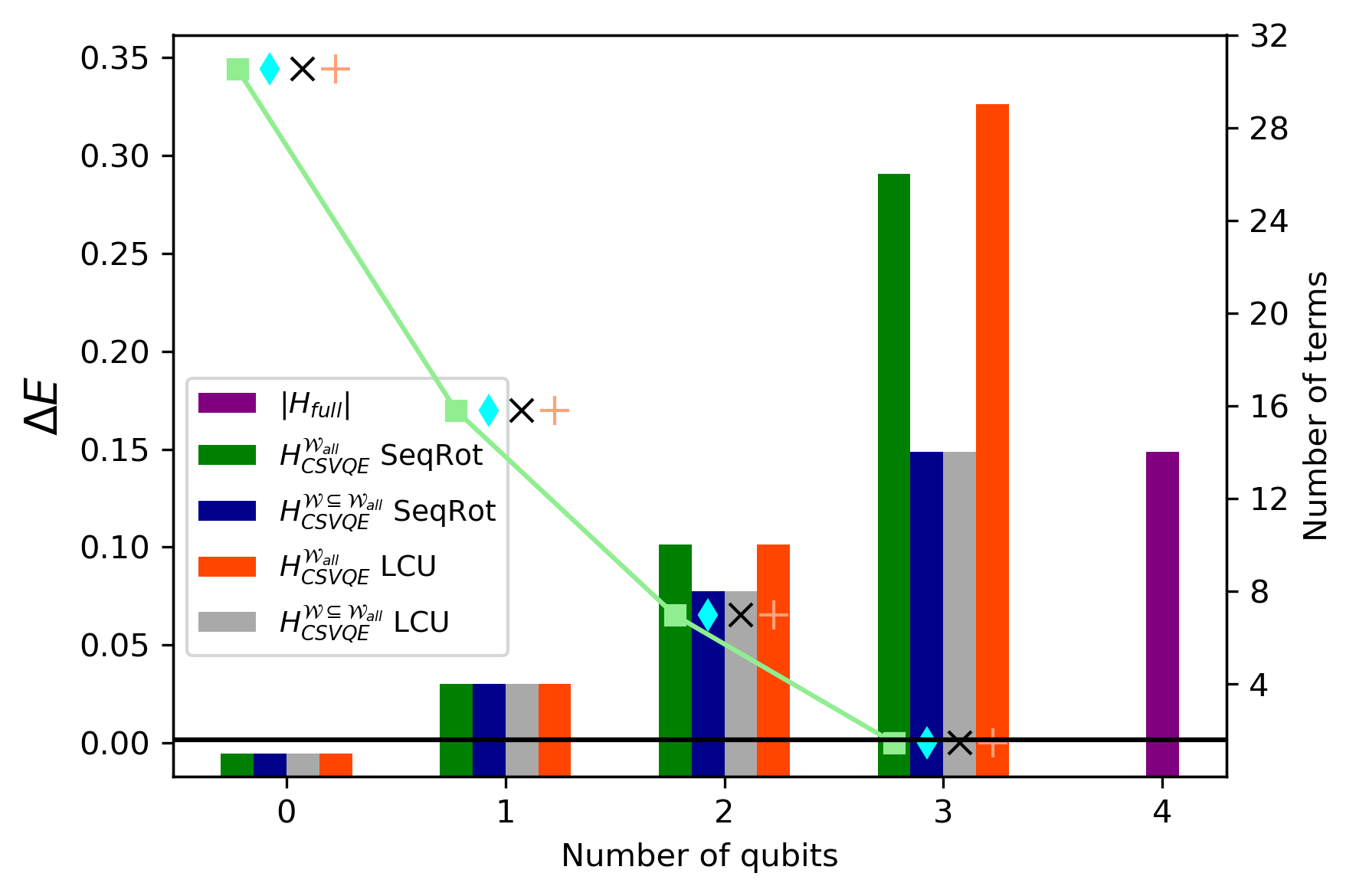}
\caption{ground-state energy and the number of terms of different contextual subspace projected Hamiltonians generated in CS-VQE. Each Hamiltonian has been transformed as $Q_{\mathcal{W}}^{\dagger}U_{\mathcal{W}}^{\dagger}HU_{\mathcal{W}}Q_{\mathcal{W}}$, apart from the $4$ qubit case, which is the full $H$. The scatter plot is associated with the left-hand y-axis and gives the energy error as: $\Delta E = |E_{approx} - E_{true}| $. The bar chart gives the number of terms in each Hamiltonian and is associated with the right-hand y-axis. From left to right the following generators are fixed:  $\{YIYI, IXYI, IIIZ, \mathcal{A}(\vec{r}_{0})\}$, $\{IXYI, IIIZ, \mathcal{A}(\vec{r}_{0}) \}$, $\{IXYI, IIIZ \}$, $\{IIIZ \}$ and $\{ \}$. The $\mathcal{W} = \{ \}$ case represents standard full VQE over the full problem. The $0$ qubit case presents the scenario where the problem is fully noncontextual and no quantum correction is made. The full details as to how each Hamiltonian is built is provided in the Supplemental Material \cite{SuppMaterial}. The horizontal black line indicates an absolute error of $1.6 \times 10^{-3}$. SeqRot, sequence of rotations}
\zlabel{fig:toy_example_plot}
\end{figure}

\begin{table}[t]
\centering
\small
\begin{tabular}{ccc}
\hline
Molecule &  Basis  & Number of gates for $R_{S}$ \\
\hline
    \ce{BeH2}& (STO-3G)  &    [90, 72] \\
    \ce{Mg}  & (STO-3G)  &  [189, 162] \\
    \ce{H3+} & (3-21G)   &  [209, 176] \\
    \ce{O2}  & (STO-3G)  &  [184, 160] \\
    \ce{OH-}  & (STO-3G)  &   [104, 80] \\
    \ce{CH4} & (STO-3G)  &  [325, 286] \\
    \ce{Be}  & (STO-3G)  &     [14, 8] \\
    \ce{NH3} & (STO-3G)  &  [299, 260] \\
    \ce{H2S} & (STO-3G)  &   [120, 96] \\
    \ce{H2}  & (3-21G)   &    [66, 48] \\
    \ce{HF}  & (3-21G)   &  [735, 672] \\
    \ce{F2}  & (STO-3G)  &  [133, 112] \\
    \ce{HCl} & (STO-3G)  &    [36, 24] \\
    \ce{HeH+}& (3-21G)   &    [88, 64] \\
    \ce{MgH2}& (STO-3G)  &  [403, 364] \\
    \ce{CO}  & (STO-3G)  &  [325, 286] \\
    \ce{LiH} & (STO-3G)  &    [36, 24] \\
    \ce{N2}  & (STO-3G)  &  [207, 180] \\
    \ce{NaH} & (STO-3G)  &  [493, 442] \\
    \ce{H2O} & (STO-3G)  &   [120, 96] \\
    \ce{H3+} & (STO-3G)  &      [3, 0] \\
    \ce{LiOH}& (STO-3G)  &  [378, 336] \\
    \ce{LiH} & (3-21G)   &  [459, 408] \\
    \ce{H2}  & (6-31G)   &    [66, 48] \\
    \ce{NH4+}& (STO-3G)  &  [325, 286] \\
    \ce{HF}  & (STO-3G)  &    [36, 24] \\         
\hline
\end{tabular}
\caption{Gate requirements to implement $R$ as a sequence of rotations in the unitary partitioning measurement reduction step. The square tuple gives the upper bound on the number of single qubit and CNOT gates  required - $[single, CNOT]$. These resource requirements  are based on the largest anticommuting clique of each Hamiltonian, as these have the largest circuit requirements for $R_{S}$.\zlabel{tab:gate_R}} 
\end{table}

We consider the qubit Hamiltonian:

\begin{equation}
    \zlabel{eq:H_full_Example}
\begin{aligned}
    H =\: &0.6\:IIYI + 0.7\: XYXI + 0.7\: XZXI + 0.6\: XZZI +\\
        &0.1\:YXYI + 0.7\: ZZZI + 0.5\:IIIZ + 0.1\: XXXI + \\
        &0.5\:XXYI + 0.2\: XXZI + 0.2\:YXXI + 0.2\: YYZI +\\
        &0.1\:YZXI + 0.1\: ZYYI,
\end{aligned}
\end{equation}
and use it to exemplify the steps of the CS-VQE algorithm. The results are reported to three decimal places and full numerical details can be found in the Supplemental Material \cite{SuppMaterial}.

Following the CS-VQE procedure \cite{kirby2021contextual}, we first split the Hamiltonian into its contextual and noncontextual parts (Equation \zref{eq:Hamilt}):

\begin{subequations}
\zlabel{eq:H_con_noncon_example}
    \begin{equation}
        \begin{aligned}
         H_{\text{noncon}} = \: &0.5\:\underbrace{IIIZ}_{\mathcal{Z}} + \\
                         &0.7\: XZXI + 0.7\:ZZZI + \\
                         &0.1\: YXYI + 0.6\: IIYI +  \\
                         &\underbrace{0.7\: XYXI + 0.6\:XZZI}_{\mathcal{T}}
        \end{aligned}
    \end{equation}
    \begin{equation}
        \begin{aligned}
             H_{\text{con}} =\: &0.1\:XXXI + 0.5\: XXYI + 0.2\: XXZI + \\
            &0.2\: YXXI + 0.2\:YYZI + 0.1\: YZXI + \\ 
            &0.1\:ZYYI.
        \end{aligned}
    \end{equation}
\end{subequations}

\noindent Each row after the first in Equation \zref{eq:H_con_noncon_example}a, is a clique of $\mathcal{T}$. From here, we define the set $\mathcal{R}$ (Equation \zref{eq:R_ind_set}):

\begin{equation}
    \zlabel{eq:R_example}
\begin{aligned}
   \mathcal{R} = \underbrace{\{ YIYI, IXYI , IIIZ\}}_{\mathcal{G}} \cup \underbrace{\{ XZXI, YXYI, XYXI \}}_{ \{ P_{0}^{(j)} | j=0,1,\hdots, N-1 \}}.
\end{aligned}
\end{equation}
Note how different combinations of the operators in Equation \zref{eq:R_example} allow all the operators in $H_{\text{noncon}}$ (Equation \zref{eq:H_con_noncon_example}a) to be inferred under the Jordan product, defined as: $P_{a} \circ P_{b} = \frac{\{P_{a}, P_{b} \}}{2}$.  Basically, the Jordan product is equal to the regular matrix product if the operators commute, and equal to zero if the operators anticommute. Next the noncontextual problem was solved.

\begin{figure*}[t]
\centering
\includegraphics[scale=0.90]{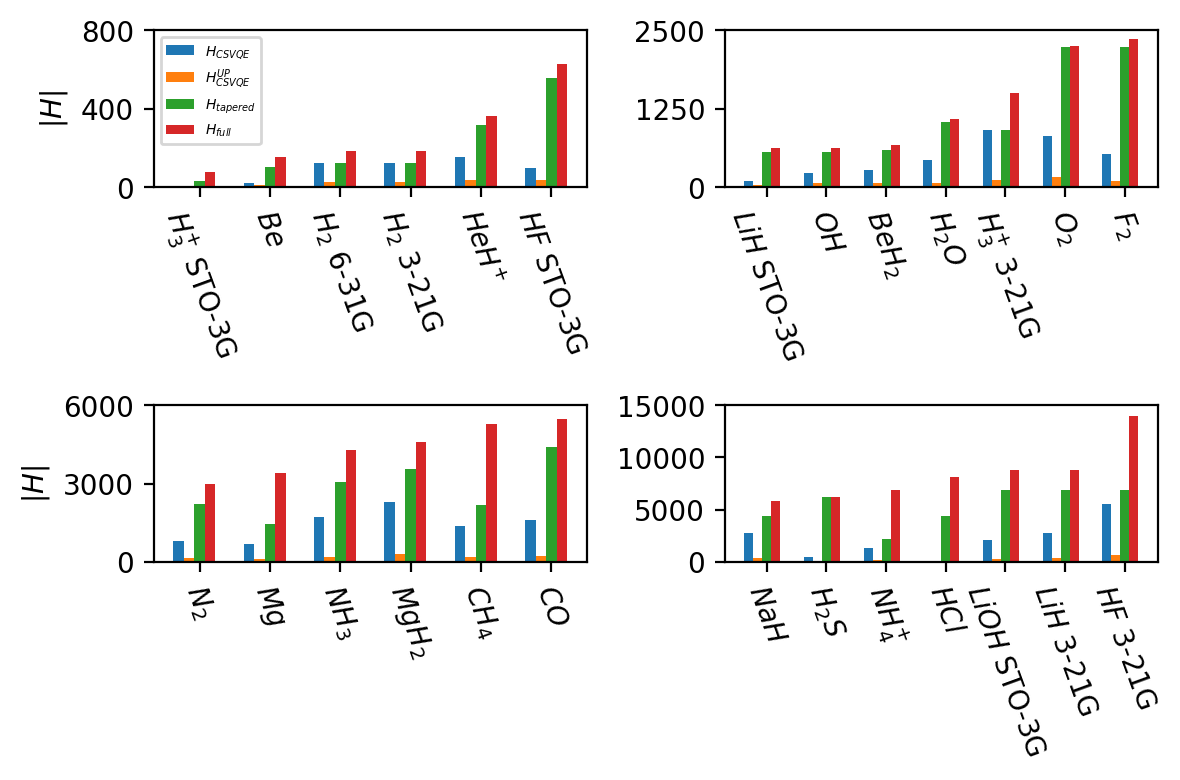}
\caption{Number of Pauli operators requiring separate measurement to determine the ground-state energy of a particular molecular Hamiltonian to within chemical accuracy. For each molecule the full Hamiltonian, tapered Hamiltonian, CS-VQE and CS-VQE with unitary partitioning measurement reduction applied are given. Full numerical numerical details of each are provided in the Supplemental Material \cite{SuppMaterial}. The size of the Hamiltonian for \ce{LiH} (3-21G singlet) with measurement reduction applied is different for the sequence of rotations and LCU  unitary partitioning methods. This is an artifact of the graph color heuristic finding different anticommuting cliques in the CS-VQE Hamiltonian.}
\zlabel{fig:N_terms_plot}
\end{figure*}

The expectation value for $H_{\text{noncon}}$ can be induced, by setting the expectation values of operators in $\mathcal{R}$ (Equation \zref{eq:R_example}), as the Pauli operators in $H_{\text{noncon}}$ are generated by $\mathcal{R}$ under the Jordan product. The expectation value of each operator in $H_{\text{noncon}}$ can therefore be inferred without contradiction. To find the ground-state of $H_{\text{noncon}}$, we 
checked all possible $\pm 1$ expectation values for each $G_{j}$ ($2^{3}=8$ possibilities). For each possible $\pm 1$ combination, the energy was minimized with respect to the unit vector $\vec{r}$, which sets the expectation value for each $ \langle P_{0}^{(j)} \rangle=r_{j}$. The vector $(\vec{q},\vec{r})$ that was found to give the lowest energy defines the noncontextual ground-state. In this case the ground-state is:

\begin{equation}
    \zlabel{eq:ground_state_example}
\begin{aligned}
    (\: \underbrace{-1, +1, -1}_{\vec{q}_{0}}, \underbrace{+0.253, -0.658, -0.709}_{\vec{r}_{0}} \:).
\end{aligned}
\end{equation}
This noncontextual state defines the operator $ \mathcal{A}(\vec{r}_{0})$:

\begin{equation}
    \zlabel{eq:script_A_example}
\begin{aligned}
    A(\vec{r}_{0}) = \: &0.253\: YXYI -0.658\:XYXI  -0.709\:XZXI.
\end{aligned}
\end{equation}
From this we can write $\mathcal{R}_{\vec{r}}$ (Equation \zref{eq:R_ind_set_scriptA})

\begin{equation}
    \zlabel{eq:R_ind_set_example}
    \begin{aligned}
    \mathcal{R}_{\vec{r}} &\equiv \{ A(\vec{r}_{0}) \} \cup \{ YIYI, IXYI, IIIZ \} 
    \end{aligned}
\end{equation}
To map $A(\vec{r}_{0})$ to a single Pauli operator we use unitary partitioning \cite{zhao2019measurement, izmaylov2019unitary, ralli2021implementation}. The required unitary can be constructed as either a sequence of rotations \cite{zhao2019measurement},

\begin{equation}
\zlabel{eq:Rs_example}
     R_{S} = \:  e^{-1i \cdot 0.788 \cdot ZYZI}  \cdot e^{+1i \cdot 1.204 \cdot ZZZI},
\end{equation}
or linear combination of unitaries \cite{zhao2019measurement},

\begin{equation}
\zlabel{eq:Rlcu_example}
     R_{LCU} =\: 0.792 \: IIII + 0.416i \: ZZZI - 0.448i \: ZYZI.
\end{equation}
These operators perform the following reduction: $R_{S} A(\vec{r}_{0})  R_{S}^{\dagger} = R_{LCU} A(\vec{r}_{0})  R_{LCU}^{\dagger} = YXYI$.

 If the eigenvalue of $A(\vec{r}_{0})$ is fixed, then we should consider $\mathcal{R}_{\vec{r}}$ (Equation \zref{eq:R_ind_set_example}) under the unitary transform $R_{LCU}$ or $R_{S}$ (Equation \zref{eq:R_ind_set_scriptA}):

\begin{equation}
    \zlabel{eq:R_ind_set_scriptA_example}
    \begin{aligned}
    \mathcal{R}^{'} &= R_{S/LCU} A(\vec{r})  R_{S/LCU}^{\dagger} \cup \{ YIYI, IXYI, IIIZ \} \\
    &= \{ YXYI \} \cup \{ YIYI, IXYI, IIIZ \}.
    \end{aligned}
\end{equation}
Equations \zref{eq:ground_state_example},  \zref{eq:R_ind_set_example} and \zref{eq:R_ind_set_scriptA_example} define the noncontextual stabilizers:
\begin{equation}
    \zlabel{eq:stabilizers_set_nonrot_example}
    \begin{aligned}
    \mathcal{W}_{all}  &\equiv \{ +1\: A(\vec{r}_{0})\:\:\:, -1\:YIYI, +1\:IXYI, -1\:IIIZ \}, \\
    \mathcal{W}_{all}' &\equiv \{ +1\:YXYI, -1\:YIYI, +1\:IXYI, -1\:IIIZ \}.
    \end{aligned}
\end{equation}

Next, we define different $U_{\mathcal{W}}$ (Equation \zref{eq:rotation_U}), depending on which stabilizers $\mathcal{W}$ we wish to fix. For this problem we found the optimal ordering of which stabilizers to fix to be $\{-1\,YIYI, +1\,IXYI, +1\,\mathcal{A}(\vec{r}_{0}), -1\,IIIZ\}$ followed by $\{+1\,IXYI, -1\,IIIZ, +1\,\mathcal{A}(\vec{r}_{0})\} $ followed by $\{+1\,IXYI, -1\,IIIZ\}$ followed by $\{-1\,IIIZ\}$. This was achieved by a brute force search over all $\sum_{i=1}^{|\mathcal{W}_{all}|} \binom{|\mathcal{W}_{all}|}{i} = 2^{4}-1 = 15$ possibilities for $\mathcal{W}$.

The members of the resulting set of four different $\mathcal{W}$ each represent different noncontextual approximations. These give four different $U_{\mathcal{W}}$ built according to Equation \zref{eq:rotation_U}. The full definition of each operator is given in the Supplemental Material \cite{SuppMaterial}. 

Taking a specific example, for \mbox{$\mathcal{W} = \{+IXYI, -IIIZ, +\mathcal{A}(\vec{r}_{0})\}$} we define $U_{\mathcal{W}}^{\dagger}$ (Equation \zref{eq:rotation_U}). This operator transforms $\mathcal{W}$ as $\mathcal{W}^{Z} = U_{\mathcal{W}}^{\dagger}\mathcal{W}U_{\mathcal{W}} = \{+IZII, -IIIZ, +IIZI\}$. The eigenvalues of the operators in $\mathcal{W}^{Z}$ are fixed by the noncontextual state to be $\langle IZII \rangle = +1$, $\langle IIZI \rangle = +1$, $\langle IIIZ \rangle = -1$. This defines the projector:

\begin{equation}
    \zlabel{eq:Q_example}
    \begin{aligned}
        Q_{\mathcal{W}} &= \Big( \underbrace{\ket{0}\bra{0}  + \ket{1}\bra{1}}_{\mathcal{I}_{(n-|\mathcal{W}^{Z}|)}} \Big) \otimes  \underbrace{\ket{0}\bra{0} \otimes  \ket{0}\bra{0}  \otimes  \ket{1}\bra{1}}_{\ket{\psi_{\text{fixed}}}} \\
        &= I \otimes \ket{001} \bra{001} .
    \end{aligned}
\end{equation}
The reduced Hamiltonian is therefore

 \begin{equation}
    \zlabel{eq:H_U3_example}
    \begin{aligned}
    H  \mapsto H_{\mathcal{W}}^{LCU} &= Q_{\mathcal{W}}^{\dagger} U_{\mathcal{W}}^{\dagger \: (LCU)}H_{\text{full}}U_{\mathcal{W}}^{(LCU)} Q_{\mathcal{W}} \\
    &= -1.827\:I -0.414\:X - 0.292\:Z + 0.648\:Y.
    \end{aligned}
\end{equation}
The Supplemental Material gives further details about this operation and provides the specifics for the other projected Hamiltonians \cite{SuppMaterial}.

\begin{figure*}[t]
\centering
\includegraphics[scale=0.70]{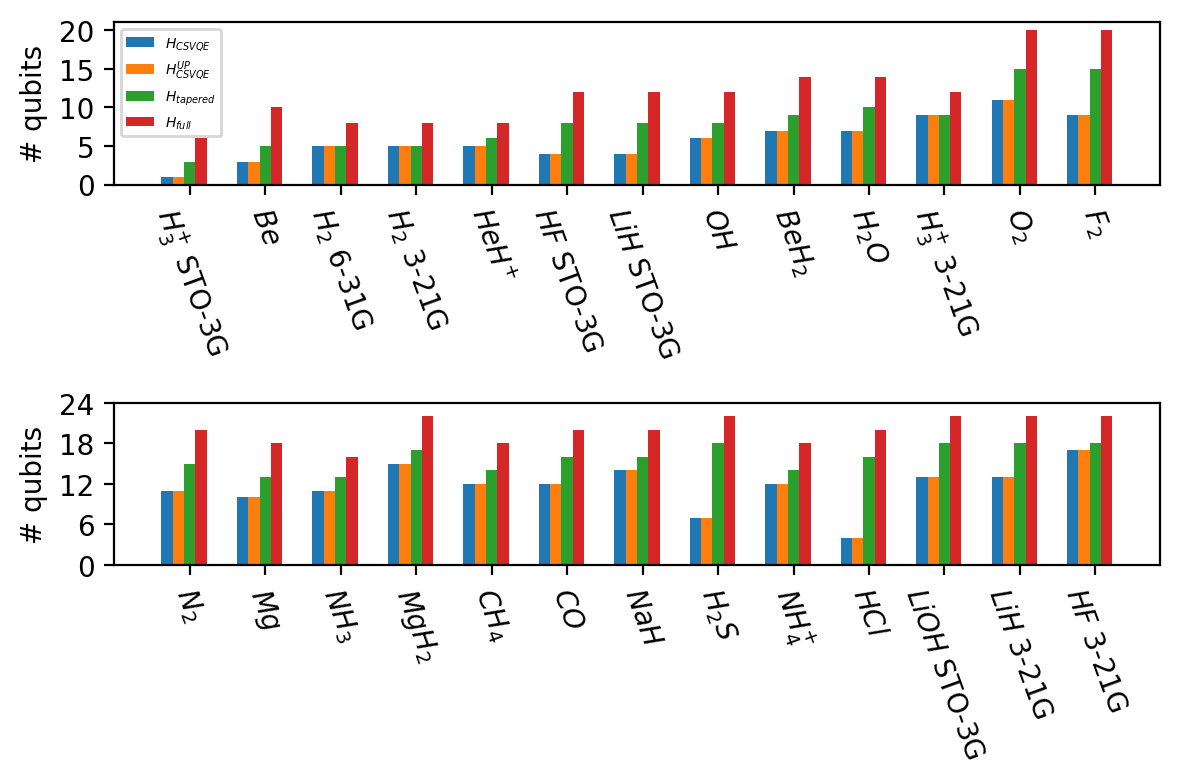}
\caption{Number of qubits required to simulate different electronic structure Hamiltonians in order to achieve chemical accuracy. For each molecule the full Hamiltonian, tapered Hamiltonian, CS-VQE and CS-VQE with unitary partitioning measurement reduction applied are given. Numerical details for each result are provided in the Supplemental Material.}
\zlabel{fig:N_qubits_plot}
\end{figure*} 

Overall four Hamiltonians are generated, representing different levels of approximation, that act on $0$, $1$, $2$ and $3$ qubits respectively. The $4$ qubit case represents the standard VQE on the full Hamiltonian. Figure \zref{fig:toy_example_plot} summarizes the error $\Delta E$ of each of these compared with the true ground-state energy (scatter plot). The number of terms in each Hamiltonian is given by the bar chart. The green and orange results have $\mathcal{W} \equiv \mathcal{W}_{all}$ for all cases and represent the old CS-VQE implementation. For these results, in the $3$ and $4$ qubit Hamiltonians have an increased number of terms due to $R_{S/LCU}$ being implemented, even though the eigenvalue of $\mathcal{A}(\vec{r}_{0})$ is not being fixed to $+1$. On the other hand, the gray and blue results in Figure \zref{fig:toy_example_plot} build $U_{\mathcal{W}}$ according to Equation \zref{eq:rotation_U}, where $\mathcal{W} \subseteq \mathcal{W}_{all}$. This approach ensures that $R_{S/LCU}$ is only applied when necessary.

\subsection{\zlabel{sec:M_reduction} Measurement reduction}

Figures \zref{fig:N_terms_plot} and \zref{fig:N_qubits_plot} summarize the results of applying the unitary partitioning measurement reduction strategy to a set of electronic structure Hamiltonians We report the number of terms and number of qubits in each Hamiltonian required to achieve chemical accuracy compared with the original problem. The Supplemental Material gives further information about each result, where the different levels of noncontextual approximation are shown \cite{SuppMaterial}. As previously discussed in \cite{kirby2021contextual}, even though CS-VQE in general is an approximate method, chemical accuracy can still be achieved using significantly fewer qubits. Applying unitary partitioning on-top of the reduced CS-VQE Hamiltonians required to achieve chemical accuracy can further reduce the number of terms by roughly an order of magnitude. This is consistent with the previous results in \cite{ralli2021implementation}.

To actually obtain a measurement reduction, one needs to show that the number of measurement required to measure the energy of a molecular system, to a certain precision $\epsilon$, is reduced. Currently, Figure \zref{fig:N_terms_plot} only shows that we have reduced the number of Pauli terms being measured. We have not commented on the variance. In the Supplemental Material \cite{SuppMaterial}, we prove that simultaneous measurement of normalized anticommuting cliques can never do worse than performing no measurement reduction and will more often than not give an improvement. The proof given is state independent. There are other measurement strategies based on grouping techniques, such as splitting a Hamiltonian into commuting or qubit-wise commuting cliques \cite{crawford2021efficient, gokhale2019n, gokhale2019minimizing, verteletskyi2020measurement, kandala2017hardware}. The measurement reduction obtained from these methods is more complicated, as the covariance of operators within a clique must be carefully accounted for \cite{mcclean2016theory, gokhale2019minimizing}. This is one of the reasons we do not analyze the performance of these strategies in this paper. Many other measurement methods have also been proposed \cite{izmaylov2019revising, cotler2020quantum, bonet2019nearly, jena2019pauli, huggins2019efficient, crawford2021efficient, huang2020predicting, hadfield2020measurements, huang2021efficient, rubin2018application, gonthier2022measurements} and their effect on the number of measurements would be interesting to investigate.

In Table \zref{tab:gate_R}, we report the upper bound on the gate count required to implement measurement reduction as a sequence of rotations. The LCU method would require ancilla qubits and analysis of the circuit depth is more complicated. Further analysis can be found in \cite{ralli2021implementation}. The number of extra coherent resources required to implement unitary partitioning measurement reduction is proportional to the size of each anticommuting clique a Hamiltonian is split into \cite{ralli2021implementation, zhao2019measurement}. The sequence of rotations circuit depth scales as $\mathcal{O}\big(N_{s}(|C|-1)\big)$ single qubit and $\mathcal{O}\big(N_{s}(|C|-1)\big)$ CNOT gates, where $N_{s}$ is the number of system qubits and $|C|$ is the size of the  anticommuting clique being measured. Table \zref{tab:gate_R} reports the gate count upper bound for the largest anticommuting clique of a given CS-VQE Hamiltonian. We do not consider possible circuit simplifications, such as gate cancellations. To decrease the depth of quantum circuit required for practical application, we suggest finding nonoptimal clique covers; for example, if anticommuting cliques are fixed to a size of $2$, the resources required to perform $R_{S}$ are experimentally realistic for current and near-term devices, as only $\mathcal{O}(N_{s})$ single qubit and $\mathcal{O}(N_{s})$ CNOT gates are required \cite{ralli2021implementation}. 

The heuristic used to determine the operators in $H_{\text{noncon}}$ selected terms in the full Hamiltonian greedily by coefficient magnitude, while keeping the set noncontextual \cite{kirby2020classical}. The Hamiltonians studied here had weights dominated by diagonal Pauli operators, as the Hartree-Fock approximation accounts for most of the energy. This heavily constrains the operators allowed in $\mathcal{A}$. For the electronic structure Hamiltonians considered in this paper, we found in all cases that $|\mathcal{A}|=2$. In general, we do expect more commuting terms in $H_{\text{noncon}}$ than anticommuting terms. This is because there are more possible commuting Pauli operators defined on $n$ qubits compared with anticommuting operators ( $2^{n}$ vs $2n+1$). $\mathcal{G}$  will therefore in general be the larger contributor to the superset $\mathcal{R}$ (Equation \zref{eq:R_ind_set}). 

In Figure \zref{fig:N_terms_plot}, the CS-VQE bars have not been split into two for the case when $R$ is constructed as $R_{LCU}$ or $R_{S}$. This is due to $|\mathcal{A}|$ being $2$ in all cases, which is the special case when these operators ($R_{LCU}$ and $R_{S}$) end up being identical. In this instance $R$ has the form $R = \alpha I + i\beta P$ and thus the number of terms will only increase for every term in the Hamiltonian that $P$ anticommutes with. However, in general $|\mathcal{A}|$ will be greater than $2$ and the effect of $R$ can dramatically affect the number of terms in the resultant rotated Hamiltonian. We observe this in Fig. \zref{fig:toy_example_plot} of the toy example, where the $2$ and $3$ qubit CS-VQE Hamiltonians have had $U_{\mathcal{W}_{all}}$ applied to them even though the eigenvalue of $A(\vec{r})$ is not fixed. In that example, for the $3$ qubit approximation the sequence-of-rotations rotated Hamiltonian (green) actually has fewer terms than the LCU rotated operator (orange). This result is an artifact of the small problem size. In the Supplemental Material we show that the scaling will favor the LCU implementation, where the number of terms in a Hamiltonian can only increase quadratically, not exponentially, when performing the unitary partitioning rotation as a LCU rather than a sequence of rotations \cite{SuppMaterial}.

In the Supplemental Material, we show the convergence of CS-VQE at different noncontextual approximations. The results illustrate that CS-VQE can converge to below chemical accuracy well before the case when no noncontextual approximation is made (full VQE). Results beyond convergence are included to show the different possible levels of approximation. In practice knowledge of the true ground-state energy is not known \textit{a priori} and so using chemical precision to motivate the noncontextual approximation will not be possible. In this setting, a way to approach quantum advantage is to note that  CS-VQE is a variational method. The quantum resources required can be expanded until the energy obtained by CS-VQE is lower than that coming from the best possible classical method. At this point, either the algorithm can be terminated or further contextual corrections can be added until the energy converges, at which point the algorithm should be stopped.

\section{Conclusion}
\zlabel{sec:conclusion}
The work presented here shows that combining the unitary partitioning measurement reduction strategy with the CS-VQE algorithm can further reduce the number of terms in the projected Hamiltonian requiring separate measurement by roughly an order of magnitude for a given molecular Hamiltonian. The number of qubits needed to achieve chemical accuracy in most cases was also dramatically decreased, for example the \ce{H2S} (STO-3G singlet) problem was reduced to $7$ qubits from $22$.

We also improve two parts of the CS-VQE algorithm. First, we avoid having to apply the unitary partitioning operator $R$ after the ansatz which averts the potential exponential increase in the number of Pauli operators of the CS-VQE Hamiltonian caused by classically computing the non-Clifford rotation of the full Hamiltonian when $R$ defined as a sequence of rotations \cite{kirby2021contextual, zhao2019measurement}. We show that applying this operation as a linear combination of unitaries \cite{zhao2019measurement}:  $H_{\text{full}} \mapsto H_{\text{full}}^{LCU'} = R_{LCU}^{\dagger} H_{\text{full}} R_{LCU}$, results in the number terms at worst increasing quadratically with the number of qubits. This result makes classically precomputing this transformation tractable and $R$ no longer needs to be performed coherently after the ansatz. Secondly, we define the unitary $U_{\mathcal{W}}$, which maps each stabilizer in $\mathcal{W}_{all}$ (equation \zref{eq:stabilizers_set_rot}) to a distinct single-qubit Pauli matrix, according to which stabilizer eigenvalues are fixed by the noncontextual state. This ensures that the non-Clifford rotation required by CS-VQE is only applied when necessary and also reduces the number of redundant Clifford operations that are classically performed.

There are still several open questions for the CS-VQE algorithm. We summarize a few here. (1) What is the best  optimization strategy to use when minimizing the energy over $(\vec{q}, \vec{r})$ in the classical noncontextual problem? (2) What heuristic is best to construct the largest $|H_{\text{noncon}}|$? (3) How can we efficiently determine which noncontextual stabilizers to fix while maintaining low errors? In this paper, the size of each electronic structure problem allowed us to classically compute the ground-state energies at each step, but if this is not possible then  VQE calculations would be required. However, as each run requires fewer qubits and  decreases the number of terms requiring separate measurement this approach may overall still be less costly than performing VQE over the whole problem, especially when combined with further measurement reduction strategies. (4) What are the most important terms to include in $H_{\text{con}}$ or equivalently in $H_{\text{noncon}}$? Currently, it is not known whether $|H_{\text{noncon}}|$ should be maximized or whether selecting high priority terms \cite{poulin2014trotter} from the whole Hamiltonian results in a better approximation for a given problem. We leave these questions to future work.

We have written an open-source CS-VQE code that includes all the updated methodology discussed in this paper. We welcome readers to make use of this, which is freely available on GitHub \cite{symmerGithub}.

\begin{acknowledgments}
A. R. and T.W. acknowledge  support from the Unitary Fund and the Engineering and Physical Sciences Research Council (Grants No. EP/L015242/1 and No. EP/S021582/1 respectively). T.W. also acknowledges support from CBKSciCon Ltd., Atos, Intel and Zapata. W.M.K. and P.J.L. acknowledge  support  by the NSF STAQ project (Grant No. PHY-1818914). W.M.K. acknowledges support from the National Science Foundation, Grant No. DGE-1842474. P.V.C. is grateful for funding from the European Commission for VECMA (800925) and EPSRC for SEAVEA (Grant No. EP/W007711/1).
\end{acknowledgments}

\bibliography{references}

\begin{thebibliography}{64}%
\makeatletter
\providecommand \@ifxundefined [1]{%
 \@ifx{#1\undefined}
}%
\providecommand \@ifnum [1]{%
 \ifnum #1\expandafter \@firstoftwo
 \else \expandafter \@secondoftwo
 \fi
}%
\providecommand \@ifx [1]{%
 \ifx #1\expandafter \@firstoftwo
 \else \expandafter \@secondoftwo
 \fi
}%
\providecommand \natexlab [1]{#1}%
\providecommand \enquote  [1]{``#1''}%
\providecommand \bibnamefont  [1]{#1}%
\providecommand \bibfnamefont [1]{#1}%
\providecommand \citenamefont [1]{#1}%
\providecommand \href@noop [0]{\@secondoftwo}%
\providecommand \href [0]{\begingroup \@sanitize@url \@href}%
\providecommand \@href[1]{\@@startlink{#1}\@@href}%
\providecommand \@@href[1]{\endgroup#1\@@endlink}%
\providecommand \@sanitize@url [0]{\catcode `\\12\catcode `\$12\catcode
  `\&12\catcode `\#12\catcode `\^12\catcode `\_12\catcode `\%12\relax}%
\providecommand \@@startlink[1]{}%
\providecommand \@@endlink[0]{}%
\providecommand \url  [0]{\begingroup\@sanitize@url \@url }%
\providecommand \@url [1]{\endgroup\@href {#1}{\urlprefix }}%
\providecommand \urlprefix  [0]{URL }%
\providecommand \Eprint [0]{\href }%
\providecommand \doibase [0]{https://doi.org/}%
\providecommand \selectlanguage [0]{\@gobble}%
\providecommand \bibinfo  [0]{\@secondoftwo}%
\providecommand \bibfield  [0]{\@secondoftwo}%
\providecommand \translation [1]{[#1]}%
\providecommand \BibitemOpen [0]{}%
\providecommand \bibitemStop [0]{}%
\providecommand \bibitemNoStop [0]{.\EOS\space}%
\providecommand \EOS [0]{\spacefactor3000\relax}%
\providecommand \BibitemShut  [1]{\csname bibitem#1\endcsname}%
\let\auto@bib@innerbib\@empty
\bibitem [{\citenamefont {Szabo}\ and\ \citenamefont
  {Ostlund}(2012)}]{szabo2012modern}%
  \BibitemOpen
  \bibfield  {author} {\bibinfo {author} {\bibfnamefont {A.}~\bibnamefont
  {Szabo}}\ and\ \bibinfo {author} {\bibfnamefont {N.~S.}\ \bibnamefont
  {Ostlund}},\ }\href@noop {} {\emph {\bibinfo {title} {Modern {Q}uantum
  {C}hemistry: {I}ntroduction to {A}dvanced {E}lectronic {S}tructure
  {T}heory}}}\ (\bibinfo  {publisher} {Macmillan, New York},\ \bibinfo {year}
  {2012})\BibitemShut {NoStop}%
\bibitem [{\citenamefont {Aspuru-Guzik}\ \emph {et~al.}(2005)\citenamefont
  {Aspuru-Guzik}, \citenamefont {Dutoi}, \citenamefont {Love},\ and\
  \citenamefont {Head-Gordon}}]{aspuru2005simulated}%
  \BibitemOpen
  \bibfield  {author} {\bibinfo {author} {\bibfnamefont {A.}~\bibnamefont
  {Aspuru-Guzik}}, \bibinfo {author} {\bibfnamefont {A.~D.}\ \bibnamefont
  {Dutoi}}, \bibinfo {author} {\bibfnamefont {P.~J.}\ \bibnamefont {Love}},\
  and\ \bibinfo {author} {\bibfnamefont {M.}~\bibnamefont {Head-Gordon}},\
  }\bibfield  {title} {\bibinfo {title} {Simulated quantum computation of
  molecular energies},\ }\href@noop {} {\bibfield  {journal} {\bibinfo
  {journal} {Science}\ }\textbf {\bibinfo {volume} {309}},\ \bibinfo {pages}
  {1704} (\bibinfo {year} {2005})}\BibitemShut {NoStop}%
\bibitem [{\citenamefont {Babbush}\ \emph {et~al.}(2018)\citenamefont
  {Babbush}, \citenamefont {Gidney}, \citenamefont {Berry}, \citenamefont
  {Wiebe}, \citenamefont {McClean}, \citenamefont {Paler}, \citenamefont
  {Fowler},\ and\ \citenamefont {Neven}}]{babbush2018encoding}%
  \BibitemOpen
  \bibfield  {author} {\bibinfo {author} {\bibfnamefont {R.}~\bibnamefont
  {Babbush}}, \bibinfo {author} {\bibfnamefont {C.}~\bibnamefont {Gidney}},
  \bibinfo {author} {\bibfnamefont {D.~W.}\ \bibnamefont {Berry}}, \bibinfo
  {author} {\bibfnamefont {N.}~\bibnamefont {Wiebe}}, \bibinfo {author}
  {\bibfnamefont {J.}~\bibnamefont {McClean}}, \bibinfo {author} {\bibfnamefont
  {A.}~\bibnamefont {Paler}}, \bibinfo {author} {\bibfnamefont
  {A.}~\bibnamefont {Fowler}},\ and\ \bibinfo {author} {\bibfnamefont
  {H.}~\bibnamefont {Neven}},\ }\bibfield  {title} {\bibinfo {title} {Encoding
  electronic spectra in quantum circuits with linear t complexity},\
  }\href@noop {} {\bibfield  {journal} {\bibinfo  {journal} {Physical Review
  X}\ }\textbf {\bibinfo {volume} {8}},\ \bibinfo {pages} {041015} (\bibinfo
  {year} {2018})}\BibitemShut {NoStop}%
\bibitem [{\citenamefont {Elfving}\ \emph {et~al.}(2020)\citenamefont
  {Elfving}, \citenamefont {Broer}, \citenamefont {Webber}, \citenamefont
  {Gavartin}, \citenamefont {Halls}, \citenamefont {Lorton},\ and\
  \citenamefont {Bochevarov}}]{elfving2020will}%
  \BibitemOpen
  \bibfield  {author} {\bibinfo {author} {\bibfnamefont {V.~E.}\ \bibnamefont
  {Elfving}}, \bibinfo {author} {\bibfnamefont {B.~W.}\ \bibnamefont {Broer}},
  \bibinfo {author} {\bibfnamefont {M.}~\bibnamefont {Webber}}, \bibinfo
  {author} {\bibfnamefont {J.}~\bibnamefont {Gavartin}}, \bibinfo {author}
  {\bibfnamefont {M.~D.}\ \bibnamefont {Halls}}, \bibinfo {author}
  {\bibfnamefont {K.~P.}\ \bibnamefont {Lorton}},\ and\ \bibinfo {author}
  {\bibfnamefont {A.}~\bibnamefont {Bochevarov}},\ }\bibfield  {title}
  {\bibinfo {title} {How will quantum computers provide an industrially
  relevant computational advantage in quantum chemistry?},\ }\href@noop {}
  {\bibfield  {journal} {\bibinfo  {journal} {arXiv preprint arXiv:2009.12472}\
  } (\bibinfo {year} {2020})}\BibitemShut {NoStop}%
\bibitem [{\citenamefont {McCaskey}\ \emph {et~al.}(2019)\citenamefont
  {McCaskey}, \citenamefont {Parks}, \citenamefont {Jakowski}, \citenamefont
  {Moore}, \citenamefont {Morris}, \citenamefont {Humble},\ and\ \citenamefont
  {Pooser}}]{mccaskey2019quantum}%
  \BibitemOpen
  \bibfield  {author} {\bibinfo {author} {\bibfnamefont {A.~J.}\ \bibnamefont
  {McCaskey}}, \bibinfo {author} {\bibfnamefont {Z.~P.}\ \bibnamefont {Parks}},
  \bibinfo {author} {\bibfnamefont {J.}~\bibnamefont {Jakowski}}, \bibinfo
  {author} {\bibfnamefont {S.~V.}\ \bibnamefont {Moore}}, \bibinfo {author}
  {\bibfnamefont {T.~D.}\ \bibnamefont {Morris}}, \bibinfo {author}
  {\bibfnamefont {T.~S.}\ \bibnamefont {Humble}},\ and\ \bibinfo {author}
  {\bibfnamefont {R.~C.}\ \bibnamefont {Pooser}},\ }\bibfield  {title}
  {\bibinfo {title} {Quantum chemistry as a benchmark for near-term quantum
  computers},\ }\href@noop {} {\bibfield  {journal} {\bibinfo  {journal} {npj
  Quantum Information}\ }\textbf {\bibinfo {volume} {5}},\ \bibinfo {pages}
  {99} (\bibinfo {year} {2019})}\BibitemShut {NoStop}%
\bibitem [{\citenamefont {Kitaev}(1995)}]{kitaev1995quantum}%
  \BibitemOpen
  \bibfield  {author} {\bibinfo {author} {\bibfnamefont {A.~Y.}\ \bibnamefont
  {Kitaev}},\ }\bibfield  {title} {\bibinfo {title} {Quantum measurements and
  the abelian stabilizer problem},\ }\href@noop {} {\bibfield  {journal}
  {\bibinfo  {journal} {arXiv preprint quant-ph/9511026}\ } (\bibinfo {year}
  {1995})}\BibitemShut {NoStop}%
\bibitem [{\citenamefont {O’Malley}\ \emph {et~al.}(2016)\citenamefont
  {O’Malley}, \citenamefont {Babbush}, \citenamefont {Kivlichan},
  \citenamefont {Romero}, \citenamefont {McClean}, \citenamefont {Barends},
  \citenamefont {Kelly}, \citenamefont {Roushan}, \citenamefont {Tranter},
  \citenamefont {Ding} \emph {et~al.}}]{o2016scalable}%
  \BibitemOpen
  \bibfield  {author} {\bibinfo {author} {\bibfnamefont {P.~J.}\ \bibnamefont
  {O’Malley}}, \bibinfo {author} {\bibfnamefont {R.}~\bibnamefont {Babbush}},
  \bibinfo {author} {\bibfnamefont {I.~D.}\ \bibnamefont {Kivlichan}}, \bibinfo
  {author} {\bibfnamefont {J.}~\bibnamefont {Romero}}, \bibinfo {author}
  {\bibfnamefont {J.~R.}\ \bibnamefont {McClean}}, \bibinfo {author}
  {\bibfnamefont {R.}~\bibnamefont {Barends}}, \bibinfo {author} {\bibfnamefont
  {J.}~\bibnamefont {Kelly}}, \bibinfo {author} {\bibfnamefont
  {P.}~\bibnamefont {Roushan}}, \bibinfo {author} {\bibfnamefont
  {A.}~\bibnamefont {Tranter}}, \bibinfo {author} {\bibfnamefont
  {N.}~\bibnamefont {Ding}}, \emph {et~al.},\ }\bibfield  {title} {\bibinfo
  {title} {Scalable quantum simulation of molecular energies},\ }\href@noop {}
  {\bibfield  {journal} {\bibinfo  {journal} {Physical Review X}\ }\textbf
  {\bibinfo {volume} {6}},\ \bibinfo {pages} {031007} (\bibinfo {year}
  {2016})}\BibitemShut {NoStop}%
\bibitem [{\citenamefont {Mohammadbagherpoor}\ \emph
  {et~al.}(2019)\citenamefont {Mohammadbagherpoor}, \citenamefont {Oh},
  \citenamefont {Singh}, \citenamefont {Yu},\ and\ \citenamefont
  {Rindos}}]{mohammadbagherpoor2019experimental}%
  \BibitemOpen
  \bibfield  {author} {\bibinfo {author} {\bibfnamefont {H.}~\bibnamefont
  {Mohammadbagherpoor}}, \bibinfo {author} {\bibfnamefont {Y.-H.}\ \bibnamefont
  {Oh}}, \bibinfo {author} {\bibfnamefont {A.}~\bibnamefont {Singh}}, \bibinfo
  {author} {\bibfnamefont {X.}~\bibnamefont {Yu}},\ and\ \bibinfo {author}
  {\bibfnamefont {A.~J.}\ \bibnamefont {Rindos}},\ }\bibfield  {title}
  {\bibinfo {title} {Experimental challenges of implementing quantum phase
  estimation algorithms on {IBM} quantum computer},\ }\href@noop {} {\bibfield
  {journal} {\bibinfo  {journal} {arXiv preprint arXiv:1903.07605}\ } (\bibinfo
  {year} {2019})}\BibitemShut {NoStop}%
\bibitem [{\citenamefont {Peruzzo}\ \emph {et~al.}(2014)\citenamefont
  {Peruzzo}, \citenamefont {McClean}, \citenamefont {Shadbolt}, \citenamefont
  {Yung}, \citenamefont {Zhou}, \citenamefont {Love}, \citenamefont
  {Aspuru-Guzik},\ and\ \citenamefont {O’brien}}]{peruzzo2014variational}%
  \BibitemOpen
  \bibfield  {author} {\bibinfo {author} {\bibfnamefont {A.}~\bibnamefont
  {Peruzzo}}, \bibinfo {author} {\bibfnamefont {J.}~\bibnamefont {McClean}},
  \bibinfo {author} {\bibfnamefont {P.}~\bibnamefont {Shadbolt}}, \bibinfo
  {author} {\bibfnamefont {M.-H.}\ \bibnamefont {Yung}}, \bibinfo {author}
  {\bibfnamefont {X.-Q.}\ \bibnamefont {Zhou}}, \bibinfo {author}
  {\bibfnamefont {P.~J.}\ \bibnamefont {Love}}, \bibinfo {author}
  {\bibfnamefont {A.}~\bibnamefont {Aspuru-Guzik}},\ and\ \bibinfo {author}
  {\bibfnamefont {J.~L.}\ \bibnamefont {O’brien}},\ }\bibfield  {title}
  {\bibinfo {title} {A variational eigenvalue solver on a photonic quantum
  processor},\ }\href@noop {} {\bibfield  {journal} {\bibinfo  {journal}
  {Nature communications}\ }\textbf {\bibinfo {volume} {5}},\ \bibinfo {pages}
  {4213} (\bibinfo {year} {2014})}\BibitemShut {NoStop}%
\bibitem [{\citenamefont {Farhi}\ \emph {et~al.}(2014)\citenamefont {Farhi},
  \citenamefont {Goldstone},\ and\ \citenamefont {Gutmann}}]{farhi2014quantum}%
  \BibitemOpen
  \bibfield  {author} {\bibinfo {author} {\bibfnamefont {E.}~\bibnamefont
  {Farhi}}, \bibinfo {author} {\bibfnamefont {J.}~\bibnamefont {Goldstone}},\
  and\ \bibinfo {author} {\bibfnamefont {S.}~\bibnamefont {Gutmann}},\
  }\bibfield  {title} {\bibinfo {title} {A quantum approximate optimization
  algorithm},\ }\href@noop {} {\bibfield  {journal} {\bibinfo  {journal} {arXiv
  preprint arXiv:1411.4028}\ } (\bibinfo {year} {2014})}\BibitemShut {NoStop}%
\bibitem [{\citenamefont {Bravo-Prieto}\ \emph {et~al.}(2019)\citenamefont
  {Bravo-Prieto}, \citenamefont {LaRose}, \citenamefont {Cerezo}, \citenamefont
  {Subasi}, \citenamefont {Cincio},\ and\ \citenamefont
  {Coles}}]{bravo2019variational}%
  \BibitemOpen
  \bibfield  {author} {\bibinfo {author} {\bibfnamefont {C.}~\bibnamefont
  {Bravo-Prieto}}, \bibinfo {author} {\bibfnamefont {R.}~\bibnamefont
  {LaRose}}, \bibinfo {author} {\bibfnamefont {M.}~\bibnamefont {Cerezo}},
  \bibinfo {author} {\bibfnamefont {Y.}~\bibnamefont {Subasi}}, \bibinfo
  {author} {\bibfnamefont {L.}~\bibnamefont {Cincio}},\ and\ \bibinfo {author}
  {\bibfnamefont {P.~J.}\ \bibnamefont {Coles}},\ }\bibfield  {title} {\bibinfo
  {title} {Variational quantum linear solver},\ }\href@noop {} {\bibfield
  {journal} {\bibinfo  {journal} {arXiv preprint arXiv:1909.05820}\ } (\bibinfo
  {year} {2019})}\BibitemShut {NoStop}%
\bibitem [{\citenamefont {Eddins}\ \emph {et~al.}(2022)\citenamefont {Eddins},
  \citenamefont {Motta}, \citenamefont {Gujarati}, \citenamefont {Bravyi},
  \citenamefont {Mezzacapo}, \citenamefont {Hadfield},\ and\ \citenamefont
  {Sheldon}}]{eddins2022doubling}%
  \BibitemOpen
  \bibfield  {author} {\bibinfo {author} {\bibfnamefont {A.}~\bibnamefont
  {Eddins}}, \bibinfo {author} {\bibfnamefont {M.}~\bibnamefont {Motta}},
  \bibinfo {author} {\bibfnamefont {T.~P.}\ \bibnamefont {Gujarati}}, \bibinfo
  {author} {\bibfnamefont {S.}~\bibnamefont {Bravyi}}, \bibinfo {author}
  {\bibfnamefont {A.}~\bibnamefont {Mezzacapo}}, \bibinfo {author}
  {\bibfnamefont {C.}~\bibnamefont {Hadfield}},\ and\ \bibinfo {author}
  {\bibfnamefont {S.}~\bibnamefont {Sheldon}},\ }\bibfield  {title} {\bibinfo
  {title} {Doubling the size of quantum simulators by entanglement forging},\
  }\href@noop {} {\bibfield  {journal} {\bibinfo  {journal} {PRX Quantum}\
  }\textbf {\bibinfo {volume} {3}},\ \bibinfo {pages} {010309} (\bibinfo {year}
  {2022})}\BibitemShut {NoStop}%
\bibitem [{\citenamefont {Huggins}\ \emph {et~al.}(2022)\citenamefont
  {Huggins}, \citenamefont {O’Gorman}, \citenamefont {Rubin}, \citenamefont
  {Reichman}, \citenamefont {Babbush},\ and\ \citenamefont
  {Lee}}]{huggins2022unbiasing}%
  \BibitemOpen
  \bibfield  {author} {\bibinfo {author} {\bibfnamefont {W.~J.}\ \bibnamefont
  {Huggins}}, \bibinfo {author} {\bibfnamefont {B.~A.}\ \bibnamefont
  {O’Gorman}}, \bibinfo {author} {\bibfnamefont {N.~C.}\ \bibnamefont
  {Rubin}}, \bibinfo {author} {\bibfnamefont {D.~R.}\ \bibnamefont {Reichman}},
  \bibinfo {author} {\bibfnamefont {R.}~\bibnamefont {Babbush}},\ and\ \bibinfo
  {author} {\bibfnamefont {J.}~\bibnamefont {Lee}},\ }\bibfield  {title}
  {\bibinfo {title} {Unbiasing fermionic quantum monte carlo with a quantum
  computer},\ }\href@noop {} {\bibfield  {journal} {\bibinfo  {journal}
  {Nature}\ }\textbf {\bibinfo {volume} {603}},\ \bibinfo {pages} {416}
  (\bibinfo {year} {2022})}\BibitemShut {NoStop}%
\bibitem [{\citenamefont {Kirby}\ \emph {et~al.}(2021)\citenamefont {Kirby},
  \citenamefont {Tranter},\ and\ \citenamefont {Love}}]{kirby2021contextual}%
  \BibitemOpen
  \bibfield  {author} {\bibinfo {author} {\bibfnamefont {W.~M.}\ \bibnamefont
  {Kirby}}, \bibinfo {author} {\bibfnamefont {A.}~\bibnamefont {Tranter}},\
  and\ \bibinfo {author} {\bibfnamefont {P.~J.}\ \bibnamefont {Love}},\
  }\bibfield  {title} {\bibinfo {title} {Contextual subspace variational
  quantum eigensolver},\ }\href@noop {} {\bibfield  {journal} {\bibinfo
  {journal} {Quantum}\ }\textbf {\bibinfo {volume} {5}},\ \bibinfo {pages}
  {456} (\bibinfo {year} {2021})}\BibitemShut {NoStop}%
\bibitem [{\citenamefont {Zhao}\ \emph {et~al.}(2020)\citenamefont {Zhao},
  \citenamefont {Tranter}, \citenamefont {Kirby}, \citenamefont {Ung},
  \citenamefont {Miyake},\ and\ \citenamefont {Love}}]{zhao2019measurement}%
  \BibitemOpen
  \bibfield  {author} {\bibinfo {author} {\bibfnamefont {A.}~\bibnamefont
  {Zhao}}, \bibinfo {author} {\bibfnamefont {A.}~\bibnamefont {Tranter}},
  \bibinfo {author} {\bibfnamefont {W.~M.}\ \bibnamefont {Kirby}}, \bibinfo
  {author} {\bibfnamefont {S.~F.}\ \bibnamefont {Ung}}, \bibinfo {author}
  {\bibfnamefont {A.}~\bibnamefont {Miyake}},\ and\ \bibinfo {author}
  {\bibfnamefont {P.~J.}\ \bibnamefont {Love}},\ }\bibfield  {title} {\bibinfo
  {title} {Measurement reduction in variational quantum algorithms},\
  }\href@noop {} {\bibfield  {journal} {\bibinfo  {journal} {Physical Review
  A}\ }\textbf {\bibinfo {volume} {101}},\ \bibinfo {pages} {062322} (\bibinfo
  {year} {2020})}\BibitemShut {NoStop}%
\bibitem [{\citenamefont {Kandala}\ \emph {et~al.}(2017)\citenamefont
  {Kandala}, \citenamefont {Mezzacapo}, \citenamefont {Temme}, \citenamefont
  {Takita}, \citenamefont {Brink}, \citenamefont {Chow},\ and\ \citenamefont
  {Gambetta}}]{kandala2017hardware}%
  \BibitemOpen
  \bibfield  {author} {\bibinfo {author} {\bibfnamefont {A.}~\bibnamefont
  {Kandala}}, \bibinfo {author} {\bibfnamefont {A.}~\bibnamefont {Mezzacapo}},
  \bibinfo {author} {\bibfnamefont {K.}~\bibnamefont {Temme}}, \bibinfo
  {author} {\bibfnamefont {M.}~\bibnamefont {Takita}}, \bibinfo {author}
  {\bibfnamefont {M.}~\bibnamefont {Brink}}, \bibinfo {author} {\bibfnamefont
  {J.~M.}\ \bibnamefont {Chow}},\ and\ \bibinfo {author} {\bibfnamefont
  {J.~M.}\ \bibnamefont {Gambetta}},\ }\bibfield  {title} {\bibinfo {title}
  {Hardware-efficient variational quantum eigensolver for small molecules and
  quantum magnets},\ }\href@noop {} {\bibfield  {journal} {\bibinfo  {journal}
  {Nature}\ }\textbf {\bibinfo {volume} {549}},\ \bibinfo {pages} {242}
  (\bibinfo {year} {2017})}\BibitemShut {NoStop}%
\bibitem [{\citenamefont {Verteletskyi}\ \emph {et~al.}(2020)\citenamefont
  {Verteletskyi}, \citenamefont {Yen},\ and\ \citenamefont
  {Izmaylov}}]{verteletskyi2020measurement}%
  \BibitemOpen
  \bibfield  {author} {\bibinfo {author} {\bibfnamefont {V.}~\bibnamefont
  {Verteletskyi}}, \bibinfo {author} {\bibfnamefont {T.-C.}\ \bibnamefont
  {Yen}},\ and\ \bibinfo {author} {\bibfnamefont {A.~F.}\ \bibnamefont
  {Izmaylov}},\ }\bibfield  {title} {\bibinfo {title} {Measurement optimization
  in the variational quantum eigensolver using a minimum clique cover},\
  }\href@noop {} {\bibfield  {journal} {\bibinfo  {journal} {The Journal of
  Chemical Physics}\ }\textbf {\bibinfo {volume} {152}},\ \bibinfo {pages}
  {124114} (\bibinfo {year} {2020})}\BibitemShut {NoStop}%
\bibitem [{\citenamefont {Izmaylov}\ \emph {et~al.}(2019)\citenamefont
  {Izmaylov}, \citenamefont {Yen},\ and\ \citenamefont
  {Ryabinkin}}]{izmaylov2019revising}%
  \BibitemOpen
  \bibfield  {author} {\bibinfo {author} {\bibfnamefont {A.~F.}\ \bibnamefont
  {Izmaylov}}, \bibinfo {author} {\bibfnamefont {T.-C.}\ \bibnamefont {Yen}},\
  and\ \bibinfo {author} {\bibfnamefont {I.~G.}\ \bibnamefont {Ryabinkin}},\
  }\bibfield  {title} {\bibinfo {title} {Revising the measurement process in
  the variational quantum eigensolver: is it possible to reduce the number of
  separately measured operators?},\ }\href@noop {} {\bibfield  {journal}
  {\bibinfo  {journal} {Chemical Science}\ }\textbf {\bibinfo {volume} {10}},\
  \bibinfo {pages} {3746} (\bibinfo {year} {2019})}\BibitemShut {NoStop}%
\bibitem [{\citenamefont {Cotler}\ and\ \citenamefont
  {Wilczek}(2020)}]{cotler2020quantum}%
  \BibitemOpen
  \bibfield  {author} {\bibinfo {author} {\bibfnamefont {J.}~\bibnamefont
  {Cotler}}\ and\ \bibinfo {author} {\bibfnamefont {F.}~\bibnamefont
  {Wilczek}},\ }\bibfield  {title} {\bibinfo {title} {Quantum overlapping
  tomography},\ }\href@noop {} {\bibfield  {journal} {\bibinfo  {journal}
  {Physical Review Letters}\ }\textbf {\bibinfo {volume} {124}},\ \bibinfo
  {pages} {100401} (\bibinfo {year} {2020})}\BibitemShut {NoStop}%
\bibitem [{\citenamefont {Bonet-Monroig}\ \emph {et~al.}(2020)\citenamefont
  {Bonet-Monroig}, \citenamefont {Babbush},\ and\ \citenamefont
  {O’Brien}}]{bonet2019nearly}%
  \BibitemOpen
  \bibfield  {author} {\bibinfo {author} {\bibfnamefont {X.}~\bibnamefont
  {Bonet-Monroig}}, \bibinfo {author} {\bibfnamefont {R.}~\bibnamefont
  {Babbush}},\ and\ \bibinfo {author} {\bibfnamefont {T.~E.}\ \bibnamefont
  {O’Brien}},\ }\bibfield  {title} {\bibinfo {title} {Nearly optimal
  measurement scheduling for partial tomography of quantum states},\
  }\href@noop {} {\bibfield  {journal} {\bibinfo  {journal} {Physical Review
  X}\ }\textbf {\bibinfo {volume} {10}},\ \bibinfo {pages} {031064} (\bibinfo
  {year} {2020})}\BibitemShut {NoStop}%
\bibitem [{\citenamefont {Gokhale}\ and\ \citenamefont
  {Chong}(2019)}]{gokhale2019n}%
  \BibitemOpen
  \bibfield  {author} {\bibinfo {author} {\bibfnamefont {P.}~\bibnamefont
  {Gokhale}}\ and\ \bibinfo {author} {\bibfnamefont {F.~T.}\ \bibnamefont
  {Chong}},\ }\bibfield  {title} {\bibinfo {title} {$ o (n^3) $ measurement
  cost for variational quantum eigensolver on molecular {H}amiltonians},\
  }\href@noop {} {\bibfield  {journal} {\bibinfo  {journal} {arXiv preprint
  arXiv:1908.11857}\ } (\bibinfo {year} {2019})}\BibitemShut {NoStop}%
\bibitem [{\citenamefont {Jena}\ \emph {et~al.}(2019)\citenamefont {Jena},
  \citenamefont {Genin},\ and\ \citenamefont {Mosca}}]{jena2019pauli}%
  \BibitemOpen
  \bibfield  {author} {\bibinfo {author} {\bibfnamefont {A.}~\bibnamefont
  {Jena}}, \bibinfo {author} {\bibfnamefont {S.}~\bibnamefont {Genin}},\ and\
  \bibinfo {author} {\bibfnamefont {M.}~\bibnamefont {Mosca}},\ }\bibfield
  {title} {\bibinfo {title} {Pauli partitioning with respect to gate sets},\
  }\href@noop {} {\bibfield  {journal} {\bibinfo  {journal} {arXiv preprint
  arXiv:1907.07859}\ } (\bibinfo {year} {2019})}\BibitemShut {NoStop}%
\bibitem [{\citenamefont {Huggins}\ \emph {et~al.}(2021)\citenamefont
  {Huggins}, \citenamefont {McClean}, \citenamefont {Rubin}, \citenamefont
  {Jiang}, \citenamefont {Wiebe}, \citenamefont {Whaley},\ and\ \citenamefont
  {Babbush}}]{huggins2019efficient}%
  \BibitemOpen
  \bibfield  {author} {\bibinfo {author} {\bibfnamefont {W.~J.}\ \bibnamefont
  {Huggins}}, \bibinfo {author} {\bibfnamefont {J.~R.}\ \bibnamefont
  {McClean}}, \bibinfo {author} {\bibfnamefont {N.~C.}\ \bibnamefont {Rubin}},
  \bibinfo {author} {\bibfnamefont {Z.}~\bibnamefont {Jiang}}, \bibinfo
  {author} {\bibfnamefont {N.}~\bibnamefont {Wiebe}}, \bibinfo {author}
  {\bibfnamefont {K.~B.}\ \bibnamefont {Whaley}},\ and\ \bibinfo {author}
  {\bibfnamefont {R.}~\bibnamefont {Babbush}},\ }\bibfield  {title} {\bibinfo
  {title} {Efficient and noise resilient measurements for quantum chemistry on
  near-term quantum computers},\ }\href@noop {} {\bibfield  {journal} {\bibinfo
   {journal} {npj Quantum Information}\ }\textbf {\bibinfo {volume} {7}},\
  \bibinfo {pages} {23} (\bibinfo {year} {2021})}\BibitemShut {NoStop}%
\bibitem [{\citenamefont {Gokhale}\ \emph {et~al.}(2019)\citenamefont
  {Gokhale}, \citenamefont {Angiuli}, \citenamefont {Ding}, \citenamefont
  {Gui}, \citenamefont {Tomesh}, \citenamefont {Suchara}, \citenamefont
  {Martonosi},\ and\ \citenamefont {Chong}}]{gokhale2019minimizing}%
  \BibitemOpen
  \bibfield  {author} {\bibinfo {author} {\bibfnamefont {P.}~\bibnamefont
  {Gokhale}}, \bibinfo {author} {\bibfnamefont {O.}~\bibnamefont {Angiuli}},
  \bibinfo {author} {\bibfnamefont {Y.}~\bibnamefont {Ding}}, \bibinfo {author}
  {\bibfnamefont {K.}~\bibnamefont {Gui}}, \bibinfo {author} {\bibfnamefont
  {T.}~\bibnamefont {Tomesh}}, \bibinfo {author} {\bibfnamefont
  {M.}~\bibnamefont {Suchara}}, \bibinfo {author} {\bibfnamefont
  {M.}~\bibnamefont {Martonosi}},\ and\ \bibinfo {author} {\bibfnamefont
  {F.~T.}\ \bibnamefont {Chong}},\ }\bibfield  {title} {\bibinfo {title}
  {Minimizing state preparations in variational quantum eigensolver by
  partitioning into commuting families},\ }\href@noop {} {\bibfield  {journal}
  {\bibinfo  {journal} {arXiv preprint arXiv:1907.13623}\ } (\bibinfo {year}
  {2019})}\BibitemShut {NoStop}%
\bibitem [{\citenamefont {Crawford}\ \emph {et~al.}(2021)\citenamefont
  {Crawford}, \citenamefont {van Straaten}, \citenamefont {Wang}, \citenamefont
  {Parks}, \citenamefont {Campbell},\ and\ \citenamefont
  {Brierley}}]{crawford2021efficient}%
  \BibitemOpen
  \bibfield  {author} {\bibinfo {author} {\bibfnamefont {O.}~\bibnamefont
  {Crawford}}, \bibinfo {author} {\bibfnamefont {B.}~\bibnamefont {van
  Straaten}}, \bibinfo {author} {\bibfnamefont {D.}~\bibnamefont {Wang}},
  \bibinfo {author} {\bibfnamefont {T.}~\bibnamefont {Parks}}, \bibinfo
  {author} {\bibfnamefont {E.}~\bibnamefont {Campbell}},\ and\ \bibinfo
  {author} {\bibfnamefont {S.}~\bibnamefont {Brierley}},\ }\bibfield  {title}
  {\bibinfo {title} {Efficient quantum measurement of pauli operators in the
  presence of finite sampling error},\ }\href@noop {} {\bibfield  {journal}
  {\bibinfo  {journal} {Quantum}\ }\textbf {\bibinfo {volume} {5}},\ \bibinfo
  {pages} {385} (\bibinfo {year} {2021})}\BibitemShut {NoStop}%
\bibitem [{\citenamefont {Huang}\ \emph {et~al.}(2020)\citenamefont {Huang},
  \citenamefont {Kueng},\ and\ \citenamefont {Preskill}}]{huang2020predicting}%
  \BibitemOpen
  \bibfield  {author} {\bibinfo {author} {\bibfnamefont {H.-Y.}\ \bibnamefont
  {Huang}}, \bibinfo {author} {\bibfnamefont {R.}~\bibnamefont {Kueng}},\ and\
  \bibinfo {author} {\bibfnamefont {J.}~\bibnamefont {Preskill}},\ }\bibfield
  {title} {\bibinfo {title} {Predicting many properties of a quantum system
  from very few measurements},\ }\href@noop {} {\bibfield  {journal} {\bibinfo
  {journal} {Nature Physics}\ }\textbf {\bibinfo {volume} {16}},\ \bibinfo
  {pages} {1050} (\bibinfo {year} {2020})}\BibitemShut {NoStop}%
\bibitem [{\citenamefont {Hadfield}\ \emph {et~al.}(2022)\citenamefont
  {Hadfield}, \citenamefont {Bravyi}, \citenamefont {Raymond},\ and\
  \citenamefont {Mezzacapo}}]{hadfield2020measurements}%
  \BibitemOpen
  \bibfield  {author} {\bibinfo {author} {\bibfnamefont {C.}~\bibnamefont
  {Hadfield}}, \bibinfo {author} {\bibfnamefont {S.}~\bibnamefont {Bravyi}},
  \bibinfo {author} {\bibfnamefont {R.}~\bibnamefont {Raymond}},\ and\ \bibinfo
  {author} {\bibfnamefont {A.}~\bibnamefont {Mezzacapo}},\ }\bibfield  {title}
  {\bibinfo {title} {Measurements of quantum hamiltonians with locally-biased
  classical shadows},\ }\href@noop {} {\bibfield  {journal} {\bibinfo
  {journal} {Communications in Mathematical Physics}\ }\textbf {\bibinfo
  {volume} {391}},\ \bibinfo {pages} {951} (\bibinfo {year}
  {2022})}\BibitemShut {NoStop}%
\bibitem [{\citenamefont {Huang}\ \emph {et~al.}(2021)\citenamefont {Huang},
  \citenamefont {Kueng},\ and\ \citenamefont {Preskill}}]{huang2021efficient}%
  \BibitemOpen
  \bibfield  {author} {\bibinfo {author} {\bibfnamefont {H.-Y.}\ \bibnamefont
  {Huang}}, \bibinfo {author} {\bibfnamefont {R.}~\bibnamefont {Kueng}},\ and\
  \bibinfo {author} {\bibfnamefont {J.}~\bibnamefont {Preskill}},\ }\bibfield
  {title} {\bibinfo {title} {Efficient estimation of pauli observables by
  derandomization},\ }\href@noop {} {\bibfield  {journal} {\bibinfo  {journal}
  {Physical Review Letters}\ }\textbf {\bibinfo {volume} {127}},\ \bibinfo
  {pages} {030503} (\bibinfo {year} {2021})}\BibitemShut {NoStop}%
\bibitem [{\citenamefont {Izmaylov}\ \emph {et~al.}(2020)\citenamefont
  {Izmaylov}, \citenamefont {Yen}, \citenamefont {Lang},\ and\ \citenamefont
  {Verteletskyi}}]{izmaylov2019unitary}%
  \BibitemOpen
  \bibfield  {author} {\bibinfo {author} {\bibfnamefont {A.~F.}\ \bibnamefont
  {Izmaylov}}, \bibinfo {author} {\bibfnamefont {T.-C.}\ \bibnamefont {Yen}},
  \bibinfo {author} {\bibfnamefont {R.~A.}\ \bibnamefont {Lang}},\ and\
  \bibinfo {author} {\bibfnamefont {V.}~\bibnamefont {Verteletskyi}},\
  }\bibfield  {title} {\bibinfo {title} {Unitary partitioning approach to the
  measurement problem in the variational quantum eigensolver method},\
  }\href@noop {} {\bibfield  {journal} {\bibinfo  {journal} {Journal of
  Chemical Theory and Computation}\ }\textbf {\bibinfo {volume} {16}},\
  \bibinfo {pages} {190} (\bibinfo {year} {2020})}\BibitemShut {NoStop}%
\bibitem [{\citenamefont {Ralli}\ \emph {et~al.}(2021)\citenamefont {Ralli},
  \citenamefont {Love}, \citenamefont {Tranter},\ and\ \citenamefont
  {Coveney}}]{ralli2021implementation}%
  \BibitemOpen
  \bibfield  {author} {\bibinfo {author} {\bibfnamefont {A.}~\bibnamefont
  {Ralli}}, \bibinfo {author} {\bibfnamefont {P.~J.}\ \bibnamefont {Love}},
  \bibinfo {author} {\bibfnamefont {A.}~\bibnamefont {Tranter}},\ and\ \bibinfo
  {author} {\bibfnamefont {P.~V.}\ \bibnamefont {Coveney}},\ }\bibfield
  {title} {\bibinfo {title} {Implementation of measurement reduction for the
  variational quantum eigensolver},\ }\href@noop {} {\bibfield  {journal}
  {\bibinfo  {journal} {Physical Review Research}\ }\textbf {\bibinfo {volume}
  {3}},\ \bibinfo {pages} {033195} (\bibinfo {year} {2021})}\BibitemShut
  {NoStop}%
\bibitem [{\citenamefont {Kochen}\ and\ \citenamefont
  {Specker}(1975)}]{kochen1975problem}%
  \BibitemOpen
  \bibfield  {author} {\bibinfo {author} {\bibfnamefont {S.}~\bibnamefont
  {Kochen}}\ and\ \bibinfo {author} {\bibfnamefont {E.~P.}\ \bibnamefont
  {Specker}},\ }\bibfield  {title} {\bibinfo {title} {The problem of hidden
  variables in quantum mechanics},\ }in\ \href@noop {} {\emph {\bibinfo
  {booktitle} {The logico-algebraic approach to quantum mechanics}}}\ (\bibinfo
   {publisher} {Springer, New York},\ \bibinfo {year} {1975})\ pp.\ \bibinfo
  {pages} {293--328}\BibitemShut {NoStop}%
\bibitem [{\citenamefont {Budroni}(2019)}]{budroni2019contextuality}%
  \BibitemOpen
  \bibfield  {author} {\bibinfo {author} {\bibfnamefont {C.}~\bibnamefont
  {Budroni}},\ }\bibfield  {title} {\bibinfo {title} {Contextuality, memory
  cost and non-classicality for sequential measurements},\ }\href@noop {}
  {\bibfield  {journal} {\bibinfo  {journal} {Philosophical Transactions of the
  Royal Society A}\ }\textbf {\bibinfo {volume} {377}},\ \bibinfo {pages}
  {20190141} (\bibinfo {year} {2019})}\BibitemShut {NoStop}%
\bibitem [{\citenamefont {Budroni}\ \emph {et~al.}(2022)\citenamefont
  {Budroni}, \citenamefont {Cabello}, \citenamefont {G{\"u}hne}, \citenamefont
  {Kleinmann},\ and\ \citenamefont {Larsson}}]{budroni2021quantum}%
  \BibitemOpen
  \bibfield  {author} {\bibinfo {author} {\bibfnamefont {C.}~\bibnamefont
  {Budroni}}, \bibinfo {author} {\bibfnamefont {A.}~\bibnamefont {Cabello}},
  \bibinfo {author} {\bibfnamefont {O.}~\bibnamefont {G{\"u}hne}}, \bibinfo
  {author} {\bibfnamefont {M.}~\bibnamefont {Kleinmann}},\ and\ \bibinfo
  {author} {\bibfnamefont {J.-{\AA}.}\ \bibnamefont {Larsson}},\ }\bibfield
  {title} {\bibinfo {title} {Kochen-specker contextuality},\ }\href@noop {}
  {\bibfield  {journal} {\bibinfo  {journal} {Reviews of Modern Physics}\
  }\textbf {\bibinfo {volume} {94}},\ \bibinfo {pages} {045007} (\bibinfo
  {year} {2022})}\BibitemShut {NoStop}%
\bibitem [{\citenamefont {de~Silva}(2017)}]{de2017graph}%
  \BibitemOpen
  \bibfield  {author} {\bibinfo {author} {\bibfnamefont {N.}~\bibnamefont
  {de~Silva}},\ }\bibfield  {title} {\bibinfo {title} {Graph-theoretic
  strengths of contextuality},\ }\href@noop {} {\bibfield  {journal} {\bibinfo
  {journal} {Physical Review A}\ }\textbf {\bibinfo {volume} {95}},\ \bibinfo
  {pages} {032108} (\bibinfo {year} {2017})}\BibitemShut {NoStop}%
\bibitem [{\citenamefont {Cabello}(2021)}]{cabello2021converting}%
  \BibitemOpen
  \bibfield  {author} {\bibinfo {author} {\bibfnamefont {A.}~\bibnamefont
  {Cabello}},\ }\bibfield  {title} {\bibinfo {title} {Converting contextuality
  into nonlocality},\ }\href@noop {} {\bibfield  {journal} {\bibinfo  {journal}
  {Physical Review Letters}\ }\textbf {\bibinfo {volume} {127}},\ \bibinfo
  {pages} {070401} (\bibinfo {year} {2021})}\BibitemShut {NoStop}%
\bibitem [{\citenamefont {Howard}\ \emph {et~al.}(2014)\citenamefont {Howard},
  \citenamefont {Wallman}, \citenamefont {Veitch},\ and\ \citenamefont
  {Emerson}}]{howard2014contextuality}%
  \BibitemOpen
  \bibfield  {author} {\bibinfo {author} {\bibfnamefont {M.}~\bibnamefont
  {Howard}}, \bibinfo {author} {\bibfnamefont {J.}~\bibnamefont {Wallman}},
  \bibinfo {author} {\bibfnamefont {V.}~\bibnamefont {Veitch}},\ and\ \bibinfo
  {author} {\bibfnamefont {J.}~\bibnamefont {Emerson}},\ }\bibfield  {title}
  {\bibinfo {title} {Contextuality supplies the ‘magic’for quantum
  computation},\ }\href@noop {} {\bibfield  {journal} {\bibinfo  {journal}
  {Nature}\ }\textbf {\bibinfo {volume} {510}},\ \bibinfo {pages} {351}
  (\bibinfo {year} {2014})}\BibitemShut {NoStop}%
\bibitem [{\citenamefont {Mermin}(1990)}]{mermin1990simple}%
  \BibitemOpen
  \bibfield  {author} {\bibinfo {author} {\bibfnamefont {N.~D.}\ \bibnamefont
  {Mermin}},\ }\bibfield  {title} {\bibinfo {title} {Simple unified form for
  the major no-hidden-variables theorems},\ }\href@noop {} {\bibfield
  {journal} {\bibinfo  {journal} {Physical review letters}\ }\textbf {\bibinfo
  {volume} {65}},\ \bibinfo {pages} {3373} (\bibinfo {year}
  {1990})}\BibitemShut {NoStop}%
\bibitem [{\citenamefont {Bell}(1964)}]{bell1964einstein}%
  \BibitemOpen
  \bibfield  {author} {\bibinfo {author} {\bibfnamefont {J.~S.}\ \bibnamefont
  {Bell}},\ }\bibfield  {title} {\bibinfo {title} {On the {E}instein {P}odolsky
  {R}osen paradox},\ }\href@noop {} {\bibfield  {journal} {\bibinfo  {journal}
  {Physics Physique Fizika}\ }\textbf {\bibinfo {volume} {1}},\ \bibinfo
  {pages} {195} (\bibinfo {year} {1964})}\BibitemShut {NoStop}%
\bibitem [{\citenamefont {Peres}(1990)}]{peres1990incompatible}%
  \BibitemOpen
  \bibfield  {author} {\bibinfo {author} {\bibfnamefont {A.}~\bibnamefont
  {Peres}},\ }\bibfield  {title} {\bibinfo {title} {Incompatible results of
  quantum measurements},\ }\href@noop {} {\bibfield  {journal} {\bibinfo
  {journal} {Physics Letters A}\ }\textbf {\bibinfo {volume} {151}},\ \bibinfo
  {pages} {107} (\bibinfo {year} {1990})}\BibitemShut {NoStop}%
\bibitem [{Sup()}]{SuppMaterial}%
  \BibitemOpen
  \href@noop {} {\bibinfo {title} {See {S}upplemental {M}aterial at
  \url{http://link.aps.org/supplemental/10.1103/PhysRevResearch.5.013095} for
  further information on the {CS-VQE} algorithm, the scaling analysis of the
  unitary partitioning rotation used in {CS-VQE}, the analysis of the variance
  of the ground-state energy obtained when using unitary partitioning for
  measurement reduction and the numerical details of the physical problems
  presented in this paper.}}\BibitemShut {Stop}%
\bibitem [{\citenamefont {Spekkens}(2016{\natexlab{a}})}]{Spekkens2016}%
  \BibitemOpen
  \bibfield  {author} {\bibinfo {author} {\bibfnamefont {R.~W.}\ \bibnamefont
  {Spekkens}},\ }\bibinfo {title} {Quasi-quantization: Classical statistical
  theories with an epistemic restriction},\ in\ \href
  {https://doi.org/10.1007/978-94-017-7303-4_4} {\emph {\bibinfo {booktitle}
  {Quantum Theory: Informational Foundations and Foils}}},\ \bibinfo {editor}
  {edited by\ \bibinfo {editor} {\bibfnamefont {G.}~\bibnamefont {Chiribella}}\
  and\ \bibinfo {editor} {\bibfnamefont {R.~W.}\ \bibnamefont {Spekkens}}}\
  (\bibinfo  {publisher} {Springer},\ \bibinfo {address} {Dordrecht},\ \bibinfo
  {year} {2016})\ pp.\ \bibinfo {pages} {83--135}\BibitemShut {NoStop}%
\bibitem [{\citenamefont {Kirby}\ and\ \citenamefont
  {Love}(2020)}]{kirby2020classical}%
  \BibitemOpen
  \bibfield  {author} {\bibinfo {author} {\bibfnamefont {W.~M.}\ \bibnamefont
  {Kirby}}\ and\ \bibinfo {author} {\bibfnamefont {P.~J.}\ \bibnamefont
  {Love}},\ }\bibfield  {title} {\bibinfo {title} {Classical simulation of
  noncontextual pauli hamiltonians},\ }\href@noop {} {\bibfield  {journal}
  {\bibinfo  {journal} {Physical Review A}\ }\textbf {\bibinfo {volume}
  {102}},\ \bibinfo {pages} {032418} (\bibinfo {year} {2020})}\BibitemShut
  {NoStop}%
\bibitem [{\citenamefont {Spekkens}(2007)}]{spekkens2007evidence}%
  \BibitemOpen
  \bibfield  {author} {\bibinfo {author} {\bibfnamefont {R.~W.}\ \bibnamefont
  {Spekkens}},\ }\bibfield  {title} {\bibinfo {title} {Evidence for the
  epistemic view of quantum states: A toy theory},\ }\href@noop {} {\bibfield
  {journal} {\bibinfo  {journal} {Physical Review A}\ }\textbf {\bibinfo
  {volume} {75}},\ \bibinfo {pages} {032110} (\bibinfo {year}
  {2007})}\BibitemShut {NoStop}%
\bibitem [{\citenamefont {Spekkens}(2016{\natexlab{b}})}]{spekkens2016quasi}%
  \BibitemOpen
  \bibfield  {author} {\bibinfo {author} {\bibfnamefont {R.~W.}\ \bibnamefont
  {Spekkens}},\ }\bibfield  {title} {\bibinfo {title} {Quasi-quantization:
  classical statistical theories with an epistemic restriction},\ }in\
  \href@noop {} {\emph {\bibinfo {booktitle} {Quantum Theory: Informational
  Foundations and Foils}}}\ (\bibinfo  {publisher} {Springer},\ \bibinfo {year}
  {2016})\ pp.\ \bibinfo {pages} {83--135}\BibitemShut {NoStop}%
\bibitem [{\citenamefont {Weaving}\ \emph {et~al.}(2023)\citenamefont
  {Weaving}, \citenamefont {Ralli}, \citenamefont {Kirby}, \citenamefont
  {Tranter}, \citenamefont {Love},\ and\ \citenamefont
  {Coveney}}]{weaving2022stabilizer}%
  \BibitemOpen
  \bibfield  {author} {\bibinfo {author} {\bibfnamefont {T.}~\bibnamefont
  {Weaving}}, \bibinfo {author} {\bibfnamefont {A.}~\bibnamefont {Ralli}},
  \bibinfo {author} {\bibfnamefont {W.~M.}\ \bibnamefont {Kirby}}, \bibinfo
  {author} {\bibfnamefont {A.}~\bibnamefont {Tranter}}, \bibinfo {author}
  {\bibfnamefont {P.~J.}\ \bibnamefont {Love}},\ and\ \bibinfo {author}
  {\bibfnamefont {P.~V.}\ \bibnamefont {Coveney}},\ }\bibfield  {title}
  {\bibinfo {title} {A stabilizer framework for the contextual subspace
  variational quantum eigensolver and the noncontextual projection ansatz},\
  }\href {https://doi.org/10.1021/acs.jctc.2c00910} {\bibfield  {journal}
  {\bibinfo  {journal} {Journal of Chemical Theory and Computation}\ ,\
  \bibinfo {pages} {1}} (\bibinfo {year} {2023})},\ \Eprint
  {https://arxiv.org/abs/https://doi.org/10.1021/acs.jctc.2c00910}
  {https://doi.org/10.1021/acs.jctc.2c00910} \BibitemShut {NoStop}%
\bibitem [{\citenamefont {Raussendorf}\ \emph {et~al.}(2020)\citenamefont
  {Raussendorf}, \citenamefont {Bermejo-Vega}, \citenamefont {Tyhurst},
  \citenamefont {Okay},\ and\ \citenamefont {Zurel}}]{raussendorf2020phase}%
  \BibitemOpen
  \bibfield  {author} {\bibinfo {author} {\bibfnamefont {R.}~\bibnamefont
  {Raussendorf}}, \bibinfo {author} {\bibfnamefont {J.}~\bibnamefont
  {Bermejo-Vega}}, \bibinfo {author} {\bibfnamefont {E.}~\bibnamefont
  {Tyhurst}}, \bibinfo {author} {\bibfnamefont {C.}~\bibnamefont {Okay}},\ and\
  \bibinfo {author} {\bibfnamefont {M.}~\bibnamefont {Zurel}},\ }\bibfield
  {title} {\bibinfo {title} {Phase-space-simulation method for quantum
  computation with magic states on qubits},\ }\href@noop {} {\bibfield
  {journal} {\bibinfo  {journal} {Physical Review A}\ }\textbf {\bibinfo
  {volume} {101}},\ \bibinfo {pages} {012350} (\bibinfo {year}
  {2020})}\BibitemShut {NoStop}%
\bibitem [{\citenamefont {Nielsen}\ and\ \citenamefont
  {Chuang}(2011)}]{nielsen2011quantum}%
  \BibitemOpen
  \bibfield  {author} {\bibinfo {author} {\bibfnamefont {M.~A.}\ \bibnamefont
  {Nielsen}}\ and\ \bibinfo {author} {\bibfnamefont {I.~L.}\ \bibnamefont
  {Chuang}},\ }\bibfield  {title} {\bibinfo {title} {Quantum computation and
  quantum information: 10th anniversary edition}\ }(\bibinfo  {publisher}
  {Cambridge University Press, Cambridge},\ \bibinfo {year} {2011})\ pp.\
  \bibinfo {pages} {454--459}\BibitemShut {NoStop}%
\bibitem [{\citenamefont {Poulin}\ \emph {et~al.}(2015)\citenamefont {Poulin},
  \citenamefont {Hastings}, \citenamefont {Wecker}, \citenamefont {Wiebe},
  \citenamefont {Doberty},\ and\ \citenamefont {Troyer}}]{poulin2014trotter}%
  \BibitemOpen
  \bibfield  {author} {\bibinfo {author} {\bibfnamefont {D.}~\bibnamefont
  {Poulin}}, \bibinfo {author} {\bibfnamefont {M.~B.}\ \bibnamefont
  {Hastings}}, \bibinfo {author} {\bibfnamefont {D.}~\bibnamefont {Wecker}},
  \bibinfo {author} {\bibfnamefont {N.}~\bibnamefont {Wiebe}}, \bibinfo
  {author} {\bibfnamefont {A.~C.}\ \bibnamefont {Doberty}},\ and\ \bibinfo
  {author} {\bibfnamefont {M.}~\bibnamefont {Troyer}},\ }\bibfield  {title}
  {\bibinfo {title} {The trotter step size required for accurate quantum
  simulation of quantum chemistry},\ }\href
  {https://doi.org/10.5555/2871401.2871402} {\bibfield  {journal} {\bibinfo
  {journal} {Quantum Info. Comput.}\ }\textbf {\bibinfo {volume} {15}},\
  \bibinfo {pages} {361–384} (\bibinfo {year} {2015})}\BibitemShut {NoStop}%
\bibitem [{\citenamefont {Lang}\ \emph {et~al.}(2021)\citenamefont {Lang},
  \citenamefont {Ryabinkin},\ and\ \citenamefont {Izmaylov}}]{lang2020unitary}%
  \BibitemOpen
  \bibfield  {author} {\bibinfo {author} {\bibfnamefont {R.~A.}\ \bibnamefont
  {Lang}}, \bibinfo {author} {\bibfnamefont {I.~G.}\ \bibnamefont
  {Ryabinkin}},\ and\ \bibinfo {author} {\bibfnamefont {A.~F.}\ \bibnamefont
  {Izmaylov}},\ }\bibfield  {title} {\bibinfo {title} {Unitary transformation
  of the electronic hamiltonian with an exact quadratic truncation of the
  {B}aker-{C}ampbell-{H}ausdorff expansion},\ }\href
  {https://doi.org/10.1021/acs.jctc.0c00170} {\bibfield  {journal} {\bibinfo
  {journal} {Journal of Chemical Theory and Computation}\ }\textbf {\bibinfo
  {volume} {17}},\ \bibinfo {pages} {66} (\bibinfo {year} {2021})},\ \bibinfo
  {note} {pMID: 33295175},\ \Eprint
  {https://arxiv.org/abs/https://doi.org/10.1021/acs.jctc.0c00170}
  {https://doi.org/10.1021/acs.jctc.0c00170} \BibitemShut {NoStop}%
\bibitem [{\citenamefont {Dehaene}\ and\ \citenamefont
  {De~Moor}(2003)}]{dehaene2003clifford}%
  \BibitemOpen
  \bibfield  {author} {\bibinfo {author} {\bibfnamefont {J.}~\bibnamefont
  {Dehaene}}\ and\ \bibinfo {author} {\bibfnamefont {B.}~\bibnamefont
  {De~Moor}},\ }\bibfield  {title} {\bibinfo {title} {Clifford group,
  stabilizer states, and linear and quadratic operations over gf (2)},\
  }\href@noop {} {\bibfield  {journal} {\bibinfo  {journal} {Physical Review
  A}\ }\textbf {\bibinfo {volume} {68}},\ \bibinfo {pages} {042318} (\bibinfo
  {year} {2003})}\BibitemShut {NoStop}%
\bibitem [{\citenamefont {Childs}\ and\ \citenamefont
  {Wiebe}(2012)}]{wiebe2012hamiltonian}%
  \BibitemOpen
  \bibfield  {author} {\bibinfo {author} {\bibfnamefont {A.~M.}\ \bibnamefont
  {Childs}}\ and\ \bibinfo {author} {\bibfnamefont {N.}~\bibnamefont {Wiebe}},\
  }\bibfield  {title} {\bibinfo {title} {Hamiltonian simulation using linear
  combinations of unitary operations},\ }\href@noop {} {\bibfield  {journal}
  {\bibinfo  {journal} {Quantum Info. Comput.}\ }\textbf {\bibinfo {volume}
  {12}},\ \bibinfo {pages} {901} (\bibinfo {year} {2012})}\BibitemShut
  {NoStop}%
\bibitem [{\citenamefont {Low}\ and\ \citenamefont
  {Chuang}(2019)}]{Low2019hamiltonian}%
  \BibitemOpen
  \bibfield  {author} {\bibinfo {author} {\bibfnamefont {G.~H.}\ \bibnamefont
  {Low}}\ and\ \bibinfo {author} {\bibfnamefont {I.~L.}\ \bibnamefont
  {Chuang}},\ }\bibfield  {title} {\bibinfo {title} {Hamiltonian {S}imulation
  by {Q}ubitization},\ }\href {https://doi.org/10.22331/q-2019-07-12-163}
  {\bibfield  {journal} {\bibinfo  {journal} {{Quantum}}\ }\textbf {\bibinfo
  {volume} {3}},\ \bibinfo {pages} {163} (\bibinfo {year} {2019})}\BibitemShut
  {NoStop}%
\bibitem [{\citenamefont {Berry}\ \emph {et~al.}(2014)\citenamefont {Berry},
  \citenamefont {Childs}, \citenamefont {Cleve}, \citenamefont {Kothari},\ and\
  \citenamefont {Somma}}]{OblivAmp14}%
  \BibitemOpen
  \bibfield  {author} {\bibinfo {author} {\bibfnamefont {D.~W.}\ \bibnamefont
  {Berry}}, \bibinfo {author} {\bibfnamefont {A.~M.}\ \bibnamefont {Childs}},
  \bibinfo {author} {\bibfnamefont {R.}~\bibnamefont {Cleve}}, \bibinfo
  {author} {\bibfnamefont {R.}~\bibnamefont {Kothari}},\ and\ \bibinfo {author}
  {\bibfnamefont {R.~D.}\ \bibnamefont {Somma}},\ }\bibfield  {title} {\bibinfo
  {title} {Exponential improvement in precision for simulating sparse
  hamiltonians},\ }in\ \href {https://doi.org/10.1145/2591796.2591854} {\emph
  {\bibinfo {booktitle} {Proceedings of the 46th Annual ACM Symposium on Theory
  of Computing}}}\ (\bibinfo  {publisher} {Association for Computing
  Machinery},\ \bibinfo {year} {2014})\ pp.\ \bibinfo {pages}
  {283--292}\BibitemShut {NoStop}%
\bibitem [{\citenamefont {Grover}(1997)}]{grover1997quantum}%
  \BibitemOpen
  \bibfield  {author} {\bibinfo {author} {\bibfnamefont {L.~K.}\ \bibnamefont
  {Grover}},\ }\bibfield  {title} {\bibinfo {title} {Quantum mechanics helps in
  searching for a needle in a haystack},\ }\href@noop {} {\bibfield  {journal}
  {\bibinfo  {journal} {Physical Review Letters}\ }\textbf {\bibinfo {volume}
  {79}},\ \bibinfo {pages} {325} (\bibinfo {year} {1997})}\BibitemShut
  {NoStop}%
\bibitem [{\citenamefont {Guerreschi}(2019)}]{guerreschi2019repeat}%
  \BibitemOpen
  \bibfield  {author} {\bibinfo {author} {\bibfnamefont {G.~G.}\ \bibnamefont
  {Guerreschi}},\ }\bibfield  {title} {\bibinfo {title} {Repeat-until-success
  circuits with fixed-point oblivious amplitude amplification},\ }\href@noop {}
  {\bibfield  {journal} {\bibinfo  {journal} {Physical Review A}\ }\textbf
  {\bibinfo {volume} {99}},\ \bibinfo {pages} {022306} (\bibinfo {year}
  {2019})}\BibitemShut {NoStop}%
\bibitem [{\citenamefont {Boyer}\ \emph {et~al.}(1998)\citenamefont {Boyer},
  \citenamefont {Brassard}, \citenamefont {Hoyer},\ and\ \citenamefont
  {Tapp}}]{boyer1998tight}%
  \BibitemOpen
  \bibfield  {author} {\bibinfo {author} {\bibfnamefont {M.}~\bibnamefont
  {Boyer}}, \bibinfo {author} {\bibfnamefont {G.}~\bibnamefont {Brassard}},
  \bibinfo {author} {\bibfnamefont {P.}~\bibnamefont {Hoyer}},\ and\ \bibinfo
  {author} {\bibfnamefont {A.}~\bibnamefont {Tapp}},\ }\bibfield  {title}
  {\bibinfo {title} {Tight bounds on quantum searching},\ }\href@noop {}
  {\bibfield  {journal} {\bibinfo  {journal} {Fortschritte der Physik: Progress
  of Physics}\ }\textbf {\bibinfo {volume} {46}},\ \bibinfo {pages} {493}
  (\bibinfo {year} {1998})}\BibitemShut {NoStop}%
\bibitem [{\citenamefont {Bravyi}\ \emph {et~al.}(2017)\citenamefont {Bravyi},
  \citenamefont {Gambetta}, \citenamefont {Mezzacapo},\ and\ \citenamefont
  {Temme}}]{bravyi2017tapering}%
  \BibitemOpen
  \bibfield  {author} {\bibinfo {author} {\bibfnamefont {S.}~\bibnamefont
  {Bravyi}}, \bibinfo {author} {\bibfnamefont {J.~M.}\ \bibnamefont
  {Gambetta}}, \bibinfo {author} {\bibfnamefont {A.}~\bibnamefont
  {Mezzacapo}},\ and\ \bibinfo {author} {\bibfnamefont {K.}~\bibnamefont
  {Temme}},\ }\bibfield  {title} {\bibinfo {title} {Tapering off qubits to
  simulate fermionic hamiltonians},\ }\href@noop {} {\bibfield  {journal}
  {\bibinfo  {journal} {arXiv preprint arXiv:1701.08213}\ } (\bibinfo {year}
  {2017})}\BibitemShut {NoStop}%
\bibitem [{\citenamefont {Kirby}(2021)}]{ContextualSubspaceVQEgithub}%
  \BibitemOpen
  \bibfield  {author} {\bibinfo {author} {\bibfnamefont {W.~M.}\ \bibnamefont
  {Kirby}},\ }\href@noop {} {\bibinfo {title} {Contextual{S}ubspace{VQE}}},\
  \bibinfo {howpublished}
  {\url{https://github.com/wmkirby1/ContextualSubspaceVQE}} (\bibinfo {year}
  {2021})\BibitemShut {NoStop}%
\bibitem [{\citenamefont {Hagberg}\ \emph {et~al.}(2008)\citenamefont
  {Hagberg}, \citenamefont {Schult},\ and\ \citenamefont
  {Swart}}]{SciPyProceedings_11}%
  \BibitemOpen
  \bibfield  {author} {\bibinfo {author} {\bibfnamefont {A.~A.}\ \bibnamefont
  {Hagberg}}, \bibinfo {author} {\bibfnamefont {D.~A.}\ \bibnamefont
  {Schult}},\ and\ \bibinfo {author} {\bibfnamefont {P.~J.}\ \bibnamefont
  {Swart}},\ }\bibfield  {title} {\bibinfo {title} {Exploring network
  structure, dynamics, and function using networkx},\ }in\ \href@noop {} {\emph
  {\bibinfo {booktitle} {Proceedings of the 7th Python in Science
  Conference}}},\ \bibinfo {editor} {edited by\ \bibinfo {editor}
  {\bibfnamefont {G.}~\bibnamefont {Varoquaux}}, \bibinfo {editor}
  {\bibfnamefont {T.}~\bibnamefont {Vaught}},\ and\ \bibinfo {editor}
  {\bibfnamefont {J.}~\bibnamefont {Millman}}}\ (\bibinfo {address} {Pasadena,
  CA USA},\ \bibinfo {year} {2008})\ pp.\ \bibinfo {pages} {11 --
  15}\BibitemShut {NoStop}%
\bibitem [{\citenamefont {Welsh}\ and\ \citenamefont
  {Powell}(1967)}]{welsh1967upper}%
  \BibitemOpen
  \bibfield  {author} {\bibinfo {author} {\bibfnamefont {D.~J.}\ \bibnamefont
  {Welsh}}\ and\ \bibinfo {author} {\bibfnamefont {M.~B.}\ \bibnamefont
  {Powell}},\ }\bibfield  {title} {\bibinfo {title} {An upper bound for the
  chromatic number of a graph and its application to timetabling problems},\
  }\href@noop {} {\bibfield  {journal} {\bibinfo  {journal} {The Computer
  Journal}\ }\textbf {\bibinfo {volume} {10}},\ \bibinfo {pages} {85} (\bibinfo
  {year} {1967})}\BibitemShut {NoStop}%
\bibitem [{\citenamefont {McClean}\ \emph {et~al.}(2016)\citenamefont
  {McClean}, \citenamefont {Romero}, \citenamefont {Babbush},\ and\
  \citenamefont {Aspuru-Guzik}}]{mcclean2016theory}%
  \BibitemOpen
  \bibfield  {author} {\bibinfo {author} {\bibfnamefont {J.~R.}\ \bibnamefont
  {McClean}}, \bibinfo {author} {\bibfnamefont {J.}~\bibnamefont {Romero}},
  \bibinfo {author} {\bibfnamefont {R.}~\bibnamefont {Babbush}},\ and\ \bibinfo
  {author} {\bibfnamefont {A.}~\bibnamefont {Aspuru-Guzik}},\ }\bibfield
  {title} {\bibinfo {title} {The theory of variational hybrid quantum-classical
  algorithms},\ }\href@noop {} {\bibfield  {journal} {\bibinfo  {journal} {New
  Journal of Physics}\ }\textbf {\bibinfo {volume} {18}},\ \bibinfo {pages}
  {023023} (\bibinfo {year} {2016})}\BibitemShut {NoStop}%
\bibitem [{\citenamefont {Rubin}\ \emph {et~al.}(2018)\citenamefont {Rubin},
  \citenamefont {Babbush},\ and\ \citenamefont
  {McClean}}]{rubin2018application}%
  \BibitemOpen
  \bibfield  {author} {\bibinfo {author} {\bibfnamefont {N.~C.}\ \bibnamefont
  {Rubin}}, \bibinfo {author} {\bibfnamefont {R.}~\bibnamefont {Babbush}},\
  and\ \bibinfo {author} {\bibfnamefont {J.}~\bibnamefont {McClean}},\
  }\bibfield  {title} {\bibinfo {title} {Application of fermionic marginal
  constraints to hybrid quantum algorithms},\ }\href@noop {} {\bibfield
  {journal} {\bibinfo  {journal} {New Journal of Physics}\ }\textbf {\bibinfo
  {volume} {20}},\ \bibinfo {pages} {053020} (\bibinfo {year}
  {2018})}\BibitemShut {NoStop}%
\bibitem [{\citenamefont {Gonthier}\ \emph {et~al.}(2022)\citenamefont
  {Gonthier}, \citenamefont {Radin}, \citenamefont {Buda}, \citenamefont
  {Doskocil}, \citenamefont {Abuan},\ and\ \citenamefont
  {Romero}}]{gonthier2022measurements}%
  \BibitemOpen
  \bibfield  {author} {\bibinfo {author} {\bibfnamefont {J.~F.}\ \bibnamefont
  {Gonthier}}, \bibinfo {author} {\bibfnamefont {M.~D.}\ \bibnamefont {Radin}},
  \bibinfo {author} {\bibfnamefont {C.}~\bibnamefont {Buda}}, \bibinfo {author}
  {\bibfnamefont {E.~J.}\ \bibnamefont {Doskocil}}, \bibinfo {author}
  {\bibfnamefont {C.~M.}\ \bibnamefont {Abuan}},\ and\ \bibinfo {author}
  {\bibfnamefont {J.}~\bibnamefont {Romero}},\ }\bibfield  {title} {\bibinfo
  {title} {Measurements as a roadblock to near-term practical quantum advantage
  in chemistry: resource analysis},\ }\href@noop {} {\bibfield  {journal}
  {\bibinfo  {journal} {Physical Review Research}\ }\textbf {\bibinfo {volume}
  {4}},\ \bibinfo {pages} {033154} (\bibinfo {year} {2022})}\BibitemShut
  {NoStop}%
\bibitem [{\citenamefont {Ralli}\ and\ \citenamefont
  {Weaving}(2022)}]{symmerGithub}%
  \BibitemOpen
  \bibfield  {author} {\bibinfo {author} {\bibfnamefont {A.}~\bibnamefont
  {Ralli}}\ and\ \bibinfo {author} {\bibfnamefont {T.~J.}\ \bibnamefont
  {Weaving}},\ }\href@noop {} {\bibinfo {title} {symmer}},\ \bibinfo
  {howpublished} {\url{https://github.com/UCL-CCS/symmer}} (\bibinfo {year}
  {2022})\BibitemShut {NoStop}%
\end{thebibliography}%

\clearpage
\appendix

\section{``Peres-Mermin Square"}
\zlabel{sec:Peres_square}

The ``Peres-Mermin square" \cite{mermin1990simple, peres1990incompatible} involves the construction of nine measurements arranged in a square. In this appendix we follow the construction given in \cite{budroni2021quantum}. Each measurement has only two possible outcomes (dichotomic) $+1$ and $-1$. In a realistic interpretation, performing each measurement on an object reveals whether the property is  present ($+1$) or absent ($-1$), yielding nine properties.

We take three measurements along a column or row to form a ``context"  - a set of measurements whose values can be jointly measured i.e. the observables commute and thus share a common eigenbasis. Table \zref{tab:square_example} gives an example.

\begin{table}[t]
\begin{tabular}{lll|l}
IZ & ZI & ZZ & $r_{0}$ \\
XI & IX & XX & $r_{1}$ \\
XZ & ZX & YY & $r_{2}$ \\ \hline
$c_{0}$  & $c_{1}$  & $c_{2}$  &  
\end{tabular}
\caption{Example Peres-Mermin square of nine possible observables for a physical system, where each measurement can be assigned a $\pm 1$ value. \zlabel{tab:square_example}} 
\end{table}

In a classical (noncontextual) model for this system, the nine measurements $\{IZ, ZI, ZZ, XI, IX, XX, XZ,ZX,YY \} $ can be assigned a definite value independent of the context the measurement is obtained in. For example if all measurements are assigned $+1$ in Table \zref{tab:square_example}, then $c_{0}=c_{1}=c_{2}=r_{0}=r_{1}=r_{2}=+1$ and six positive products are obtained. If a single entry in Table \zref{tab:square_example} is changed it will affect two products (a row and column product). We consider the following Equation in this setting:

\begin{equation}
    \zlabel{eq:inequality}
    \begin{aligned}
    \langle PM \rangle \equiv &\langle IZ \cdot ZI \cdot ZZ \rangle + \langle XI \cdot IX \cdot XX \rangle +  \\
    &\langle XZ \cdot ZX \cdot YY \rangle + \langle IZ \cdot XI \cdot XZ \rangle +  \\
    &\langle ZI \cdot IX \cdot ZX \rangle - \langle ZZ \cdot XX \cdot YY \rangle \\
    = &r_{0} + r_{1} + r_{2} + c_{0}+ c_{1} - c_{2}.
    \end{aligned}
\end{equation}
We find that classically we get an inequality $\langle PM \rangle \leq 4$. We reiterate that this is the setting of eight $+1$ assignments and a single $-1$ assignment. This inequality is saturated when the $-1$ value is assigned to one of the observables in the last column of Table \zref{tab:square_example}.

The significance of this inequality is that it can be violated by quantum systems. Thinking of this in a quantum setting, the operators in rows and columns of Table \zref{tab:square_example} commute. If we multiply along the rows and columns we get $+II$ apart from the last column where $c_{2} = -II$ (see Table \zref{tab:square_example2}). This is the case regardless of what quantum state is considered . Using the expectation values of the product of these operators in Equation \zref{eq:inequality}, we find $\langle PM \rangle =6$, violating the classical bound.

\begin{table}[b]
\begin{tabular}{lll|l}
IZ & ZI & ZZ & $\langle +II \rangle=+1$ \\
XI & IX & XX & $\langle +II \rangle=+1$ \\
XZ & ZX & YY & $\langle +II \rangle=+1$ \\ \hline
 $\langle +II \rangle=+1$  & $\langle +II \rangle=+1$  & $\langle -II \rangle=-1$  &  
\end{tabular}
\caption{Example Peres-Mermin square of nine Hermitian operators, all with $\pm 1$ eigenvalues - representing observables. \zlabel{tab:square_example2}} 
\end{table}

Classically Equation \zref{eq:inequality} is bounded as $\langle PM \rangle \leq 4$ due to the assumption that the nine observables of the object can be assigned a value consistently. Violation of this bound implies that either the value assignment must depend on which context (row or column) the observable appears in or there is no value assignment. This phenomenon is known as quantum contextuality \cite{budroni2021quantum}.

In VQE, a Hamiltonian is defined by a linear combination of Pauli operators. The expectation value is obtained by measuring each Pauli operator in a separate experiment and combining the results. Different groups of commuting operators form contexts. In general there will be incompatible contexts where it is impossible to consistently assign joint outcomes. In other words, different inference relations will lead to contradictions. Outcomes assigned to individual measurements are therefore context-dependent and the problem is contextual. If not, then the problem is noncontextual and a noncontextual (classical) hidden variable model can be used to solve such systems.

\end{document}


\preprint{APS/123-QED}

\title{Supplemental Material \\Unitary Partitioning and the Contextual Subspace \\ Variational Quantum Eigensolver}

\author{Alexis Ralli}
\email{alexis.ralli.18@ucl.ac.uk}
\affiliation{
Centre for Computational Science, Department of Chemistry, University College London, WC1H 0AJ, United Kingdom
}
\author{Tim Weaving}
\email{timothy.weaving.20@ucl.ac.uk}
\affiliation{
Centre for Computational Science, Department of Chemistry, University College London, WC1H 0AJ, United Kingdom
}
\author{Andrew Tranter}
\email{tufts@atranter.net}
\affiliation{
 Department of Physics and Astronomy, Tufts University, Medford, MA 02155, USA
}
\affiliation{Cambridge Quantum Computing, 9a Bridge Street Cambridge, CB2 1UB, United Kingdom
}
\author{William M. Kirby}
\email{william.kirby@tufts.edu}
\affiliation{
 Department of Physics and Astronomy, Tufts University, Medford, MA 02155, USA
}
\author{Peter J. Love}
\email{peter.love@tufts.edu}
\affiliation{
 Department of Physics and Astronomy, Tufts University, Medford, MA 02155, USA
}
\affiliation{Computational Science Initiative, Brookhaven National Laboratory, Upton, NY 11973, USA}
\author{Peter V. Coveney}
\email{p.v.coveney@ucl.ac.uk}
\affiliation{
Centre for Computational Science, Department of Chemistry, University College London, WC1H 0AJ, United Kingdom
}
\affiliation{
Informatics Institute, University of Amsterdam, Amsterdam, 1098 XH, Netherlands
}


\date{\today}


\maketitle

\tableofcontents

\onecolumngrid

\section{\label{sec:CS_VQE_review}Contextual-Subspace VQE overview}

In the following subsections we summarise all the details required to implement the CS-VQE algorithm, where a problem Hamiltonian is split  into a contextual and noncontextual part \cite{kirby2021contextual}:

\begin{equation}
    \label{eq:Hamilt_appendix}
    H_{\text{full}} = H_{\text{con}} + H_{\text{noncon}}.
\end{equation}



\subsection{\label{sec:NoncontextualPart}Noncontextual Part}

\subsubsection{\label{sec:nonconTest}Testing for contexuality}

Let $\mathcal{S}^{H_{\text{full}}}$ be the set of Pauli operators, in the full system Hamiltonian, requiring measurement in a VQE experiment. It was shown in \cite{raussendorf2020phase, kirby2019contextuality}, that a set of four Pauli operators $\{A, B, C, D\}$ is strongly contextual if $A$ commutes with both $B$ and $C$, but $B$ anticommutes with $C$. If this condition is not present, then the set is noncontextual.

Given an arbitrary set of Pauli operators $\mathcal{P}$, this gives an algorithm to check for contextuality \cite{kirby2019contextuality}. A pseudo algorithm is given in algorithm \ref{alg:contextuality_test}. First an $\mathcal{O}(|\mathcal{P}|^{2})$ routine is used to remove completely commuting operators $\mathcal{Z}$, leaving the remaining set $\mathcal{T}$. Then a procedure taking $\mathcal{O}(|\mathcal{T}|^{3})$ steps is used to determine whether $\mathcal{P}$ is contextual. This check for contextuality is implemented in OpenFermion \cite{mcclean2020openfermion}.

\begin{figure}
\begin{algorithm}[H]
\caption{Test for strong contextuality in a given set of Pauli operators \cite{kirby2019contextuality}}
\label{alg:contextuality_test}
\begin{algorithmic}
\State $\textbf{Input: }\mathcal{P} = \{P_{0}, P_{1}, P_{2}, ... \}$  \Comment{Input $\mathcal{P}$ is a set of Pauli operators.}
\State $\textbf{Output: } contextual\: (True/False)$ \Comment{Whether the set $\mathcal{P}$ is strongly contextual.}
\item[]
\State $\mathcal{Z} \gets \{ \}$
\State $\mathcal{T} \gets \{ \}$
\State $contextual \gets False$
\item[]
\For{$i= 0$ to $|\mathcal{P}|-1$} 
    \If{$[P_{i}, P_{j}]=0$ $\forall j \neq i$ where $j=0$ to $|\mathcal{P}|-1$}
        \State $\mathcal{Z} \gets \mathcal{Z} \cup \{P_{i}\}$
    \Else
        \State $\mathcal{T} \gets \mathcal{T} \cup \{P_{i}\}$
        \EndIf
\EndFor
\item[]
\For{$i= 0$ to $|\mathcal{T}|-3$}
    \For{$j = i+1$ to $|\mathcal{T}|-2$}
        \For{$k = j+1$ to $|\mathcal{T}|-1$}
            \item[]
            \If{$[P_{i}, P_{j}]=0$, $[P_{i}, P_{k}]=0$ and $\{P_{j}, P_{k}\}=0$}
                \State $contextual \gets True$
                \State $\textbf{return } contextual$
            \Else
                \State $\textbf{continue}$
            \EndIf
            \item[]
        \EndFor
    \EndFor
\EndFor
\State $\textbf{return } contextual$
\end{algorithmic}
\end{algorithm}
\end{figure}

\subsubsection{\label{sec:getHnoncon}Obtaining Noncontexual Hamiltonian}

To begin CS-VQE,  we first need to define the contextual and noncontextual parts. The task of finding the largest noncontextual subset of Pauli operators in $\mathcal{S}^{H_{\text{full}}}$ is a generalization of the disjoint cliques problem \cite{wagner8linear, kirby2020classical}, which is NP-complete. However, different heuristics can be used to approximately solve this problem.

To date, VQE experiments have mainly focused on chemistry Hamiltonians, where Hartree-Fock accounts for most of the energy. Such Hamiltonians contain Pauli operators that $l_{1}$ norms are dominated by diagonal terms - Pauli operators made up of tensor products of single qubit $I$ and Pauli $Z$ matrices. To find a noncontextual set in such a scenario, a greedy heuristic selecting high weight terms from the full Hamiltonian first can be used, while checking the set remains noncontextual using algorithm \ref{alg:contextuality_test}. This gives a noncontextual set containing mainly diagonal terms, with some additional operators \cite{kirby2020classical}. Alternative procedures to find the largest noncontextual subsets remain an open question for the CS-VQE algorithm.

\subsubsection{\label{sec:HnonconHiddenVar}Noncontextual hidden variable model}

Once the noncontextual Hamiltonian $H_{\text{noncon}}$ is determined, we can define the set $\mathcal{S}^{H_{\text{noncon}}}$ to be the Pauli operators in $H_{\text{noncon}}$. We split $\mathcal{S}^{H_{\text{noncon}}}$ into two subsets $\mathcal{Z}$ and $\mathcal{T}$ - representing the set of universally commuting Pauli operators $\mathcal{Z}$ and their complement respectively \cite{kirby2020classical, kirby2021contextual}:

\begin{equation}
    \label{eq:non_con_set}
    \mathcal{S}^{H_{\text{noncon}}} = \mathcal{Z} \cup \mathcal{T} =  \Bigg\{ \bigcup_{\substack{i=0 \\  \forall P_{i} \in \mathcal{Z}} \\ }^{| \mathcal{Z}|-1} P_{i} \Bigg\} \cup \Bigg\{ \bigcup_{\substack{i=0 \\  \forall P_{i} \in \mathcal{T}} \\ }^{| \mathcal{T}|-1} P_{i} \Bigg\}.
\end{equation}
Slight modifications to Algorithm \ref{alg:contextuality_test} achieve this - where $\mathcal{P}$ would be set to be  $\mathcal{S}^{H_{\text{noncon}}}$ and both $\mathcal{Z}, \mathcal{T}$ should be returned.

The operators in $\mathcal{Z}$ are noncontextual, as by definition they are universally commuting and represent symmetries of $H_{\text{noncon}}$. For the overall super-set $\mathcal{S}^{H_{\text{noncon}}}$ to be noncontextual, the remaining operators in $\mathcal{T}$ must be made up of $N$ disjoint cliques $C_{j}$ \cite{kirby2020classical}, where operators within a clique must all commute with each other and operators between cliques pairwise anticommute. This is because commutation forms an equivalence relation on $\mathcal{T}$ if and only if $\mathcal{S}^{H_{\text{noncon}}}$ is noncontextual \cite{kirby2020classical, kirby2021contextual, raussendorf2020phase}. We can write $\mathcal{T}$ as:

\begin{equation}
    \label{eq:set_T}
    \mathcal{T} = \bigcup_{\substack{i=0 \\  \forall P_{i} \in \mathcal{T}}}^{| \mathcal{T}|-1} P_{i} = \bigcup_{j=0}^{N-1}  C_{j}  = \bigcup_{j=0}^{N-1} \Big( \bigcup_{\substack{k=0 \\  \text{where} \\ [P_{k}, P_{l}]=0 \\ \forall P_{k}, P_{l} \in C_{j}}}^{|  C_{j} |-1} P_{k}^{(j)}\Big).
\end{equation}
We re-define each clique $C_{j}$ using the identity operation defined by the first operator of the $j$th clique, $P_{0}^{(j)} P_{0}^{(j)} = \mathcal{I}$, which represents of the first operator in each of the $N$ cliques. We  write the $j$th clique as \cite{kirby2020classical}:

\begin{equation}
    \label{eq:clique_redef}
    C_{j} = \bigcup_{\forall P_{k} \in C_{j} } P_{k}^{(j)} =  \bigcup_{\forall P_{k} \in C_{j} }  P_{k}^{(j)} P_{0}^{(j)} P_{0}^{(j)} =  \bigcup_{\forall P_{k} \in C_{j} } \Big(P_{k}^{(j)} P_{0}^{(j)} \Big) P_{0}^{(j)} =  \bigcup_{k=0}^{|C_{j}|-1} A_{k}^{(j)} P_{0}^{(j)}
\end{equation}
The new operators $A_{k}^{(j)} = P_{k}^{(j)} P_{0}^{(j)}$ are just Pauli operators up to a complex phase. The new operators  $A_{k}^{(j)}$ must still commute with the universally commuting operators in $\mathcal{Z}$, but now must also commute with all the other terms in the $N-1$ cliques $C_{j}$ \cite{kirby2020classical}. Using this, the noncontextual set (Equation \ref{eq:non_con_set}) can be rewritten as:

\begin{subequations} \label{eq:Hamiltnoncon_full_set}
    \begin{align}
       \mathcal{S}^{H_{\text{noncon}}} &=  \Bigg\{ \bigcup_{\substack{i=0 \\  \forall P_{i} \in \mathcal{Z}}}^{| \mathcal{Z}|-1} P_{i} \Bigg\} \cup \Bigg\{ \bigcup_{j=0}^{N-1} C_{j} \Bigg\}  \\
      &=   \underbrace{\Bigg\{ \bigcup_{\substack{i=0 \\  \forall P_{i} \in \mathcal{Z}}}^{| \mathcal{Z}|-1} P_{i} \Bigg\}}_{\mathcal{Z}} \cup    \underbrace{\Bigg\{ \bigcup_{j=0}^{N-1} \Big( \bigcup_{\substack{k=0 \\  \text{where} \\ [P_{k}, P_{l}]=0 \\ \forall P_{k}, P_{l} \in C_{j}}}^{|  C_{j} |-1} A_{k}^{(j)} P_{0}^{(j)}\Big) \Bigg\}}_{\mathcal{T}}.
    \end{align}
\end{subequations}

So far we have considered the noncontextual set of Pauli operators $\mathcal{S}^{H_{\text{noncon}}}$, which in general will be a dependent set. By this we mean that some operators in the set can be written as a product of other commuting operators in the set. We need to reduce this set  $\mathcal{S}^{H_{\text{noncon}}}$ to an independent set of Pauli operators, where all operators in the noncontextual Hamiltonian can be inferred from the values of other operators in the set under the Jordan product. 

To obtain an independent set from $\mathcal{S}^{H_{\text{noncon}}}$, we first take the completely commuting Pauli operators:

\begin{equation}
\label{eq:G_non_indpt_set}
      \begin{aligned}
     \mathcal{G}' &\equiv \mathcal{Z} \cup \Bigg\{ \bigcup\limits_{j=0}^{N-1} \{A_{k}^{(j)} | k=1,2,\hdots, |C_{j}|-1 \} \Bigg\}\\
     &\equiv \Bigg\{ \bigcup_{\substack{i=0 \\  \forall P_{i} \in \mathcal{Z}}}^{| \mathcal{Z}|-1} P_{i} \Bigg\} \cup \Bigg\{ \bigcup\limits_{j=0}^{N-1} \{A_{k}^{(j)} | k=1,2,\hdots, |C_{j}|-1 \} \Bigg\}, 
    \end{aligned}
\end{equation}
and using the procedure in \cite{nielsen2011quantum} find an independent subset $\mathcal{G}$:
\begin{equation}
    \label{eq:G_indpt_set}
     \mathcal{G} \equiv \{ P_{i} | i=0,1,\hdots, |\mathcal{G}|-1 \} .
\end{equation}
Appendix C in \cite{kirby2020classical} gives all steps required. 

Finally, we need to consider the $N$ pairwise anticommuting $P_{0}^{(j)}$ operators defined by the $N$ anticommuting cliques. As the operators in $\mathcal{G}$ universally commute with all operators in the noncontextual Hamiltonian, each operator in the set  $\{ P_{0}^{(j)} | j=0,1,\hdots, (N-1) \}$ must be independent of $\mathcal{G}$ under the Jordan product \cite{kirby2020classical}. Combining these results, we get the set $\mathcal{R}$:

\begin{equation}
    \label{eq:R_indpt_set_anti}
     \mathcal{R} \equiv \{ P_{0}^{(j)} | j=0,1,\hdots, N-1 \} \cup \mathcal{G}
\end{equation}
Inspecting the properties of $\mathcal{R}$, one can bound its size. The set $\mathcal{G}$ has size at most $n-1$, as $n$ independent commuting Pauli operators form a complete commuting set of observables for $n$ qubits. In other words, as  $\mathcal{G}$ is a universally commuting set, if its size was $n$ (or more) then taking  $\mathcal{G}$ and one operator $P_{0}^{(j)}$ (the set $\mathcal{G} \cup \{P_{0}^{(j)} \}$) is also a fully commuting set - and would be a commuting set of size $n+1$ (or more) \cite{kirby2020classical}. The maximum number of independent anticomuting operators on $n$ qubits was shown in \cite{raussendorf2020phase} to be $2n+1$. This actually bounds the size of $\mathcal{R}$, which occurs when the set $\mathcal{G}$ (and thus $\mathcal{Z}$) is empty \cite{kirby2020classical}. 


Looking at the noncontextual set of Pauli operators Equation \ref{eq:Hamiltnoncon_full_set}, making up the $H_{\text{noncon}}$, we see that the subset $\mathcal{G}$ in $\mathcal{R}$ (Equation \ref{eq:R_indpt_set_anti}) includes all the generators for the terms in $\mathcal{Z}$ and each Pauli $A_{k}^{(j)}$ operator. Any operator in $\mathcal{Z}$ and each Pauli $A_{k}^{(j)}$ operator can therefore be found by a finite combination of operators in $\mathcal{G}$. Each operator in $\mathcal{T}$ can also be generated by a combination of one $P_{0}^{(j)}$ operator and some combination of operators in  $\mathcal{G}$. Again $\mathcal{R}$ (Equation \ref{eq:R_indpt_set_anti}) contains all the operators required. To summarise, the set $\mathcal{R}$ contains all the required terms to reproduce the expectation value of any operator in $\mathcal{S}^{H_{\text{noncon}}}$  under the Jordan product.

We have shown how the Pauli operators making $H_{\text{noncon}}$ can be simultaneously assigned definite values without contradiction. This allows the introduction of a phase-space description of the eigenspace of $H_{\text{noncon}}$ \cite{kirby2020classical}. Next we will introduce what this phase-space model is in the context of this work.

A joint value assignment of $\pm 1$ to each operator in $\mathcal{R}$ represents the ontic state of the physical system. Probability distributions corresponding to valid quantum states must obey an uncertainty relation \cite{spekkens2016quasi, kirby2020classical}. To enforce these two conditions are sufficient. (1) The commuting generators $\mathcal{G}$ have definite values and (2) the expectation values for the $P_{0}^{(j)}$ terms form a unit vector \cite{kirby2020classical}. In this frame, our noncontextual state is defined as \cite{kirby2020classical}:

\begin{equation}
    \label{eq:state}
       (\vec{q}, \vec{r}) =  (q_{0}, q_{1}, \hdots, q_{|\mathcal{G}|-1},  \: \: r_{0}, r_{1}, \hdots, r_{N-1} ).
\end{equation}
With respect to the phase-space model \cite{kirby2020classical}, a valid noncontextual state $(\vec{q}, \vec{r})$ sets the expectation value of the operators in $\mathcal{R}$, where $\langle G_{i} \rangle=q_{i}=\pm 1$ and $ \langle P_{0}^{(j)} \rangle=r_{j}$ such that $\big(\sum_{j=0}^{N-1} |r_{j}|^{2}\big)^{1/2}=1$. 

It was shown in \cite{kirby2020classical} that probabilities for outcomes $G_{i}$ and $P_{0}^{(j)}$ should be obtained as the marginals of:

\begin{equation}
    \label{eq:marginals}
       P(p_{j=0}, p_{j=1},  \hdots,  p_{j=(N-1)}, g_{0}, g_{1}, \hdots, g_{|\mathcal{G}|-1}) = \Bigg(\prod_{i=0}^{|\mathcal{G}|-1} \delta_{g_{i}, q_{i}} \Bigg)  \Bigg( \prod_{j=0}^{N-1} \frac{1}{2} | p_{j} + r_{j}| \Bigg).
\end{equation}
Further analysis is given in \cite{kirby2020classical}.

In summary, a noncontextual state is fully defined by $(\vec{q}, \vec{r})$ which determines all the expectation values of the operators in $\mathcal{R}$ (Equation \ref{eq:R_indpt_set_anti}), where $\langle G_{i} \rangle = q_{i} \in \{-1,+1 \}$ and $\langle  P_{0}^{(j)} \rangle = r_{j}$. We summarise this as:

\begin{equation}
    \label{eq:summary_R_set}
    \underbrace{ \{ \; \underbrace{\langle P_{0}^{(0)} \rangle}_{r_{0}}, \underbrace{\langle P_{0}^{(1)} \rangle}_{r_{1}} , \hdots, \underbrace{\langle P_{0}^{(N-1)} \rangle}_{r_{N-1}} \; \} }_{\vec{r}} \text{and} \underbrace{ \{ \; \underbrace{\langle G_{0} \rangle}_{q_{0}}, \underbrace{\langle G_{1} \rangle}_{q_{1}} , \hdots, \underbrace{\langle G_{|\mathcal{G}|-1}\rangle}_{q_{|\mathcal{G}|-1}} \; \} }_{\vec{q}}.
\end{equation}
The expectation value of all the operators in $\mathcal{S}^{H_{\text{noncon}}}$ are generated from some finite combination of terms in $\mathcal{R}$ under the Jordan product. This by extension will induce the expectation value for $H_{\text{noncon}}$. Explicitly, let $P_{i}^{\mathcal{Z}} \in \mathcal{Z} \subseteq \mathcal{S}^{H_{\text{noncon}}} $ then if we let $\mathcal{J}^{\mathcal{G}}_{P_{i}^{\mathcal{Z}}}$ be the set of indices such that $P_{i}^{\mathcal{Z}} = \prod_{i \in \mathcal{J}^{\mathcal{G}}_{P_{i}^{\mathcal{Z}}}} G_{i}$; then \cite{kirby2020classical}:

\begin{equation}
    \label{eq:expPz}
    \begin{aligned}
       \langle P_{i}^{\mathcal{Z}} \rangle =\prod_{i \in \mathcal{J}^{\mathcal{G}}_{P_{i}^{\mathcal{Z}}}} \langle G_{i} \rangle = \prod_{i \in \mathcal{J}^{\mathcal{G}}_{P_{i}^{\mathcal{Z}}}} q_{i}.
    \end{aligned}
\end{equation}
In words, we combine the expectation value of some finite set of Pauli operators in the independent set $\mathcal{G}$ - given by $\mathcal{J}^{\mathcal{G}}_{P_{i}^{\mathcal{Z}}}$ - to reproduce the expectation value for $\langle P_{i}^{\mathcal{Z}} \rangle$.

Similarly, the expectation value for each $A_{k}^{(j)} P_{0}^{(j)} \in \mathcal{T} \subseteq \mathcal{S}^{H_{\text{noncon}}}$ (Equation \ref{eq:Hamiltnoncon_full_set}) term is given by:

\begin{equation}
    \label{eq:expAPj}
    \begin{aligned}
       \langle A_{i}^{(j), \in \mathcal{G}} P_{0}^{(j), \in \mathcal{T}} \rangle = \Big( \prod_{i \in \mathcal{J}^{\mathcal{G}}_{A_{i}^{(j)}}} \langle G_{i} \rangle \Big)  r_{j} =   \Big( \prod_{i \in \mathcal{J}^{\mathcal{G}}_{A_{i}^{(j)}}} q_{i} \Big) r_{j},
    \end{aligned}
\end{equation}
where $\mathcal{J}^{\mathcal{G}}_{A_{i}^{(j)}}$ are the set of indices such that $ \langle A_{i}^{(j)} \rangle = \prod_{i \in \mathcal{J}^{\mathcal{G}}_{A_{i}^{(j)}}} \langle G_{i}\rangle$ and $\langle  P_{0}^{(j)} \rangle = r_{j}$ \cite{kirby2020classical}.

We can write the noncontexual Hamiltonian as:

\begin{equation}
    \label{eq:nonconHrewritten}
    \begin{aligned}
       H_{\text{noncon}} = \Big( \sum_{i=0}^{|\mathcal{Z}|-1} c_{i}  P_{i}^{\mathcal{Z}}  \Big) + \sum_{j=0}^{N-1} \Bigg[ \sum_{k=0}^{|C_{j}|-1} a_{k}A_{k}^{(j)} P_{0}^{(j)} \Bigg]
    \end{aligned}
\end{equation}

and find the energy by:

\begin{equation}
    \label{eq:nonconEnergy}
    \begin{aligned}
       \langle H_{\text{noncon}} \rangle = E_{noncon}(\vec{q}, \vec{r}) = \Big( \sum_{i=0}^{|\mathcal{Z}|-1} \beta_{i}  \langle P_{i}^{\mathcal{Z}}  \rangle \Big)  + \sum_{j=0}^{N-1} \Bigg[ \sum_{k=0}^{|C_{j}|-1} \beta_{k} \langle A_{k}^{(j)} P_{0}^{(j)} \rangle \Bigg],
    \end{aligned}
\end{equation}

where $\beta_{i}$ and $\beta_{k}$ are real coefficients and each expectation value is given by Equation \ref{eq:expPz} and  \ref{eq:expAPj} \cite{kirby2020classical}.

\subsubsection{\label{sec:SolveHnoncon}Solving the Noncontexual Hamiltonian}

To find the ground state of $H_{\text{noncon}}$, we minimize Equation \ref{eq:nonconEnergy} via a brute-force search as described in \cite{kirby2020classical}. Algorithm \ref{alg:Variational_Noncon} summarises the steps. This could be done in the work presented here, because  $\mathcal{R}$ was small for all molecular systems considered. First a trial $\vec{q}$ is defined.  This is a set of $\pm 1$ expectation values for each $G_{j}$. An initial guess of the amplitudes $r_{i}$ of the unit vector $\vec{r}$ is made and the energy (Equation \ref{eq:nonconEnergy}) is minimized over this continuous parameterization of $\vec{r}$ for a fixed trial $\vec{q}$, until the energy converges to a minimum. These steps are repeated for all the $2^{|\mathcal{G}|}$ assignments of $\vec{q}$. The $(\vec{q}, \vec{r})$ combination that gives the lowest overall energy represents the noncontextual ground state of the physical system. In the main text we denote this parameterization as $(\vec{q}_{0}, \vec{r}_{0})$. Note for a fixed $\vec{q}$, we optimize over $\vec{r}$. This can be thought of as optimizing a function defined on a hypersphere. Currently we haven't explored the properties of this function.

It remains an open question for the CS-VQE algorithm if alternate optimization strategies are possible, for example using chemical intuition during optimization. This brute force approach of searching over all $2^{|\mathcal{G}|}$ possibilities for $\vec{q}$ may not be necessary. In the next section, we discuss how to map the contextual problem into a subspace consistent with a defined noncontextual state $(\vec{q}, \vec{r})$.

\begin{algorithm}[H]
\caption{Brute force method to solve noncontexual problem}
\label{alg:Variational_Noncon}
\begin{algorithmic}[2]
\State $\mathcal{Q} \gets \{q_{0}, {q_{1}, ..., q_{|\mathcal{G}|-1}} \}^{2^{|\mathcal{G}|}}$ \Comment{Set $\mathcal{Q}$ contains all possible $\vec{q}$ vectors, where $q_{i} \in \{+1, -1 \}$.}
\item[]
\State $\vec{q}_{0} \gets \{ \}$
\State $\vec{r}_{0} \gets \{ \}$
\State $E_{\text{noncon}}^{0} \gets 0$
\item[]
\For{$\vec{q}_{\text{test}}$ in $\mathcal{Q}$} 
    \State $\vec{r}_{opt}, E_{\text{noncon}}^{opt} \gets \underset{\vec{r}}{argmin}[E_{\text{noncon}}(\vec{q}_{\text{test}}, \vec{r})]$ \Comment{for a given $\vec{q}_{\text{test}}$, minimize the energy (Equation \ref{eq:nonconEnergy})with respect to $\vec{r}$.}
    \item[]
    \If{$E_{\text{noncon}}^{opt} < E_{\text{noncon}}^{0}$}
        \State $\vec{q}_{0} \gets \vec{q}_{\text{test}}$
        \State $\vec{r}_{0} \gets \vec{r}_{opt}$
        \State $E_{\text{noncon}}^{0} \gets E_{\text{noncon}}^{opt}$
    \Else
        \State $\textbf{continue}$
        \EndIf
\EndFor
\item[]
\State $\textbf{return } \vec{q}_{0}, \vec{r}_{0}, E_{\text{noncon}}^{0}$
\end{algorithmic}
\end{algorithm}

\subsection{\label{sec:ContextualmappinpTransform}Mapping to contextual subspace}

In section \ref{sec:NonMappingConSub} of the main text, the full Hamiltonian is mapped contextual subspace consistent with the noncontextual ground state by implementing $U_{\mathcal{W}}$ (Equation \ref{eq:rotation_U}) followed by projecting the rotated Hamiltonian with $Q_{\mathcal{W}}$. However, the definitions for the operators making up $U_{\mathcal{W}}$ were omitted. This subsection gives these details and is split into three parts. The first two parts consider how $R$ is constructed. This is the problem of mapping a linear combination of pairwise anticommuting Pauli operators to a single Pauli operator and is known as unitary partitioning \cite{zhao2019measurement, izmaylov2019unitary}. In the context of this work, we use this to define $R$ such that $RA(\vec{r})R^{\dagger} \mapsto P_{0}^{(k)}$. In the original formulation of CS-VQE only the sequence of rotations construction of $R$ is used in the algorithm. We provide an alternative approach using the linear combination of unitaries construction proposed in \cite{zhao2019measurement}, which results in superior scaling. We show the effect each conjugate rotation $R$ has on the number of terms in a given qubit Hamiltonian. The last subsection gives the unitary rotations required to map a commuting set of Pauli operators to single qubit Pauli $Z$ operators. 

In each of these subsections, we use the notation that Pauli operators with multiple indices represent the multiplication of Pauli operators: $P_{a}P_{b}P_{c} = P_{abc}$. These terms will also be Pauli operators up to a complex phase.

\subsubsection{\label{sec:R_seqrot} Unitary partitioning via a sequence of rotations} 

In this subsection, we show how $A(\vec{r}) \mapsto P_{0}^{(k)} = R_{S}  A(\vec{r})  R_{S} ^{\dagger}$, where $R_{S}$ is defined by a sequence of rotations \cite{zhao2019measurement, ralli2021implementation}. Given the set of anticommuting operators $A(\vec{r})$ (Equation \ref{eq:anticomm}), we can define the following self-inverse operators:

\begin{equation}
    \label{eq:chi_sk}
    \{\mathcal{X}_{kj} = i P_{0}^{(k)} P_{0}^{(j)} \;\; \forall P_{0}^{(j)} \in \mathcal{A} \text{ where } j \neq k \},
\end{equation}
where $P_{0}^{(k)} \in \mathcal{A}$. To simplify the notation we drop the subscript $0$ (denoting the first operator in a clique) and write each $ P_{0}^{(k)}$, $ P_{0}^{(j)}$ as $ P_{k}$ and $ P_{j}$ respectively.

The adjoint rotation generated by one of these operator $\mathcal{X}_{kj}$ operators will be:

\begin{equation}
    \label{eq:R_sk_action}
\begin{aligned}
e^{(-i \frac{\theta_{kj}}{2}\mathcal{X}_{kj})}  A(\vec{r})  e^{(+i \frac{\theta_{kj}}{2}\mathcal{X}_{kj})} &=  R_{S_{kj}}(\theta_{kj}) A(\vec{r}) R_{S_{kj}}^{\dagger}(\theta_{kj}) \\ 
&= \big( r_{j} \cos\theta_{kj} - r_{k} \sin\theta_{kj} \big) P_{j} + \big( \beta_{j} \sin\theta_{kj} + r_{k} \cos\theta_{kj} \big) P_{k} + \sum_{\substack{P_{l} \in \mathcal{A} \\ \forall l \neq k,j}} \beta_{l} P_{l} .
\end{aligned}
\end{equation}

\noindent The coefficient of $P_{j}$ can be made to go to $0$, by setting $r_{j} \cos \theta_{kj} = r_{k} \sin\theta_{kj}$. This approach removes the term with index $j$ and increases the coefficient of $P_{k}$ from $r_{k} \mapsto \sqrt{r_{k}^{2} + r_{j}^{2}}$ \cite{zhao2019measurement}. This process is repeated over all indices excluding $j = k$ until only the $P_{k}$ term remains. This procedure can be concisely written using the following operator \cite{zhao2019measurement}:

\begin{equation}
    \label{eq:R_SeqRot}
    \begin{aligned}
    R_{S} &=   \prod_{\substack{j=0 \\  \forall j \neq k}}^{|\mathcal{A}|-1} e^{(-i \frac{\theta_{kj}}{2})  \mathcal{X}_{kj}}  =   \prod_{\substack{j=0 \\  \forall j \neq k}}^{|\mathcal{A}|-1} R_{S_{kj}} (\theta_{kj})  = \prod_{\substack{j=0 \\  \forall j \neq k}}^{|\mathcal{A}|-1} \Big[ \cos \big( \frac{\theta_{kj}}{2} \big) \mathcal{I} - i \sin \big( \frac{\theta_{kj}}{2} \big)  \mathcal{X}_{kj} \Big],
    \end{aligned}
\end{equation}

\noindent which is simply a sequence of rotations. The angle $\theta_{kj}$  is defined recursively at each step of the removal process, as  the coefficient  of $P_{k}$ increases at each step and thus must be taken into account. The correct solution for $\theta_{kj}$ must be chosen given the signs of $r_{k}$ and $r_{k}$ \cite{zhao2019measurement}. The overall action of this sequence of rotations is:

\begin{equation}
 \label{eq:RsHsRs}
	R_{S}  A(\vec{r})  R_{S} ^{\dagger} =  P_{k}.
\end{equation}

Looking at Equation \ref{eq:R_SeqRot}, expanding the product of rotations results in $R_{S}$ containing $\mathcal{O}(2^{|\mathcal{A}|-1})$  Pauli operators. We write this operator as:

\begin{equation}
    \label{eq:R_s_expanded}
     R_{S} = \sum_{b}^{\mathcal{O}(2^{|\mathcal{A}|-1})} \delta_{b} P_{b}.
\end{equation}
The adjoint rotation of $R_{S}$ on a general Hamiltonian $H_{q} = \sum_{a}^{|H_{q}|} c_{a}P_{a}$ is:

\begin{equation}
    \label{eq:H_rot_SeqRot_part1}
\begin{aligned}
    R_{S} H_{q} R_{S}^{\dagger} &= \bigg(\sum_{b}^{\mathcal{O}(2^{|\mathcal{A}|-1})} \delta_{b}P_{b} \bigg) \sum_{a}^{|H_{q}|} c_{a}P_{a}
    \bigg( \sum_{c}^{\mathcal{O}(2^{|\mathcal{A}|-1})} \delta_{c}^{*} P_{c} \bigg) \\
    &= \sum_{b}^{\mathcal{O}(2^{|\mathcal{A}|-1})} \sum_{a}^{|H_{q}|} \sum_{c}^{\mathcal{O}(2^{|\mathcal{A}|-1})} (\delta_{b}c_{a}\delta_{c}^{*}) P_{b}P_{a}P_{c}.
\end{aligned}
\end{equation}
We see that the number of terms increases as $\mathcal{O}(2^{|\mathcal{A}|}|H_{q}|)$ which was previously shown in  \cite{kirby2020classical}. What we show next is additional structure in $R_{S}$ - due to the $X_{kj}$ operators - means the base of the exponent can be slightly lower; however, it still remains exponential in $|\mathcal{A}|$.

Consider the adjoint rotation of a particular $X_{kj}$ in $R_{S}$ (Equation \ref{eq:R_SeqRot}):

\begin{equation}
    \begin{aligned}
    R_{S_{kj}} &=   \cos \big( \frac{\theta_{kj}}{2} \big) \mathcal{I} +  \sin \big( \frac{\theta_{kj}}{2} \big)  P_{kj}, \\
   R_{S_{kj}}^{\dagger} &= \cos \big( \frac{\theta_{kj}}{2} \big) \mathcal{I} + \sin \big( \frac{\theta_{kj}}{2} \big)  P_{jk}.
    \end{aligned}
\end{equation}

Performing the adjoint rotation on $H_{q}$ results in the following:

\begin{equation}
\label{eq:R_sj_on_H}
    \begin{aligned}
    R_{S_{kj}} H_{q} R_{S_{kj}}^{\dagger}  &=  \Bigg[ \alpha_{kj}\mathcal{I} +  \beta_{kj}P_{kj}\Bigg] \sum_{a}c_{a}P_{a} \Bigg[ \alpha_{kj} \mathcal{I} + \beta_{kj} P_{jk}\Bigg] \\
    &= \sum_{a} c_{a} \big( \alpha_{kj}P_{a} + \beta_{kj}P_{kj}P_{a} \big) \Bigg[ \alpha_{kj} \mathcal{I} + \beta_{kj} P_{jk}\Bigg] \\
    &= \sum_{a} c_{a} \big( \alpha_{kj}^{2} P_{a} + \alpha_{kj} \beta_{kj} P_{a}P_{jk} + \alpha_{kj} \beta_{kj} P_{\underline{kj}}P_{a} + \beta_{kj}^{2} P_{kj}P_{a} P_{jk}\big) \\
     &= \sum_{a} c_{a} \big( \alpha_{kj}^{2} P_{a} + \alpha_{kj} \beta_{kj} \underline{P_{a}P_{jk}} - \alpha_{kj} \beta_{kj} \underline{P_{jk}P_{a}} + \beta_{kj}^{2} P_{kj}P_{a} P_{jk}\big) \\
    &= \sum_{a} c_{a} \big( \alpha_{kj}^{2} P_{a} + \alpha_{kj} \beta_{kj} [ P_{a}, P_{jk} ] + \beta_{kj}^{2} P_{kj}P_{a} P_{jk}\big)  \\
    &= \sum_{a} c_{a} \begin{cases}
       \big( \alpha_{kj}^{2} P_{a} + \beta_{kj}^{2} P_{kj}P_{a} P_{jk}\big), & \text{if}\ [ P_{a}, P_{jk} ]=0 \\
       \big( \alpha_{kj}^{2} P_{a} + 2\alpha_{kj} \beta_{kj} P_{a}P_{jk}  + \beta_{kj}^{2} P_{kj} P_{a} P_{jk} \big), & \text{else } \{ P_{a}, P_{jk} \} =0
    \end{cases}
    \end{aligned}
\end{equation}
When $[P_{a}, P_{jk}]=0$, we get:

\begin{equation}
      \begin{aligned}
\sum_{a} c_{a} \big( \alpha_{kj}^{2} P_{a} + \beta_{kj}^{2} P_{kj} \underline{P_{a} P_{jk}}\big) &= \sum_{a} c_{a}\big( \alpha_{kj}^{2} P_{a} + \beta_{kj}^{2} \underline{P_{kj}  P_{jk}} P_{a}\big) \\
      &= \sum_{a} c_{a} \big( \alpha_{kj}^{2} P_{a} + \beta_{kj}^{2} P_{a}\big) \\
      &= \sum_{a} c_{a} \big( \alpha_{kj}^{2} + \beta_{kj}^{2} \big)  P_{a}\\
      &= \sum_{a} c_{a} P_{a}.
    \end{aligned}
\end{equation}
When $\{ P_{a}, P_{jk} \}=0$,  we find:

\begin{equation}
      \begin{aligned}
     \sum_{a} c_{a} \big( \alpha_{kj}^{2} P_{a} + 2\alpha_{kj} \beta_{kj} P_{a}P_{jk}  + \beta_{kj}^{2} P_{kj} \underline{P_{a} P_{jk}} \big) &= \sum_{a} c_{a} \big( \alpha_{kj}^{2} P_{a} + 2\alpha_{kj} \beta_{kj} P_{a}P_{jk}  - \beta_{kj}^{2} \underline{P_{kj} P_{jk}} P_{a}  \big)\\
     &= \sum_{a} c_{a} \big( \alpha_{kj}^{2} P_{a} + 2\alpha_{kj} \beta_{kj} P_{a}P_{jk}  - \beta_{kj}^{2} P_{a}  \big)\\
     &= \sum_{a} c_{a} \big( \alpha_{kj}^{2} P_{a} + sin(\theta_{kj}) P_{a}P_{jk}  - \beta_{kj}^{2} P_{a}  \big)\\
     &= \sum_{a} c_{a} \big( (\alpha_{kj}^{2} - \beta_{kj}^{2} )P_{a} + sin(\theta_{kj}) P_{a}P_{jk} \big)\\
          &= \sum_{a} c_{a} \big( cos(\theta_{kj})P_{a} + sin(\theta_{kj}) P_{a}P_{jk} \big).
    \end{aligned}
\end{equation}
Both cases use the following identities:

\begin{equation}
    \begin{aligned}
     \alpha_{kj}^{2} - \beta_{kj}^{2} &= \cos^{2} \big( \frac{\theta_{kj}}{2} \big) - \sin^{2} \big( \frac{\theta_{kj}}{2} \big) = \cos \big( \theta_{kj} \big), \\
     \alpha_{kj}^{2} + \beta_{kj}^{2} &= 1, \\
     2\alpha_{kj}\beta_{kj} &=2  \cos \big( \frac{\theta_{kj}}{2} \big) \sin \big( \frac{\theta_{kj}}{2} \big)  = \sin \big( \theta_{kj} \big),
    \end{aligned}
\end{equation}
where $\alpha_{kj} =  \cos \big( \frac{\theta_{kj}}{2} \big)$ and $\beta_{kj} =  \sin \big( \frac{\theta_{kj}}{2} \big)$. Using these results Equation \ref{eq:R_sj_on_H} reduces to:

\begin{equation}
\label{eq:first_R_skj}
    \begin{aligned}
    R_{S_{kj}} H_{q} R_{S_{kj}}^{\dagger}  &=  \sum_{\substack{a\\  \forall [P_{a}, P_{jk}]=0}} c_{a} P_{a} + \sum_{\substack{a\\  \forall \{P_{a}, P_{jk}\}=0}} c_{a} \Bigg( cos(\theta_{kj})P_{a} + sin(\theta_{kj}) P_{a}P_{jk} \Bigg)\\
    &=  \sum_{a} \eta_{a}  P_{a} + \sum_{\substack{a\\  \forall \{P_{a}, P_{jk}\}=0}} \eta_{a} \big( P_{jk}P_{a} \big),
    \end{aligned}
\end{equation}
where $\eta_{a}$ represent the new real coefficients.

Consider the application of the next rotation operator $R_{kl}$ in $R_{S}$ (note $k$ index represents the same Pauli operator $P_{k}$):

\begin{equation}
    \label{eq:H_rot_second_rotation}
    \begin{aligned}
    R_{S_{kl}} R_{S_{kj}} H_{q} R_{S_{kj}}^{\dagger}R_{S_{kl}}^{\dagger} 
    &=  R_{S_{kl}} \bigg( \sum_{a} \eta_{a}  P_{a} \bigg) R_{S_{kl}}^{\dagger} +   \underline{R_{S_{kl}} \Bigg( \sum_{\substack{a\\  \forall \{P_{a}, P_{jk}\}=0}} \eta_{a} \big(P_{jk}P_{a} \big) \Bigg) R_{S_{kl}}^{\dagger}}  
    \end{aligned}
\end{equation}

Focusing on the last term in Equation \ref{eq:H_rot_second_rotation}:
\begin{equation}
    \label{eq:last_part_double_rot_Rs_1}
    \begin{aligned}
      R_{S_{kl}} & \Bigg( \sum_{\substack{a\\  \forall \{P_{a}, P_{jk}\}=0}} \eta_{a} \big(P_{jk}P_{a} \big)  \Bigg) R_{S_{kl}}^{\dagger} =\Bigg[ \gamma_{kl}\mathcal{I} + \delta_{kl}P_{kl}\Bigg] \sum_{\substack{a\\  \forall \{P_{a}, P_{jk}\}=0}} \eta_{a} \big( P_{jk}P_{a} \big) \Bigg[ \gamma_{kl} \mathcal{I} + \delta_{kl} P_{lk}\Bigg] \\
     &= \sum_{\substack{a\\  \forall \{P_{a}, P_{jk}\}=0}} \eta_{a} \Bigg(  \gamma_{kl} P_{jk}P_{a} + \delta_{kl}  P_{kl} P_{jk}P_{a}   \Bigg)  \Bigg[ \gamma_{kl} \mathcal{I} + \delta_{kl} P_{lk}\Bigg] \\
     &= \sum_{\substack{a\\  \forall \{P_{a}, P_{jk}\}=0}} \eta_{a} \Bigg(  \gamma_{kl}^{2} P_{jk}P_{a} + \gamma_{kl} \delta_{kl}  P_{jk}P_{a} P_{lk} +   \gamma_{kl} \delta_{kl}   P_{k\underline{l}} P_{\underline{j}k}P_{a}   + \delta_{kl}^{2}  P_{kl} P_{jk}P_{a}P_{lk} \Bigg)\\
     &= \sum_{\substack{a\\  \forall \{P_{a}, P_{jk}\}=0}} \eta_{a} \Bigg(  \gamma_{kl}^{2} P_{jk}P_{a} + \gamma_{kl} \delta_{kl}  P_{jk}P_{a} P_{lk} -   \gamma_{kl} \delta_{kl}   P_{\underline{kj}}P_{lk}P_{a}   + \delta_{kl}^{2}  P_{kl} P_{jk}P_{a}P_{lk} \Bigg)\\
      &= \sum_{\substack{a\\  \forall \{P_{a}, P_{jk}\}=0}} \eta_{a} \Bigg(  \gamma_{kl}^{2} P_{jk}P_{a} + \gamma_{kl} \delta_{kl}  P_{jk}P_{a} P_{lk} + \gamma_{kl} \delta_{kl}   P_{jk} P_{lk}P_{a}   + \delta_{kl}^{2}  P_{kl} P_{jk}P_{a}P_{lk} \Bigg)\\
      &= \sum_{\substack{a\\  \forall \{P_{a}, P_{jk}\}=0}} \eta_{a} \Bigg(  \gamma_{kl}^{2} P_{jk}P_{a} + \gamma_{kl} \delta_{kl}  P_{jk} \{P_{a}, P_{lk} \} + \delta_{kl}^{2}  P_{kl} P_{jk}P_{a}P_{lk} \Bigg)\\
     &= \sum_{\substack{a\\  \forall \{P_{a}, P_{jk}\}=0}} \eta_{a} \begin{cases}
    \Bigg(  \gamma_{kl}^{2} P_{jk}P_{a} + 2\gamma_{kl} \delta_{kl}  P_{jk} P_{a}P_{lk} + \delta_{kl}^{2}  P_{kl} P_{jk}P_{a}P_{lk} \Bigg), & \text{if}\ [P_{a}, P_{lk} ]=0   \\
    \Bigg(  \gamma_{kl}^{2} P_{jk}P_{a} + \delta_{kl}^{2}  P_{kl} P_{jk}P_{a}P_{lk} \Bigg), & \text{if}\ \{P_{a}, P_{lk} \}=0
\end{cases}
     \end{aligned}
\end{equation}

For the case $\{P_{a}, P_{lk} \}=0$:

\begin{equation}
    \label{eq:last_part_double_rot_Rs_2}
    \begin{aligned}
     \sum_{\substack{a\\  \forall \{P_{a}, P_{jk}\}=0}} \eta_{a} \Bigg(  \gamma_{kl}^{2} P_{jk}P_{a} + \delta_{kl}^{2}  P_{kl} P_{jk}\underline{P_{a}P_{lk}} \Bigg) &= \sum_{\substack{a\\  \forall \{P_{a}, P_{jk}\}=0}} \eta_{a} \Bigg(  \gamma_{kl}^{2} P_{jk}P_{a} - \delta_{kl}^{2}  \underline{P_{kl} P_{jk}} P_{lk} P_{a} \Bigg) \\
&= \sum_{\substack{a\\  \forall \{P_{a}, P_{jk}\}=0}} \eta_{a} \Bigg(  \gamma_{kl}^{2} P_{jk}P_{a} + \delta_{kl}^{2}   P_{jk} P_{kl} P_{lk} P_{a} \Bigg) \\
  &= \sum_{\substack{a\\  \forall \{P_{a}, P_{jk}\}=0}} \eta_{a} \Bigg(  \gamma_{kl}^{2} P_{jk}P_{a} + \delta_{kl}^{2}  P_{jk} P_{a} \Bigg) \\
  &= \sum_{\substack{a\\  \forall \{P_{a}, P_{jk}\}=0}} \eta_{a} \Bigg(  P_{jk}P_{a}\Bigg).
     \end{aligned}
\end{equation}
We observe that there is no increase in the number of terms and the weight of each Pauli operator changes.

For the case $[P_{a}, P_{lk} ]=0$:

\begin{equation}
    \label{eq:last_part_double_rot_Rs_3}
    \begin{aligned}
     \sum_{\substack{a\\  \forall \{P_{a}, P_{jk}\}=0}} \eta_{a}  \big(  \gamma_{kl}^{2} P_{jk}P_{a} + 2\gamma_{kl} \delta_{kl}  P_{jk} P_{a}P_{lk} + \delta_{kl}^{2}  P_{kl} P_{jk}\underline{P_{a}P_{lk}} \big)
     &= \sum_{\substack{a\\  \forall \{P_{a}, P_{jk}\}=0}} \eta_{a}  \big(  \gamma_{kl}^{2} P_{jk}P_{a} + 2\gamma_{kl} \delta_{kl}  P_{jk} P_{a}P_{lk} + \delta_{kl}^{2}  \underline{P_{kl} P_{jk} }P_{lk} P_{a}\big) \\
      &= \sum_{\substack{a\\  \forall \{P_{a}, P_{jk}\}=0}} \eta_{a}  \big(  \gamma_{kl}^{2} P_{jk}P_{a} + 2\gamma_{kl} \delta_{kl}  P_{jk} P_{a}P_{lk} - \delta_{kl}^{2}   P_{jk}P_{kl}P_{lk} P_{a}\big)\\
      &= \sum_{\substack{a\\  \forall \{P_{a}, P_{jk}\}=0}} \eta_{a}  \big(  \gamma_{kl}^{2} P_{jk}P_{a} + 2\gamma_{kl} \delta_{kl}  P_{jk} P_{a}P_{lk} - \delta_{kl}^{2}   P_{jk} P_{a}\big) \\ 
    &= \sum_{\substack{a\\  \forall \{P_{a}, P_{jk}\}=0}} \eta_{a}  \big(  (\gamma_{kl}^{2} - \delta_{kl}^{2} ) P_{jk}P_{a} + 2\gamma_{kl} \delta_{kl}  P_{jk} P_{a}P_{lk}\big) \\ 
        &= \sum_{\substack{a\\  \forall \{P_{a}, P_{jk}\}=0}} \eta_{a}  \big(  \cos \big( \theta_{kl} \big) P_{jk}P_{a} + \sin \big( \theta_{kl} \big)  P_{jk} P_{a}P_{lk}\big) .
     \end{aligned}
\end{equation}
The number of terms in the resulting operator has increased for each case where $[P_{a}, P_{lk} ]=0$. The action of two rotations of $R_{S}$ on the whole Hamiltonian results in: 

\begin{equation}
\label{eq:double_Rs_rot}
    \begin{aligned}
    R_{S_{kl}} R_{S_{kj}} H R_{S_{kj}}^{\dagger}R_{S_{kl}}^{\dagger} 
    =  &R_{S_{kl}} \bigg( \sum_{a} \eta_{a}  P_{a} \bigg) R_{S_{kl}}^{\dagger} +  R_{S_{kl}} \Bigg( \sum_{\substack{a\\  \forall \{P_{a}, P_{jk}\}=0}} \eta_{a} \big( P_{jk}P_{a} \big) \Bigg) R_{S_{kl}}^{\dagger}  \\
    = &R_{S_{kl}} \bigg( \sum_{a} \eta_{a}  P_{a} \bigg) R_{S_{kl}}^{\dagger} + \sum_{\substack{a \\  \forall \{ P_{a}, P_{jk} \}=0 \\  \forall \{P_{a}, P_{lk}\}=0}} \eta_{a} P_{jk}P_{a} + \\ &\sum_{\substack{a\\  \forall \{P_{a}, P_{jk}\}=0 \\ [P_{a}, P_{lk}]=0 }} \eta_{a}  \big(  \cos \big( \theta_{kl} \big) P_{jk}P_{a} + \sin \big( \theta_{kl} \big)  P_{jk} P_{a}P_{lk}\big) \\
    = &R_{S_{kl}} \bigg( \sum_{a} \eta_{a}  P_{a} \bigg) R_{S_{kl}}^{\dagger} + \sum_{\substack{a \\  \forall \{ P_{a}, P_{jk} \}=0}} \mu_{a} P_{jk}P_{a} + \sum_{\substack{a\\  \forall \{P_{a}, P_{jk}\}=0 \\ [P_{a}, P_{lk}]=0 }} \mu_{a} P_{jk} P_{a}P_{lk} \\
    = &\sum_{a} \nu_{a}  P_{a} + \sum_{\substack{a\\  \forall \{P_{a}, P_{lk}\}=0}} \nu_{a} \big( P_{lk}P_{a} \big) + \sum_{\substack{a \\  \forall \{ P_{a}, P_{jk} \}=0}} \mu_{a} P_{jk}P_{a} + \sum_{\substack{a\\  \forall \{P_{a}, P_{jk}\}=0 \\ [P_{a}, P_{lk}]=0 }} \mu_{a} P_{jk} P_{a}P_{lk}.
    \end{aligned}
\end{equation}
where Greek letters are new coefficients according to the expansion.  We use the results of Equations \ref{eq:last_part_double_rot_Rs_2} and \ref{eq:last_part_double_rot_Rs_3} to determine what occurs to the second term of Equation \ref{eq:double_Rs_rot}. We have applied the result in Equation \ref{eq:first_R_skj} to the first term ($R_{S_{kl}} \bigg( \sum_{i} \eta_{i}  P_{i} \bigg) R_{S_{kl}}^{\dagger}$) in Equation \ref{eq:double_Rs_rot}.

From these results we can infer how the terms in $H_{q}$ will scale for a general sequence of rotations of size $|R_{S}|$ (Equation \ref{eq:R_SeqRot}), which in general change as:

\begin{equation}
\label{eq:seq_rot_scaling}
    \begin{aligned}
    |H_{q}| \sum_{g=0}^{|R_{S}|} \binom{|R_{S}|}{g} = 2^{|R_{S}|} |H_{q}|
    \end{aligned}
\end{equation}
This operation increases the number of terms in $H_{q}$ to $\mathcal{O}\big( 2^{(|\mathcal{A}|-1)} |H_{q}|  \; \big)$. However, the structure of the sequence of rotation operator actually requires $2^{g}$ commuting/anticommuting conditions to be met for new Pauli operators to be generated by subsequent rotations. We therefore need to consider the probability that a given Pauli operator will either commute or anticommute with another. For the case of single qubit Pauli matrices $\sigma_{a}, \sigma_{b} \in \{I, X, Y, Z \}$ by a simple counting argument $P([\sigma_{a}, \sigma_{b}]=0)=\frac{5}{8}$ and $P(\{\sigma_{a}, \sigma_{b}\}=0)=\frac{3}{8}$, for Pauli matrices selected uniformly at random. Generalising this to tensor products of Pauli matrices on $n$ qubits, for a Pauli operator to anticommute with another there needs to be an odd number of anticommuting tensor factors. First consider the binomial distribution:

\begin{equation}
\label{eq:binomial_dist}
    P(x) = \binom{n}{x} p^{x}q^{n-x},
\end{equation}
where $n$ is the number of trials (repeated experiments), $p$ is the probability of success - here the probability a single Pauli matrix anticommutes with another ($p=\frac{3}{8}$) - and $q$ is the probability of failure - here the probability a single Pauli matrix commutes with another ($q=\frac{5}{8}$). Under these conditions, $P(x)$ gives the probability that two $n$-fold Pauli operators, selected uniformly at random, anticommute in $x$-many tensor factors. Therefore, the probability of two uniformly random Pauli operators anticommuting (commuting) is given as a sum over odd (even) values of $x \leq n$:
\begin{equation}
    P\big(\{P_{a}, P_{b}\}=0\big) = \sum_{c=1}^{\lceil n/2 \rceil} P(2c-1).
\end{equation}


Now, the binomial theorem states
\begin{equation}
\label{eq:binomial_theorem}
    (p+q)^{n} = \sum_{c=0}^{n}\binom{n}{c} p^{c}q^{n-c}
\end{equation}
for any $p,q \in \mathbb{R}$ and define the following difference:
\begin{equation}
\label{eq:binomial_theorem_diff}
    (p+q)^{n} - (-p+q)^{n} = \sum_{c=0}^{n}\binom{n}{c} \underbrace{\big[1-(-1)^{c} \big]}_{=\begin{cases}
  2, & \text{if}\ c \text{ odd} \\
  0, & \text{if}\ c \text{ even}
\end{cases}} p^{c}q^{n-c} = 2\sum_{c=1}^{\lceil n/2 \rceil}\binom{n}{2c-1} p^{2c-1}q^{n-(2c-1)}.
\end{equation}
Overall we find the probability that two $n$-fold Pauli operators anticommute to be:

\begin{equation}
\label{eq:prob_of_commutation}
\begin{aligned}
    P\big(\{P_{a}, P_{b}\}=0\big) 
    &= \sum_{c=1}^{\lceil n/2 \rceil} P(2c-1)\\
    &=\sum_{c=1}^{\lceil n/2 \rceil} \binom{n}{2c-1} \cdot \bigg(\frac{3}{8}\bigg)^{2c-1} \cdot \bigg(\frac{5}{8}\bigg)^{n-(2c-1)} \\
    &= \frac{1}{2} \bigg[\big(\frac{3}{8} + \frac{5}{8} \big)^{n} - \big(-\frac{3}{8} + \frac{5}{8} \big)^{n}\bigg] \\
    &= \frac{1}{2}\bigg[ 1- \bigg(\frac{1}{4}\bigg)^{n}\bigg],
\end{aligned}
\end{equation}
when each operator $P_a, P_b$ is chosen uniformly at random. The $n$ choose $2c-1$ term in equation \ref{eq:prob_of_commutation} counts all the possible ways an odd number of single qubit pairs of Pauli tensor factors can differ on $n$ qubits, the first fraction gives the probability that there are $2c-1$ anticommuting terms on each pair of qubits and the final fraction gives the probability that the remaining $n-(2c-1)$ qubit positions pairwise commute on each qubit. The penultimate line of equation \ref{eq:prob_of_commutation} uses the definition in \ref{eq:binomial_theorem_diff}, with the factor of two taken into account. Through equation \ref{eq:prob_of_commutation}, it can be seen that the probability of two $n$-fold Pauli operators anticommuting quickly converges to $0.5$ as the number of qubits $n$ increases. The motivation for \ref{eq:binomial_theorem_diff} arises from observing that the quantity we subtract, $(1/4)^n$, is the probability of obtaining an $n$-fold identity operator, which has the unique property of commuting universally. The complement $1-(1/4)^n$ therefore corresponds with the probability of selecting uniformly at random a Pauli operator with at least one non-trivial tensor factor. After discounting identity operators from consideration, the probabilities of anticommuting or commuting coincide, hence each occurs half of the time, explaining the $1/2$ factor in \ref{eq:prob_of_commutation}; the probability bias towards commutation is a consequence of the identity operator commuting universally, whereas there is no such operator that can anticommute universally.

If we consider how the number of terms in $H_{q}$ changes upon the sequence of rotations transformation: $H_{q} \mapsto R_{S}H_qR_{S}^{\dagger}$ where terms either commute or anticommute with a probability of $0.5$, then the scaling is as follows:

\begin{equation}
\label{eq:seq_rot_scaling_with_prefactor}
    \begin{aligned}
     \sum_{g=0}^{|R_{S}|} \frac{|H_{q}|}{2^{g}} \binom{|R_{S}|}{g} = \big(\frac{3}{2}\big)^{|R_{S}|} |H_{q}|.
    \end{aligned}
\end{equation}
Equation \ref{eq:seq_rot_scaling} is modified to have a constant factor of $2^{-g}$, where $g$ represents the number of commuting or anticommuting conditions required for operators in $H_{q}$ to obey in order to increase the number of terms upon a rotation of $R_{S}$.  Here each condition is assumed to occur with a probability of $0.5$. This operation increases the number of terms in $H_{q}$ to $\mathcal{O}\big( 1.5^{(|\mathcal{A}|-1)} |H_{q}|  \; \big)$. Note $|R_{S}| = |\mathcal{A}|-1$. In general, the scaling will be $\mathcal{O}\big( x^{(|\mathcal{A}|-1)} |H_{q}|  \; \big)$ where $1 \leq x \leq 2$, depending on how each rotation in the sequence of rotations commutes with terms in $H_{q}$. The $x=1$ case occurs if each rotation in $R_{S}$ commutes with the whole Hamiltonian. Apart from this special case, the number of terms in $H_{q}$ will increase exponentially with the size of $\mathcal{A}$ or equivalently with the number of qubits $n$ (as $|\mathcal{A}| \leq 2n+1$ \cite{kirby2020classical}) when $R$ is defined by a sequence of rotations. 


\subsubsection{\label{sec:LinComb_Paulis}  Unitary partitioning via a linear combination of unitaries}

Here we show how $A(\vec{r}) \mapsto P_{0}^{(k)} = R_{LCU}  A(\vec{r})  R_{LCU} ^{\dagger}$, where $R_{LCU}$ is defined by a linear combination of Pauli operators. We consider the set of anticommuting Pauli operators making up $A(\vec{r})$ (Equation  \ref{eq:anticomm}). We can re-write this Equation, with the term we are reducing to ($r_{k} P_{0}^{(k)}$) outside the sum:

\begin{equation}
    \label{eq:Script_A_re_arranged}
A(\vec{r}) =  r_{k} P_{0}^{(k)} + \sum_{\substack{j=0 \\  \forall j \neq k}}^{N-1} r_{j} P_{0}^{(j)}.
\end{equation}
To simplify the notation we drop the subscript $0$ (denoting the first operator in a clique) and write each $ P_{0}^{(k)}$, $ P_{0}^{(j)}$ as $ P_{k}$ and $ P_{j}$ respectively.

A re-normalization can be performed on the remaining sum yielding:

\begin{equation}
    \label{eq:Script_A_excluding_P_n_RENORMALISED}
\begin{aligned}
A(\vec{r}) &= r_{k} P_{k} +  \Omega \sum_{\substack{j=0 \\  \forall j \neq k}}^{N-1} \delta_{j} P_{j} \\ 
&=  r_{k} P_{k} +  \Omega H_{\mathcal{A} \backslash \{ r_{k} P_{k} \}},
\end{aligned}
\end{equation}
where:
\begin{subequations} 
    \label{eq:Script_A_excluding_P_n_RENORMALISED_conditions}
    \begin{align}
     \sum_{\substack{j=0 \\  \forall j \neq k}}^{N-1} |\delta_{j}|^{2} =1, \\
     r_{j} = \Omega \delta_{j}, \\
     H_{\mathcal{A} \backslash \{ r_{k} P_{k} \}}=\sum_{\substack{j=0 \\  \forall j \neq k}}^{N-1} \delta_{j} P_{j}.
    \end{align}
\end{subequations} 

Using the Pythagorean trigonometric identity: $\sin^{2}(x)+\cos^{2}(x)=1$, $A(\vec{r})$ can be re-written as: 

\begin{equation}
    \label{eq:re_defined_script_A}
\begin{aligned}
A(\vec{r}) &= \cos(\phi_{k}) P_{k} +  \sin(\phi_{k}) \sum_{\substack{j=0 \\  \forall j \neq k}}^{N-1} \delta_{i} P_{j} \\ 
&= \cos(\phi_{k}) P_{k} +  \sin(\phi_{k}) H_{\mathcal{A} \backslash \{ r_{k} P_{k} \}}.
\end{aligned}
\end{equation}
Comparing Equations \ref{eq:Script_A_excluding_P_n_RENORMALISED} and  \ref{eq:re_defined_script_A}, it is clear that $\cos(\phi_{k})=r_{k}$ and $\sin(\phi_{k})=\Omega$.

It was shown in \cite{zhao2019measurement} that one can consider rotations of $A(\vec{r})$ around an axis that is Hilbert-Schmidt orthogonal to both $H_{\mathcal{A} \backslash \{ r_{k} P_{k} \}}$ and $P_{k}$ :

\begin{equation}
 \label{eq:X}
	\mathcal{X}= \frac{i}{2} \big[ H_{\mathcal{A} \backslash \{ r_{k} P_{k} \}} , P_{k} \big] = i \sum_{\substack{j=0 \\  \forall j \neq k}}^{|H_{\mathcal{A} \backslash \{ r_{k} P_{k} \}}|-1} \delta_{j} P_{j}P_{k}.
\end{equation}
$\mathcal{X}$  anticommutes with $\mathcal{A}$ and  is self-inverse \cite{zhao2019measurement}:

\begin{subequations} \label{eq:self_inv}
    \begin{align}
      \mathcal{X}^{2} &= \Bigg( i \sum_{j=0}^{|H_{\mathcal{A} \backslash \{ r_{k} P_{k} \}}|-1} \delta_{j} P_{j}P_{k} \Bigg) \Bigg( i \sum_{l=0}^{|H_{\mathcal{A} \backslash \{ r_{k} P_{k} \}}|-1} \delta_{l} P_{l}P_{k} \Bigg), \\
      &=  - \sum_{j=0}^{|H_{\mathcal{A} \backslash \{ r_{k} P_{k} \}}|-1} \sum_{l=0}^{|H_{\mathcal{A} \backslash \{ r_{k} P_{k} \}}|-1} \delta_{j}\delta_{l} P_{j}P_{k} P_{l}P_{k}, \\
      &=  - \sum_{k=0}^{|H_{\mathcal{A} \backslash \{ r_{k} P_{k} \}}|-1} \sum_{\substack{l=0 \\  \forall l = j}}^{|H_{\mathcal{A} \backslash \{ r_{k} P_{k} \}}|-1} \delta_{j}\delta_{j} P_{j}P_{k} P_{j}P_{k} -  \sum_{j=0}^{|H_{\mathcal{A} \backslash \{ r_{k} P_{k} \}}|-1} \sum_{\substack{l=0 \\  \forall l \neq j}}^{|H_{\mathcal{A} \backslash \{ r_{k} P_{k} \}}|-1} \delta_{j}\delta_{l} P_{j}P_{k} P_{l}P_{k}, \\
      &=  + \sum_{k=0}^{|H_{\mathcal{A} \backslash \{ r_{k} P_{k} \}}|-1} \delta_{j}^{2} \underbrace{P_{k} P_{j}}_{\text{order change}} P_{j}P_{k} -  \sum_{j=0}^{|H_{\mathcal{A} \backslash \{ r_{k} P_{k} \}}|-1} \sum_{l>j}^{|H_{\mathcal{A} \backslash \{ r_{k} P_{k} \}}|-1} \delta_{j}\delta_{l} \underbrace{\{P_{j}P_{k}, P_{l}P_{k}\}}_{=0 \text{ when } j \neq l}, \\
      &=  + \sum_{j=0}^{|H_{\mathcal{A} \backslash \{ r_{k} P_{k} \}}|-1} \delta_{j}^{2} \mathcal{I} , \\
      &= \mathcal{I}
    \end{align}
\end{subequations}

and has the following action \cite{zhao2019measurement}:

\begin{equation}
 \label{eq:X_action_on_scriptA}
	\mathcal{X} A(\vec{r}) = i \big( -\sin \phi_{k} P_{k} +  \cos \phi_{k} H_{\mathcal{A} \backslash \{ r_{k} P_{k} \}} \big).
\end{equation}

One can also define the rotation \cite{zhao2019measurement, ralli2021implementation}:

\begin{subequations} \label{eq:R_LCU}
    \begin{align}
      R_{LCU} &= e^{ \big( -i \frac{\alpha}{2} \mathcal{X} \big)} = \cos \big( \frac{\alpha}{2} \big) \mathcal{I} - i \sin \big( \frac{\alpha}{2} \big)  \mathcal{X} \\
      &=  \cos \big( \frac{\alpha}{2} \big) \mathcal{I} - i \sin \big( \frac{\alpha}{2} \big) \bigg(   i \sum_{\substack{j=0 \\  \forall j \neq k}}^{|H_{\mathcal{A} \backslash \{ r_{k} P_{k} \}}|-1} \delta_{j} P_{j} P_{k} \bigg) \\
      &=  \cos \big( \frac{\alpha}{2} \big) \mathcal{I} + \sin \big( \frac{\alpha}{2} \big)  \sum_{\substack{j=0 \\  \forall j \neq k}}^{|H_{\mathcal{A} \backslash \{ r_{k} P_{k} \}}|-1} \delta_{j} P_{jk}, \\
      &=   \delta_{\mathcal{I}} \mathcal{I} + \sum_{\substack{j=0 \\  \forall j \neq k}}^{|H_{\mathcal{A} \backslash \{ r_{k} P_{k} \}}|-1} \delta_{j} P_{jk}.
    \end{align}
\end{subequations}
The conjugate rotation will be:

\begin{subequations} \label{eq:R_LCU_dagger}
    \begin{align}
      R_{LCU}^{\dagger} &= \cos \big( \frac{\alpha}{2} \big) \mathcal{I} + i\sin \big( \frac{\alpha}{2} \big) i \sum_{\substack{j=0 \\  \forall j \neq k}}^{|H_{\mathcal{A} \backslash \{ r_{k} P_{k} \}}|-1} \delta_{j}  P_{j}P_{k}, \\
      &=  \cos \big( \frac{\alpha}{2} \big) \mathcal{I} + \sin \big( \frac{\alpha}{2} \big)  \sum_{\substack{j=0 \\  \forall j \neq k}}^{|H_{\mathcal{A} \backslash \{ r_{k} P_{k} \}}|-1} \delta_{j} \underbrace{P_{k} P_{j}}_{\text{order change}}, \\
      &=   \delta_{\mathcal{I}} \mathcal{I} + \sum_{\substack{j=0 \\  \forall j \neq k}}^{|H_{\mathcal{A} \backslash \{ r_{k} P_{k} \}}|-1} \delta_{j} P_{kj}.
    \end{align}
\end{subequations}
Note the different order of $j$ and $k$ for $R_{LCU}$ and $ R_{LCU}^{\dagger}$. The adjoint action of $R_{LCU}$ on $A(\vec{r})$ is:

\begin{equation}
      R_{LCU}A(\vec{r})R_{LCU}^{\dagger} = \cos \big( \phi_{k} - \alpha \big)P_{k} + \sin \big( \phi_{k} - \alpha  \big)H_{\mathcal{A} \backslash \{ r_{k} P_{k} \}}.
\end{equation}
By choosing $\alpha = \phi_{k}$, the following transformation occurs $ R_{LCU}A(\vec{r})R_{LCU}^{\dagger}=P_{k}$ \cite{zhao2019measurement, ralli2021implementation}. This fully defines the $R_{LCU}$ operator required by unitary partitioning. Next we need to consider the use of this operator in CS-VQE.

The adjoint action of $R_{LCU}$ on a general Hamiltonian $H_{q} = \sum_{i}^{|H_{q}|} c_{i}P_{i}$ is:

\begin{subequations} \label{eq:R_LCU_dagger_op}
    \begin{align}
    R_{LCU} H_{q} R_{LCU}^{\dagger} = &\Bigg( \delta_{\mathcal{I}} \mathcal{I} + \sum_{j}^{|R_{LCU}|-1} \delta_{j} P_{jk} \Bigg) \sum_{i}^{|H_{q}|} c_{i}P_{i}
    \Bigg(\delta_{\mathcal{I}} \mathcal{I} + \sum_{l}^{|R_{LCU}|-1} \delta_{l} P_{kl} \Bigg) \\
    = &\Bigg( \delta_{\mathcal{I}} \sum_{i}^{|H_{q}|} c_{i}P_{i} +  \sum_{j}^{|R_{LCU}|-1}\sum_{i}^{|H_{q}|} \delta_{j} c_{i} P_{jk}P_{i} \Bigg) \Bigg(\delta_{\mathcal{I}} \mathcal{I} + \sum_{l}^{|R_{LCU}|-1} \delta_{l} P_{kl} \Bigg) \\
    = &\: \: \delta_{\mathcal{I}}^{2} \sum_{i}^{|H_{q}|} c_{i}P_{i}  \label{eq:R_LCU_dagger_op_1}\\ 
    &\; + \sum_{l}^{|R_{LCU}|-1}  \sum_{i}^{|H_{q}|} \delta_{\mathcal{I}} c_{i}\delta_{l} P_{i} P_{kl} + \sum_{j}^{|R_{LCU}|-1}\sum_{i}^{|H_{q}|} \delta_{\mathcal{I}} \delta_{j} c_{i} P_{jk}P_{i} \label{eq:R_LCU_dagger_op_2}\\
    &\; + \sum_{j}^{|R_{LCU}|-1}\sum_{i}^{|H_{q}|}  \sum_{l}^{|R_{LCU}|-1} \delta_{j} c_{i} \delta_{l}  P_{jk}P_{i} P_{kl} \label{eq:R_LCU_dagger_op_3}
    \end{align}
\end{subequations}

We can rewrite the final term (Equation \ref{eq:R_LCU_dagger_op_3}) as:

\begin{subequations} \label{eq:R_LUC_dagger_final_term}
    \begin{align}
    \sum_{j}^{|R_{LCU}|-1}\sum_{i}^{|H_{q}|}  \sum_{l}^{|R_{LCU}|-1} \delta_{j} c_{i} \delta_{l}  P_{jk}P_{i} P_{kl} &=  \sum_{j}^{|R_{LCU}|-1} \sum_{i}^{|H_{q}|} \sum_{\substack{l=j \\ [P_{jk},P_{i}]=0}}^{|R_{LCU}|-1} (c_{i}\delta_{j}\delta_{j}^{*}) P_{i} \\
    & +  \sum_{j}^{|R_{LCU}|-1} \sum_{i}^{|H_{q}|} \sum_{\substack{l=j \\ \{P_{jk},P_{i}\}=0}}^{|R_{LCU}|-1} (-c_{i}\delta_{j}\delta_{j}^{*}) P_{i} \\
    & + \sum_{j}^{|R_{LCU}|-1}\sum_{i}^{|H_{q}|} \sum_{\substack{l \neq j}}^{|R_{LCU}|-1} (\delta_{j}c_{i}\delta_{l})  P_{jk} P_{i} P_{kl} \label{eq:R_LUC_dagger_final_term_3}.
    \end{align}
\end{subequations}
Here we have applied the identity of conjugating a Pauli operator $P_{u}$ with another Pauli operator $P_{v}$ resulting in two cases:

\begin{equation}
 \label{eq:triple_P}
P_{v} P_{u} P_{v}=
\begin{cases}
  P_{u}, & \text{if}\ [P_{v},P_{u}]=0 \\
  -P_{u}, & \text{otherwise}\ \{P_{v},P_{u}\}=0
\end{cases}
\end{equation}
Focusing on the last term of Equation \ref{eq:R_LUC_dagger_final_term}, we can simplify \ref{eq:R_LUC_dagger_final_term_3} as $j$ and $l$ run over the same indices we can re-write each $l \neq j$ sum as $l > j$ and expand into two terms:

\begin{equation}
\begin{aligned}
    \sum_{j}^{|R_{LCU}|-1}\sum_{i}^{|H_{q}|} \sum_{\substack{l \neq j}}^{|R_{LCU}|-1} (\delta_{j}c_{i}\delta_{l})  P_{jk} P_{i} P_{kl} &=  \sum_{j}^{|R_{LCU}|-1}\sum_{i}^{|H_{q}|} \sum_{\substack{l > j}}^{|R_{LCU}|-1} (\delta_{j}c_{i}\delta_{l})  \Big( P_{jk} P_{i} P_{kl} +  P_{lk} P_{i} P_{kj} \Big)
\end{aligned}
\end{equation}
We can expand then expand this equation into the four cases for when:
\begin{enumerate}
    \item $[P_{jk}, P_{i}]=0$ and  $[P_{lk}, P_{i}]=0$
    \item $[P_{jk}, P_{i}]=0$ and  $\{P_{lk}, P_{i}\}=0$
    \item $\{P_{jk}, P_{i}\}=0$ and  $[P_{lk}, P_{i}]=0$
    \item $\{P_{jk}, P_{i}\}=0$ and  $\{P_{lk}, P_{i}\}=0$
\end{enumerate}

For the first case and last case:
\begin{equation}
\begin{aligned}
    \sum_{j}^{|R_{LCU}|-1}\sum_{i}^{|H_{q}|} \sum_{\substack{l > j}}^{|R_{LCU}|-1} (\delta_{j}c_{i}\delta_{l})  \Big( \underline{P_{jk} P_{i}} P_{kl} +  \underline{P_{lk} P_{i}} P_{kj} \Big) &=  \sum_{j}\sum_{i}^{|H_{q}|} \sum_{\substack{l > j}} (\delta_{j}c_{i}\delta_{l})  \Big( \pm P_{i} \underline{P_{jk}  P_{kl}} \pm  P_{i} \underline{P_{lk}  P_{kj}} \Big)  \\
    &= \sum_{j}\sum_{i}^{|H_{q}|} \sum_{\substack{l > j}} (\delta_{j}c_{i}\delta_{l})  \Big( \pm P_{i} P_{j}  P_{l} \pm  P_{i} P_{l}  P_{j} \Big)\\
    &= \sum_{j}\sum_{i}^{|H_{q}|} \sum_{\substack{l > j}} (\delta_{j}c_{i}\delta_{l})  \pm P_{i} \{P_{j},  P_{l}\}\\
    &= 0
\end{aligned}
\end{equation}

Whereas, for the second and third cases:
\begin{equation}
\begin{aligned}
    \sum_{j}^{|R_{LCU}|-1}\sum_{i}^{|H_{q}|} \sum_{\substack{l > j}}^{|R_{LCU}|-1} (\delta_{j}c_{i}\delta_{l})  \Big( \underline{P_{jk} P_{i}} P_{kl} +  \underline{P_{lk} P_{i}} P_{kj} \Big) &=  \sum_{j}\sum_{i}^{|H_{q}|} \sum_{\substack{l > j}} (\delta_{j}c_{i}\delta_{l})  \Big( \pm P_{i} \underline{P_{jk}  P_{kl}} \mp  P_{i} \underline{P_{lk}  P_{kj}} \Big)  \\
    &= \sum_{j}\sum_{i}^{|H_{q}|} \sum_{\substack{l > j}} (\delta_{j}c_{i}\delta_{l})  \Big( \pm P_{i} P_{j}  P_{l} \mp  P_{i} \underline{P_{l}  P_{j}} \Big)\\
    &= \sum_{j}\sum_{i}^{|H_{q}|} \sum_{\substack{l > j}} (\delta_{j}c_{i}\delta_{l})  \Big( \pm P_{i} P_{j}  P_{l} \pm  P_{i} P_{j}  P_{l} \Big)\\
    &= \sum_{j}\sum_{i}^{|H_{q}|} \sum_{\substack{l > j}} (\delta_{j}c_{i}\delta_{l})  \pm 2 P_{i} P_{j} P_{l}
\end{aligned}
\end{equation}

We can rewrite Equation \ref{eq:R_LUC_dagger_final_term} using this result:

\begin{equation}
\begin{aligned}
\label{eq:R_LUC_dagger_final_term_reduced}
    \sum_{j}^{|R_{LCU}|-1}\sum_{i}^{|H_{q}|}  \sum_{l}^{|R_{LCU}|-1} \delta_{j} c_{i} \delta_{l}  P_{jk}P_{i} P_{kl} =  &\sum_{j} \sum_{i}^{|H_{q}|} \sum_{\substack{l=j \\ [P_{jk},P_{i}]=0}} (c_{i}\delta_{j}\delta_{j}^{*}) P_{i}  +  \sum_{j} \sum_{i}^{|H_{q}|} \sum_{\substack{l=j \\ \{P_{jk},P_{i}\}=0}} (-c_{i}\delta_{j}\delta_{j}^{*}) P_{i} + \\
     &\sum_{j}\sum_{i}^{|H_{q}|} \sum_{\substack{l > j \\ \forall [P_{jk}, P_{i}]=0 \\  \{P_{lk}, P_{i}\}=0}} (\delta_{j}c_{i}\delta_{l})  2 P_{i} P_{j} P_{l} - \sum_{j}\sum_{i}^{|H_{q}|} \sum_{\substack{l > j \\ \forall \{P_{jk}, P_{i}\}=0 \\ [P_{lk}, P_{i}]=0}} (\delta_{j}c_{i}\delta_{l})  2 P_{i} P_{j} P_{l} \\
    = &\sum_{i}^{|H_{q}|} \nu_{i} P_{i} + \sum_{j}^{|R_{LCU}|-1}\sum_{i}^{|H_{q}|} \sum_{\substack{l > j \\ \forall \{P_{i}, P_{j}P_{l} \}=0 }}^{|R_{LCU}|-1}  \nu_{ijl} P_{i} P_{j} P_{l}
\end{aligned}
\end{equation}
where we have combined the second and third conditions into a single condition of $\{P_{a}, P_{jk}P_{kl} \} = \{P_{a}, P_{j}P_{l} \}=0$ and combined the new coefficients into one coefficient denoted $\nu$.

Next consider the \ref{eq:R_LCU_dagger_op_2} term of equation \ref{eq:R_LCU_dagger_op}.  One can use the fact that $j$ and $l$ run over the same indices:

\begin{equation} \label{eq:R_LCU_second_term}
\begin{aligned}
    \underbrace{\sum_{l}^{|R_{LCU}|-1}  \sum_{i}^{|H_{q}|} \delta_{\mathcal{I}} c_{i}\delta_{l} P_{i} P_{kl}}_{\text{re-write using }l=j} + \sum_{j}^{|R_{LCU}|-1}\sum_{i}^{|H_{q}|} \delta_{\mathcal{I}} \delta_{j} c_{i} P_{jk}P_{i} &= \sum_{j}^{|R_{LCU}|-1}  \sum_{i}^{|H_{q}|} \delta_{\mathcal{I}} c_{i}\delta_{j} P_{i} P_{kj} + \sum_{j}^{|R_{LCU}|-1}\sum_{i}^{|H_{q}|} \delta_{\mathcal{I}} \delta_{j} c_{i} P_{jk}P_{i} \\ 
    &= \sum_{j}^{|R_{LCU}|-1}  \sum_{i}^{|H_{q}|} -\delta_{\mathcal{I}} c_{i}\delta_{j} P_{i} P_{jk} + \sum_{j}^{|R_{LCU}|-1}\sum_{i}^{|H_{q}|} \delta_{\mathcal{I}} \delta_{j} c_{i} P_{jk}P_{i} \\ 
    &= \sum_{j}^{|R_{LCU}|-1}  \sum_{i}^{|H_{q}|} \delta_{\mathcal{I}} c_{i}\delta_{j} 
    \big(P_{jk} P_{i} - P_{i}P_{jk} \big)\\ 
    &= \sum_{j}^{|R_{LCU}|-1}  \sum_{i}^{|H_{q}|} \delta_{\mathcal{I}} c_{i}\delta_{j} 
    [P_{jk}, P_{i}]\\ 
    &= \sum_{j}^{|R_{LCU}|-1}  \sum_{\substack{i \\  \forall \{P_{jk}, P_{i} \} =0}}^{|H_{q}|} 2\delta_{\mathcal{I}} c_{i}\delta_{j} 
    P_{jk} P_{i}
\end{aligned}
\end{equation}

Overall we can re-write equation \ref{eq:R_LCU_dagger_op} using these results, yielding:

\begin{equation} \label{eq:R_LCU_final_form}
\begin{aligned}
R_{LCU} H_{q} R_{LCU}^{\dagger} = &\underbrace{\delta_{\mathcal{I}}^{2} \sum_{i}^{|H_{q}|} c_{i}P_{i}}_{\ref{eq:R_LCU_dagger_op_1}} + \underbrace{\sum_{j}^{|R_{LCU}|-1}  \sum_{\substack{i \\  \forall \{P_{jk}, P_{i} \} =0}}^{|H_{q}|} 2\delta_{\mathcal{I}} c_{i}\delta_{j} P_{jk} P_{i}}_{\ref{eq:R_LCU_dagger_op_2}  \text{ using } \ref{eq:R_LCU_second_term}} + \\
&\underbrace{\sum_{i}^{|H_{q}|} \nu_{i} P_{i} + \sum_{j}^{|R_{LCU}|-1}\sum_{i}^{|H_{q}|} \sum_{\substack{l > j \\ \forall \{P_{i}, P_{j}P_{l} \}=0 }}^{|R_{LCU}|-1}  \nu_{ijl} P_{i} P_{j} P_{l}}_{\ref{eq:R_LCU_dagger_op_3} \text{ using } \ref{eq:R_LUC_dagger_final_term_reduced}} \\
= & \sum_{i}^{|H_{q}|} (\delta_{\mathcal{I}}^{2} c_{i} + \nu_{i}) P_{i} + \sum_{j}^{|R_{LCU}|-1}  \sum_{\substack{i \\  \forall \{P_{jk}, P_{i} \} =0}}^{|H_{q}|} 2\delta_{\mathcal{I}} c_{i}\delta_{j} P_{jk} P_{i} \: + \\
&\sum_{j}^{|R_{LCU}|-1}\sum_{i}^{|H_{q}|} \sum_{\substack{l > j \\ \forall \{P_{i}, P_{j}P_{l} \}=0 }}^{|R_{LCU}|-1}  \nu_{ijl} P_{i} P_{j} P_{l} 
\end{aligned}
\end{equation}
We observe that the number of terms in $R_{LCU} H_{q} R_{LCU}^{\dagger}$ at worst scales as $|H_{q}| + |H_{q}| \cdot (|R_{LCU}|-1) + |H_{q}|\big(\frac{(|R_{LCU}|-1) (|R_{LCU}|-2)}{2} \big)$ or $\mathcal{O}\big( \; |H_{q}|\cdot |\mathcal{A}|^{2} \; \big) $. The total number of qubits $n$ bounds the size of $|\mathcal{A}| \leq 2n+1$ \cite{kirby2020classical}, and thus the number of terms in $H_{q}$ will increase quadratically with the size of $\mathcal{A}$ or number of qubits $n$ when $R$ is defined by a linear combination of unitaries.


\subsubsection{\label{sec:Vops} Mapping Pauli operators to single-qubit Pauli Z operators}

In Appendix A of \cite{kirby2021contextual} a proof is given on how to map a completely commuting set of Pauli operators to a single qubit Pauli $Z$ operator. We summarise the operation required and omit the proof. We denote a given Pauli operator on $n$ qubits as: $P=\bigotimes_{i=0}^{n-1}\sigma_{i}^{P}$, where $\sigma_{i}$ are a single qubit Pauli operators. There are two cases we need to consider (diagonal and non-diagonal), with the goal to reduce the operators in $\mathcal{R}^{'} \equiv \{ P_{0}^{(k)} \} \cup \mathcal{G}$ (Equation \ref{eq:R_ind_set_scriptA}) to single-qubit $Z$ Pauli operators.

For a non-diagonal Pauli operators $P_{a} \in \mathcal{R}'$, there must be at least one single qubit Pauli operator indexed by qubit $k$ such that: $\sigma_{k}^{P_{a}} \in \{ X, Y \}$. We can use this to define operator $P_{b}$ that must anticommute with $P_{a}$:

\begin{equation}
\label{eq:P_aligned}
  \begin{aligned}
  P_{a}&=\bigg( \bigotimes_{i=0}^{k-1} \sigma_{i}^{P_{a}} \bigg) \otimes \sigma_{k}^{P_{a}}  \otimes  \bigg( \bigotimes_{i=k+1}^{n-1} \sigma_{i}^{P_{a}} \bigg) \\
  P_{b}&=\bigg( \bigotimes_{i=0}^{k-1} \sigma_{i}^{P_{a}} \bigg) \otimes \sigma_{k}'  \otimes  \bigg( \bigotimes_{i=k+1}^{n-1} \sigma_{i}^{P_{a}} \bigg)
\end{aligned}  \text{ where } \{P_{a}, P_{b} \} = 0 \Leftrightarrow \sigma_{k}'=
    \begin{cases}
      X, & \text{if}\ \sigma_{k}^{P_{a}} =Y \\
      Y, & \text{if}\ \sigma_{k}^{P_{a}} =X 
    \end{cases}
\end{equation}

These two Pauli operators differ by exactly one Pauli operator on qubit index $k$. We can define the rotation:

\begin{equation}
\label{eq:B_op}
    B = \exp\bigg(i\frac{\pi}{4} P_{b}\bigg) 
\end{equation}

Conjugating $P_{a}$ with this operator results in: 

\begin{equation}
\label{eq:conj_B}
    B  P_{a} B^{\dagger} = \pm 1 \bigg( \bigotimes_{i=0}^{k-1} I_{i} \bigg) \otimes Z_{k}  \otimes  \bigg( \bigotimes_{i=k+1}^{n-1} I_{i} \bigg) = P_{a}',
\end{equation}

\noindent and $P_{a}$ has been mapped to a single qubit Pauli $Z$ operator.

For diagonal operators $P_{c} \in \mathcal{R}'$, all the $n$-fold tensor products of single qubit Pauli operators must be either $Z$ or $I$: $P_{c}=\bigotimes_{i=0}^{n-1}\sigma_{i}^{P_{c}}$ where $\sigma_{i}^{P_{c}} \in \{I,Z \} \; \forall i$. Since $\mathcal{R}'$ is an independent set, for all the rotated $P_{a}'$ there must be at least one index $l$ such that $\sigma_{l}^{P_{a}'}=I$ and $\sigma_{l}^{P_{c}}=Z$. We denote this operator $P_{c}$. We also define a new operator $P_{d}$ from this, which only acts non-trivially on the $l$-th qubit with a single qubit $Y$. To summarise:

\begin{equation}
\label{eq:P_aligned_diag}
  \begin{aligned}
  P_{c} &= \bigg( \bigotimes_{i=0}^{l-1} \sigma_{i}^{P_{c}} \bigg) \otimes Z_{l}  \otimes  \bigg( \bigotimes_{l=k+1}^{n-1} \sigma_{i}^{P_{c}} \bigg) \text{ where } \sigma_{i}^{P_{c}} \in \{I,Z \} \; \forall i \\
  P_{d} &=\bigg( \bigotimes_{i=0}^{l-1} I_{i} \bigg) \otimes Y_{l}  \otimes  \bigg( \bigotimes_{i=l+1}^{n-1} I_{i} \bigg)  \\
  P_{a}'&=\bigg( \bigotimes_{i=0}^{l-1} \sigma_{i}^{P_{a}'} \bigg) \otimes I_{l}  \otimes  \bigg( \bigotimes_{i=l+1}^{n-1} \sigma_{i}^{P_{a}'} \bigg) \text{ where } [P_{d}, P_{a}'] = 0 \text{ and } \sigma_{i}^{P_{a}'} \in \{ I,Z \} \forall \; i 
\end{aligned}
\end{equation}

We can define the rotation:

\begin{equation}
\label{eq:D_op}
    D = \exp\bigg(i\frac{\pi}{4} P_{d}\bigg) 
\end{equation}

Conjugating $P_{c}$ with this operator results in: 

\begin{equation}
\label{eq:conj_E}
    D  P_{c} D^{\dagger} = \pm 1 \bigg( \bigotimes_{i=0}^{l-1} \sigma_{i}^{P_{c}} \bigg) \otimes X_{l}  \otimes  \bigg( \bigotimes_{i=l+1}^{n-1} \sigma_{i}^{P_{c}} \bigg) = P_{c}'.
\end{equation}
$P_{c}'$ is now a non-diagonal Pauli operator (contains a single qubit $X$ acting on qubit $l$). This operator $P_{c}'$ can now be mapped to a single qubit $Z$ operator using a further $\frac{\pi}{2}$-rotation following the previously given procedure for non-diagonal Pauli operators. 

The operators $V_{i}$ in the main text (equation \ref{eq:rotation_U}) are defined by these $\frac{\pi}{2}$-rotations, such that each $q_{i}G_{i}$ and  $P_{0}^{(k)}$ is mapped to a single qubit Pauli Z term. At worst, two $\frac{\pi}{2}$-rotations are needed for every operator in $\mathcal{R}'$ (Equation \ref{eq:R_ind_set_scriptA}), which occurs when all operators in $\mathcal{R}'$ are diagonal.

\section{\label{sec:Unitary_Partitioning_Measurement_Reduction}Unitary Partitioning Measurement Reduction}

In unitary partitioning, the Hamiltonian is partitioned into groups of operators that's linear combination are unitary Hermitian operators. This is done by forming normalized groups of Pauli operators that pairwise anticommute. We can write this as:

 \begin{equation}
\label{eq:H_grouped}
\begin{aligned}
H &= \sum_{i}c_{i}P_{i} \\
&= \sum_{j} \gamma_{j} C_{j} \\
 &= \sum_{j}^{N_{C}} \gamma_{j}   \bigg( \sum_{\substack{k \\ \{P_{a}, P_{b} \}=0 \\ \forall P_{a}, P_{b} \in C_{j} \\ a \neq b}}^{|C_{j}|} \frac{c_{k}}{\gamma_{j}} P_{k} \bigg), 
\end{aligned}
\end{equation}
where $\gamma_{j} = \big( \sum_{k}^{C_{j}}c_{k}^{2} \big)^{0.5}$. The complete approach is provided in \cite{zhao2019measurement, izmaylov2019unitary, ralli2021implementation}.  We follow the analysis of Crawford \textit{et al.} to determine the measurement cost to determine $\expec{H}$ to a certain precision \cite{crawford2021efficient}.   The measurement requirement for measuring the Hamiltonian in terms of  grouped terms to precision $\epsilon$ is \cite{crawford2021efficient}:

 \begin{equation}
\label{eq:measurement_grouped}
M_{g} =   \frac{1}{\epsilon_{\expec{H}}^{2}}   \Bigg( \sum_{j}^{N_{C}} \sqrt{Var[C_{j}]} \Bigg)^{2},
\end{equation}
We can use this to determine the number of measurements when no partitioning has being done \cite{crawford2021efficient}:

 \begin{equation}
\label{eq:measurement_ungrouped}
M_{u} =   \frac{1}{\epsilon_{\expec{H}}^{2}}   \Bigg( \sum_{j}^{N_{C}} \bigg[ \sum_{k=0}^{|C_{j}|-1}  |c_{k}^{(j)}| \sqrt{Var[P_{k}^{(j)}} \bigg] \Bigg)^{2} .
\end{equation}
This can be thought of as each clique is of size one. The subscript $u$ is to denote no grouping. A natural metric to evaluate the the measurement cost of a particular grouping of Pauli operators is therefore given by the ratio $R$ of these two terms:

 \begin{equation}
\label{eq:measurement_ratio}
R = \frac{M_{u}}{M_{g} } = \Bigg(\frac{  \sum_{j}^{N_{C}} \bigg[ \sum_{k=0}^{|C_{j}|-1}    |c_{k}^{(j)}| \sqrt{Var[P_{k}^{(j)}} \bigg] }{\sum_{j}^{N_{C}} \sqrt{Var[C_{j}]}} \Bigg)^{2}
\end{equation}
where the greater the value of $R$, the better the measurement saving is by assembling these operators into a particular group. 

Next,  our analysis diverges from Crawford \textit{et al.}, where we consider groups of anticommuting operators (rather than commuting operators)\cite{crawford2021efficient}. First we consider the covariance of two anticommuting Pauli operators.

The amount two random variables vary together (co-vary) is measured by their covariance.  Consider the results of random variables $x$ and $y$,  one can obtain a set of  $M$ paired measurements:

\begin{equation}
\begin{aligned}
\{&(x_{0}, y_{0}), \\
&(x_{1}, y_{1}),  \\
&\hdots, \\
& (x_{M-1}, y_{M-1}) \}.
\end{aligned}
\end{equation}
A positive covariance indicates that higher than average values of one variable tend to be paired with higher than average values of the other variable.  A negative covariance indicates that a higher than average value of one variable tend to be paired with lower than average values of the other.  If two random variables are independent, then their covariance will be zero. However,  a covariance of zero does not mean two random variables are independent,  as nonlinear relationships can result in a covariance of zero.
 
In the context of measuring a quantum state in the Pauli basis on a quantum computer this would be a set of paired single shot samples $\{ s_{i}^{a}, s_{i}^{b} | i=0,1, \hdots, N-1\}$, where $s_{i}^{a}, s_{i}^{b} \in \{-1, +1\}$. Experimentally, each pair is the (single shot) measurement outcome for $P_{a}$ followed by the (single shot) measurement outcome for $P_{b}$. Taking simultaneous projective measurements, without re-preparing the quantum state is a meaningful operation if the operators share a common eigenbasis. The order of measurement does not effect measurement outcomes, but the paired samples will be statistically correlated and have a certain covariance.  However, for anticommuting Pauli operators this is not the case,  as these operators do not share a common eigenbasis.   Projective measurement means the expectation value of these operators cannot be known simultaneously.  We consider the covariance in this scenario. 

Consider the the spectral decomposition of two anticommuting Pauli operators $\{P_{a}, P_{b} \}$:

\begin{equation}
P_{a} = +1 \ket{\kappa_{0}}\bra{\kappa_{0}} -  1  \ket{\kappa_{1}}\bra{\kappa_{1}},
\end{equation}
and

\begin{equation}
P_{b} = +1 \ket{\Omega_{0}}\bra{\Omega_{0}}  -1  \ket{\Omega_{1}}\bra{\Omega_{1}},
\end{equation}
where for Pauli operators:

\begin{subequations}
    \begin{equation} 
   \bra{\kappa_{0}}\kappa_{0} \rangle = \bra{\kappa_{1}}\kappa_{1} \rangle = \bra{\Omega_{0}}\Omega_{0} \rangle = \bra{\Omega_{1}}\Omega_{1} \rangle = 1,
    \end{equation}
        \begin{equation}
   \bra{\kappa_{0}}\kappa_{1} \rangle = \bra{\Omega_{0}}\Omega_{1} \rangle = 0,
    \end{equation}
    \begin{equation}
   |\bra{\kappa_{0}}\Omega_{0} \rangle |^{2} = |\bra{\kappa_{0}}\Omega_{1} \rangle |^{2} = |\bra{\kappa_{1}}\Omega_{0} \rangle |^{2} =   |\bra{\kappa_{1}}\Omega_{1} \rangle |^{2}  = 0.5.
    \end{equation}
\end{subequations}
Without loss of generality,  assume $P_{a}$ is measured first on a general normalized quantum state $ \ket{\psi} = \gamma  \ket{\kappa_{0}} + \delta  \ket{\kappa_{1}}$.  The only possible post measurement outcomes are $ \ket{\kappa_{0}}$  or $ \ket{\kappa_{1}}$, with probabilities $|\gamma|^{2}$ or $|\delta|^{2}$ respectively.  Consider the result of subsequently measuring $P_{b}$.  The expectation value in each scenario will be:

\begin{subequations}
    \begin{equation} 
    \begin{aligned}
   \bra{\kappa_{0}} P_{b} \ket{\kappa_{0}} &=  \bra{\kappa_{0}} \big(  \ket{\Omega_{0}}\bra{\Omega_{0}}  - \ket{\Omega_{1}}\bra{\Omega_{1}} \big) \ket{\kappa_{0}} \\
   &=  \bra{\kappa_{0}}  \Omega_{0} \rangle \langle \Omega_{0} \ket{\kappa_{0}} - \bra{\kappa_{0}}  \Omega_{1} \rangle \langle \Omega_{1} \ket{\kappa_{0}}  \\
   &= | \bra{\kappa_{0}}  \Omega_{0} \rangle|^{2} - |\bra{\kappa_{0}}  \Omega_{1} \rangle |^{2} \\
   &= \underbrace{0.5}_{\mathbb{P}({\Omega_{0}}|{\kappa_{0}})} - \underbrace{0.5}_{\mathbb{P}({\Omega_{1}}|{\kappa_{0}})} = 0
   \end{aligned}
    \end{equation}
        \begin{equation}
    \begin{aligned}
   \bra{\kappa_{1}} P_{b} \ket{\kappa_{1}} &=  \bra{\kappa_{1}} \big(  \ket{\Omega_{0}}\bra{\Omega_{0}}  - \ket{\Omega_{1}}\bra{\Omega_{1}} \big) \ket{\kappa_{1}} \\
   &=  \bra{\kappa_{1}}  \Omega_{0} \rangle \langle \Omega_{0} \ket{\kappa_{1}} - \bra{\kappa_{1}}  \Omega_{1} \rangle \langle \Omega_{1} \ket{\kappa_{1}}  \\
   &= | \bra{\kappa_{1}}  \Omega_{0} \rangle|^{2} - |\bra{\kappa_{1}}  \Omega_{1} \rangle |^{2} \\
&= \underbrace{0.5}_{\mathbb{P}({\Omega_{0}}|{\kappa_{1}})} - \underbrace{0.5}_{\mathbb{P}({\Omega_{1}}|{\kappa_{1}})} = 0
   \end{aligned}
    \end{equation}
\end{subequations}
Overall,  we find the probabilities of all possible combinations of measurement outcomes to be:
\begin{equation}
\label{eq:independence_stat}
\begin{aligned}
   &\mathbb{P}\big(P_{b}=\ket{\Omega_{0}}|P_{a}=\ket{\kappa_{0}}\big)= \mathbb{P}\big(P_{b}=\ket{\Omega_{1}}|P_{a}=\ket{\kappa_{0}}\big) = 0.5\\
   &\mathbb{P}\big(P_{b}=\ket{\Omega_{0}}|P_{a}=\ket{\kappa_{1}}\big) = \mathbb{P}\big(P_{b}=\ket{\Omega_{1}}|P_{a}=\ket{\kappa_{1}}\big) = 0.5
\end{aligned}
\end{equation}
This result shows that the probability of obtaining $\ket{\Omega_{0}}$ or $\ket{\Omega_{1}}$ is not affected by the probability of obtaining $\ket{\kappa_{0}}$ or $\ket{\kappa_{1}}$ in the first measurement.  The variables are therefore statistically independent \footnote{This analysis is strictly for the case of subsequent measurement of anticommuting Pauli operators.}. We find the covariance of  $P_{a}$ and  $P_{b}$, where $\{P_{a}, P_{b} \}=0$,  to be:

\begin{equation}
\begin{aligned}
Cov[{P_{a}}, {P_{b}}] &=  \mathbb{E}\Bigg[ \bigg(p_{a} - \expec{P_{a}} \bigg) \bigg(p_{b} - {\expec{P_{b}}} \bigg) \Bigg]\\
&=   \mathbb{E}\Bigg[ \bigg(p_{a}p_{b}- p_{a}\expec{P_{b}} -  \expec{P_{a}}p_{b} +   \expec{P_{a}} \expec{P_{b}} \bigg) \Bigg]\\
&=  \bigg( \mathbb{E}[p_{a}p_{b}]- \mathbb{E}[p_{a}\expec{P_{b}}] -  \mathbb{E}[\expec{P_{a}}p_{b}] +   \mathbb{E}[\expec{P_{a}} \expec{P_{b}}] \bigg) \\
&= \expec{P_{a}P_{b}}  -   \expec{P_{a}} \expec{P_{b}} -   \expec{P_{a}} \expec{P_{b}} +   \expec{P_{a}} \expec{P_{b}} \\
&= \expec{P_{a}P_{b}}  -   \expec{P_{a}} \expec{P_{b}} \\
&= \expec{P_{a}} \expec{P_{b}}  -   \expec{P_{a}} \expec{P_{b}} = 0
\end{aligned}
\end{equation}
where under independence: $\expec{P_{a}P_{b}} = \expec{P_{a}} \expec{P_{b}}$.  Intuitively, this result makes sense.  The projective measurement of the first Pauli operator maximally randomizes the expectation value of the other Pauli operator and thus the covariance will be zero.   Interestingly,  the projective measurement causes the underlying distribution of the quantum state to change and so subsequent measurements generating paired samples are not well defined in this setting (for anticommuting operators).  This phenomenon is not present in classical experiments.  However,  the same statistical analysis can be done if we just take pairs of subsequent measurements and only do a statistical analysis on these random variables. We note that our analysis did not have to account for $\expec{P_{a}P_{b}}$ not being a valid observable, as for anticommuting Pauli operators this operator is not Hermitian.

Given the covariance of two anticommuting Pauli operators is zero, we find the variance of a normalized anticommuting clique $\gamma_{j} C_{j}$ to be:

\begin{equation}
\label{eq:Var_anticommuting_case}
\begin{aligned}
Var[\gamma_{j}{C_{j}}]= \gamma_{j}^{2}Var[{C_{j}}] &= \gamma_{j}^{2} Var[\sum_{i}^{|C_{j}|} \frac{c_{i}}{\gamma_{j}} {P_{i}}] = \gamma_{j}^{2} \sum_{i}^{|C_{j}|} \sum_{k}^{|C_{j}|}   Cov[{\frac{c_{i}}{\gamma_{j}} P_{i}}, {\frac{c_{k}}{\gamma_{j}}P_{k}}] \\ 
&=  \gamma_{j}^{2}\sum_{i}^{|C_{j}|} \frac{c_{i}^{2}}{\gamma_{j}^{2}}  Var[{P_{i}}] +  \gamma_{j}^{2}\sum_{i}^{|C_{j}|} \sum_{\substack{k \\ \forall k \neq i}}^{|C_{j}|}  \frac{c_{i}}{\gamma_{j}} \frac{c_{k}}{\gamma_{j}}\underbrace{Cov[{P_{i}}, {P_{k}}]}_{=0} \\
&= \sum_{i}^{|C_{j}|} c_{i}^{2}  Var[{P_{i}}]
\end{aligned}
\end{equation}
We use this to obtain the following $R$ ratio (equation \ref{eq:measurement_ratio}):

\begin{equation}
\label{eq:covar_pauli_anticomm_groups}
\begin{aligned}
    R = \frac{M_{u}}{M_{g} } &= \Bigg(\frac{  \sum_{j}^{N_{C}} \bigg[ \sum_{k=0}^{|C_{j}|-1}   \sqrt{ Var[c_{k}^{(j)} P_{k}^{(j)} ]} \bigg] }{\sum_{j}^{N_{C}}  \sqrt{Var[\gamma_{j} C_{j}]}} \Bigg)^{2} \\
    &= \Bigg(\frac{  \sum_{j}^{N_{C}} \bigg[ \sum_{k=0}^{|C_{j}|-1}    |c_{k}^{(j)}|\sqrt{ Var[P_{k}^{(j)} ]} \bigg] }{\sum_{j}^{N_{C}}  \sqrt{\sum_{k=0}^{|C_{j}|-1}|c_{k}^{(j)}|^{2}  Var[{P_{k}^{(j)}}]}} \Bigg)^{2} \\
   &= \Bigg(\frac{  \sum_{j}^{N_{C}} \bigg[ \sum_{k=0}^{|C_{j}|-1}   |x_{k}^{(j)}| \bigg] }{\sum_{j}^{N_{C}}  \sqrt{\sum_{k=0}^{|C_{j}|-1} |x_{k}^{(j)}|^{2} }} \Bigg)^{2} \\
      &= \Bigg(\frac{  \sum_{j}^{N_{C}} \| \vec{x}_{j} \|_{1} }{\sum_{j}^{N_{C}}  \| \vec{x}_{j} \|_{2} }  \Bigg)^{2}
\end{aligned}
\end{equation}
where $x_{k}^{(j)} = |c_{k}^{(j)}| \sqrt{Var[P_{k}^{(j)} ]}$ and $\vec{x}_{j} = (x_{0}^{(j)}, x_{1}^{(j)}, \hdots, x_{|C_{j}|-1}^{(j)})$. Minkowski inequality ensures $\| \vec{x}_{j}  \|_{2} \leq \| \vec{x}_{j}  \|_{1}$.  At worst unitary partitioning will achieve the same number of measurements as no grouping and will more often achieve an improvement. However, we can actually bound the improvement in general as:

\begin{equation}
\label{eq:L1_L2_inequality}
\begin{aligned}
\| u \|_{1} = \sum_{i}^{n} |u_{i}| = \sum_{i}^{n} |u_{i}| \cdot 1 \leq  \bigg(\sum_{i}^{n} |u_{i}|^{2} \bigg)^{0.5} \cdot \bigg(\sum_{i}^{n} 1^{2} \bigg)^{0.5} = \sqrt{n} \| u \|_{2}
\end{aligned}
\end{equation}
where the  Cauchy-Schwarz inequality has been utilized.  Overall,  we find $\| u \|_{2} \leq  \| u \|_{1} \leq \sqrt{n} \| u \|_{2}$  and thus:

\begin{equation}
\label{eq:M_bounded_UP}
\begin{aligned}
1 \leq R = \frac{M_{u}}{M_{g}} = \Bigg(\frac{  \sum_{j}^{N_{C}} \| \vec{x}_{j} \|_{1} }{\sum_{j}^{N_{C}}  \| \vec{x}_{j} \|_{2} }  \Bigg)^{2} \leq  \Bigg(\frac{\sum_{j}^{N_{C}} \sqrt{|C_{j}|} \cdot \| \vec{x}_{j}   \|_{2}}{\sum_{j}^{N_{C}} \| \vec{x}_{j}   \|_{2}}\Bigg)^{2}. 
\end{aligned}
\end{equation}

\section{\label{sec:example_full}Numerical details of the toy example}

This section provides all the details for the Toy problem described in Section \ref{sec:example_will}. The full noncontextual ground state is:

\begin{equation}
    \label{eq:ground_state_example_full}
\begin{aligned}
(\: \underbrace{-1, +1, -1}_{\vec{q}_{0}}, \underbrace{0.25318483, -0.65828059, -0.70891756}_{\vec{r}_{0}} \:).
\end{aligned}
\end{equation}
This defines the $A(\vec{r}_{0})$:

\begin{equation}
    \label{eq:script_A_example_full}
\begin{aligned}
    A(\vec{r}_{0}) = \: &0.25318483\:YXYI -0.65828059\: XYXI -0.70891756\:XZXI .
\end{aligned}
\end{equation}
The operators to map $A(\vec{r}_{0})$ to a single Pauli operator are:

\begin{equation}
    \label{eq:R_S_example_full}
\begin{aligned}
         R_{S} = \:  e^{+1i \cdot -0.7879622757719398 \cdot ZYZI}
         \cdot e^{+1i \cdot 1.2036225088338255 \cdot ZZZI},
\end{aligned}
\end{equation}
and
\begin{equation}
    \label{eq:R_LCU_example_full}
\begin{aligned}
         R_{LCU} =\: 0.79157591 \: IIII  +0.41580383i \: ZZZI - 0.44778874i \: ZYZI.
\end{aligned}
\end{equation}
Their action results in: $R_{S} A(\vec{r}_{0})  R_{S}^{\dagger} = R_{LCU} A(\vec{r}_{0})  R_{LCU}^{\dagger} = YXYI$.

We then defined $U$ depending on which generators we wish to fix. We found the optimal ordering of stabilizers (supplied in Equation \ref{eq:stabilizers_set_nonrot_example}) to fix via a brute force search over all $\sum_{i=1}^{|\mathcal{W}_{all}|} \binom{|\mathcal{W}_{all}|}{i} = 2^{4}-1 = 15$ possibilities for $\mathcal{W}$. The following optimal ordering was obtained:
\begin{enumerate}
    \item $\{-1\:IIIZ\}$
    \item $\{+1\:IXYI, -1\:IIIZ\}$
    \item $\{+1\:IXYI, -1\:IIIZ, +1\:\mathcal{A}(\vec{r}_{0})\}$
    \item $\{-1\:YIYI, +1\:IXYI, -1\:IIIZ, +1\:\mathcal{A}(\vec{r}_{0})\}$.
\end{enumerate}
This defines all the information required to implement CS-VQE. Table \ref{tab:SeqRot_example_tab} summarises the stabilizers fixed, the rotation $U_{\mathcal{W}}$, required projection $Q_{\mathcal{W}}$ and final projected Hamiltonian $Q_{\mathcal{W}}^{\dagger}U_{\mathcal{W}}^{\dagger}HU_{\mathcal{W}}Q_{\mathcal{W}}$ for this ordering.

The old approach of applying $U_{\mathcal{W}_{all}}^{\dagger}HU_{\mathcal{W}_{all}}$ and then fixing certain stabilizer eigenvalues are summarised in Table  \ref{tab:old_approach_example_tab}. It can be seen from these results, that always implementing the unitary partitioning rotation $R$ can unnecessarily increase the number of terms in the Hamiltonian and thus should only be applied if the eigenvalue for $\langle A(\vec{r}) \rangle $ is fixed.

\begin{table*}[t]
\centering
\begin{adjustbox}{width=1\textwidth}
\small
\begin{tabular}{cccccl}
\hline
$\mathcal{W}$                                                                                        & $U_{\mathcal{W}}^{\dagger}$                                                              & $\mathcal{W}^{Z}= U_{\mathcal{W}}^{\dagger}\mathcal{W}U_{\mathcal{W}}$                                    & $Q_{\mathcal{W}}$                                                                             & \multicolumn{2}{c}{$Q_{\mathcal{W}}^{\dagger}U_{\mathcal{W}}^{\dagger}HU_{\mathcal{W}}Q_{\mathcal{W}}$}                                                                                                                                                                                                                                                                                                                \\ \hline
\begin{tabular}[c]{@{}c@{}}-1 YIYI \\ +1 IXYI \\  -1 IIIZ\\  $\mathcal{A}(\vec{r}_{0})$\end{tabular} & $e^{1i \frac{\pi}{4} XIYI} e^{1i \frac{\pi}{4} IYYI}  R_{S/LCU}$ & \begin{tabular}[c]{@{}c@{}}- ZIII \\ + IZII \\ -  IIIZ \\ + IIZI\end{tabular} & $\ket{1}\bra{1} \otimes \ket{0}\bra{0} \otimes \ket{1}\bra{1} \otimes \ket{0}\bra{0}$                                                  & \multicolumn{2}{c}{-2.475+0.000j}                                                                                                                                                                                                                                                                                                                              \\
\begin{tabular}[c]{@{}c@{}}+1 IXYI \\ -1 IIIZ\\  $\mathcal{A}(\vec{r}_{0})$\end{tabular}             & $ e^{1i \frac{\pi}{4} IYYI} R_{S/LCU}$                           & \begin{tabular}[c]{@{}c@{}}+ IZII \\ - IIIZ\\ + IIZI\end{tabular}             & $ I \otimes  \ket{0}\bra{0} \otimes  \ket{0}\bra{0}  \otimes  \ket{1}\bra{1}  $ & \multicolumn{1}{c|}{\begin{tabular}[c]{@{}c@{}}SeqRot\\ -1.827+0.000j I +\\ -0.198+0.000j X +\\ -0.467+0.000j Z +\\ 0.648+0.000j Y\end{tabular}}                                          & \begin{tabular}[c]{@{}c@{}}LCU\\ -1.827+0.000j I +\\ -0.414+0.000j X +\\ -0.292+0.000j Z +\\ 0.648+0.000j Y\end{tabular}                                         \\
\begin{tabular}[c]{@{}c@{}}+1 IXYI \\ -1 IIIZ\end{tabular}                                           & $ e^{1i \frac{\pi}{4} IYYI}$                                     & \begin{tabular}[c]{@{}c@{}}+ IZII \\ - IIIZ\end{tabular}                      & $ I \otimes  \ket{0}\bra{0} \otimes  I  \otimes  \ket{1}\bra{1}  $              & \multicolumn{2}{c}{\begin{tabular}[c]{@{}c@{}}-0.500+0.000j II +\\ 0.500+0.000j XI +\\ 0.700+0.000j XX +\\ 0.100+0.000j YI +\\ -0.100+0.000j YX +\\ 1.300+0.000j XZ +\\ 0.600+0.000j IY +\\ 0.700+0.000j ZZ\end{tabular}}                                                                                                                                      \\
-1 IIIZ                                                                                              & $IIII$                                                           & -IIIZ                                                                         & $ I \otimes I \otimes  I  \otimes  \ket{1}\bra{1}  $                            & \multicolumn{2}{c}{\begin{tabular}[c]{@{}c@{}}-0.500+0.000j III +\\ 0.100+0.000j XXX +\\ 0.200+0.000j YXX +\\ 0.700+0.000j XZX +\\ 0.700+0.000j XYX +\\ 0.100+0.000j YZX +\\ 0.200+0.000j XXZ +\\ 0.600+0.000j IIY +\\ 0.500+0.000j XXY +\\ 0.100+0.000j YXY +\\ 0.600+0.000j XZZ +\\ 0.700+0.000j ZZZ +\\ 0.200+0.000j YYZ +\\ 0.100+0.000j ZYY\end{tabular}} \\ \hline
\end{tabular}
\end{adjustbox}
\caption{Different contextual subspace Hamiltonians defined from $H$ (Equation \ref{eq:H_full_Example}). $R_{S}$ and $R_{LCU}$ are defined in Equations \ref{eq:R_S_example_full} and \ref{eq:R_LCU_example_full}. \label{tab:SeqRot_example_tab}} 
\end{table*}

\clearpage
\newpage
\begin{table*}[h!]
\centering
\begin{adjustbox}{width=1\textwidth}
\small
\begin{tabular}{ll|ccll}
\hline
\multicolumn{1}{c}{$H_{SeqRot}=U_{\mathcal{W}_{all}}^{\dagger, SeqRot} H U_{\mathcal{W}_{all}}^{SeqRot} $}                                                                                                                                                                                                                                                                                                                                                                                                                                                                                                                                                              & \multicolumn{1}{c|}{$  H_{LCU} =U_{\mathcal{W}_{all}}^{\dagger, LCU} H U_{\mathcal{W}_{all}}^{LCU} $}                                                                                                                                                                                                                                                                                                                                                                                                                                                                                                                                                                                                                                      & \multicolumn{1}{c}{$\mathcal{W}^{Z}$}                                         & $Q_{\mathcal{W}}$                                                                             & \multicolumn{1}{c}{$Q_{\mathcal{W}}^{\dagger}H_{SeqRot}Q_{\mathcal{W}}$}                                                                                                                                                                                                                                                                                                                                                                                                                                                                                                                                                                     & \multicolumn{1}{c}{$Q_{\mathcal{W}}^{\dagger}H_{LCU}Q_{\mathcal{W}}$}                                                                                                                                                                                                                                                                                                                                                                                                                                                                                                                                                                                                                                        \\ \hline
\multirow{4}{*}{\begin{tabular}[c]{@{}l@{}}0.932-0.000j ZIII +\\ -0.056+0.000j YIII +\\ -0.025+0.000j ZXII +\\ -0.025+0.000j YXII +\\ 0.057-0.000j ZIXI +\\ -0.197+0.000j YIXI +\\ -0.051+0.000j ZXXI +\\ 0.051+0.000j YXXI +\\ 0.560-0.000j XZII +\\ 0.395+0.000j ZZII +\\ 0.397-0.000j YZII +\\ 0.141-0.000j ZZXI +\\ 0.142-0.000j YZXI +\\ 0.345-0.000j XIZI +\\ 0.093-0.000j XXZI +\\ 0.467-0.000j IIYI +\\ -0.187+0.000j IXYI +\\ -0.496+0.000j ZIZI +\\ 0.494-0.000j YIZI +\\ 0.215+0.000j XZZI +\\ -0.200+0.000j IYZI +\\ -0.152+0.000j ZZZI +\\ -0.153+0.000j YZZI +\\ 0.071+0.000j ZYYI +\\ 0.071-0.000j YYYI +\\ -0.500+0.000j IIIZ\end{tabular}} & \multirow{4}{*}{\begin{tabular}[c]{@{}l@{}}0.261-0.000j XIII +\\ 0.932-0.000j ZIII +\\ -0.230+0.000j YIII +\\ -0.025+0.000j ZXII +\\ -0.071+0.000j YXII +\\ 0.295-0.000j ZIXI +\\ -0.197+0.000j YIXI +\\ -0.142+0.000j ZXXI +\\ 0.051+0.000j YXXI +\\ 0.395-0.000j XZII +\\ 0.037-0.000j IYXI +\\ 0.395-0.000j ZZII +\\ 0.223-0.000j YZII +\\ 0.120-0.000j ZZXI +\\ 0.142-0.000j YZXI +\\ 0.263-0.000j XIZI +\\ 0.066-0.000j XXZI +\\ 0.366-0.000j IIYI +\\ -0.132+0.000j IXYI +\\ -0.496+0.000j ZIZI +\\ 0.419-0.000j YIZI +\\ 0.393+0.000j XZZI +\\ -0.200+0.000j IYZI +\\ -0.074+0.000j IZYI +\\ -0.152+0.000j ZZZI +\\ -0.425+0.000j YZZI +\\ 0.060+0.000j ZYYI +\\ 0.071-0.000j YYYI +\\ -0.500+0.000j IIIZ\end{tabular}} & \begin{tabular}[c]{@{}c@{}}- ZIII \\ + IZII \\ -  IIIZ \\ + IIZI\end{tabular} & $\ket{1}\bra{1} \otimes \ket{0}\bra{0} \otimes \ket{1}\bra{1} \otimes \ket{0}\bra{0}$                                                  & -2.475+0.000j                                                                                                                                                                                                                                                                                                                                                                                                                                                                                                                                                                                                    & -2.475+0.000j                                                                                                                                                                                                                                                                                                                                                                                                                                                                                                                                                                                                                                                                    \\ \cline{3-6} 
                                                                                                                                                                                                                                                                                                                                                                                                                                                                                                                                                                                                                                                            &                                                                                                                                                                                                                                                                                                                                                                                                                                                                                                                                                                                                                                                                                                                                & \begin{tabular}[c]{@{}c@{}}+ IZII \\ - IIIZ\\ + IIZI\end{tabular}             & $ I \otimes  \ket{0}\bra{0} \otimes  \ket{0}\bra{0}  \otimes  \ket{1}\bra{1}  $ & \begin{tabular}[c]{@{}l@{}}-1.827+0.000j I +\\ -0.198+0.000j X +\\ 0.648+0.000j Z +\\ 0.467+0.000j Y\end{tabular}                                                                                                                                                                                                                                                                                                                                                                                                                                                                                                & \begin{tabular}[c]{@{}l@{}}-1.827+0.000j I +\\ -0.414+0.000j X +\\ 0.648+0.000j Z +\\ 0.292+0.000j Y\end{tabular}                                                                                                                                                                                                                                                                                                                                                                                                                                                                                                                                                                \\ \cline{3-6} 
                                                                                                                                                                                                                                                                                                                                                                                                                                                                                                                                                                                                                                                            &                                                                                                                                                                                                                                                                                                                                                                                                                                                                                                                                                                                                                                                                                                                                & \begin{tabular}[c]{@{}c@{}}+ IZII \\ - IIIZ\end{tabular}                      & $ I \otimes  \ket{0}\bra{0} \otimes  I  \otimes  \ket{1}\bra{1}  $              & \begin{tabular}[c]{@{}l@{}}-0.500+0.000j II +\\ 0.560+0.000j XI +\\ 1.327+0.000j ZI +\\ 0.341+0.000j YI +\\ 0.198+0.000j ZX +\\ -0.056+0.000j YX +\\ 0.560+0.000j XZ +\\ 0.467+0.000j IY +\\ -0.648+0.000j ZZ +\\ 0.341+0.000j YZ\end{tabular}                                                                                                                                                                                                                                                                                                                                                                   & \begin{tabular}[c]{@{}l@{}}-0.500+0.000j II +\\ 0.656+0.000j XI +\\ 1.327+0.000j ZI +\\ -0.006+0.000j YI +\\ 0.414+0.000j ZX +\\ -0.056+0.000j YX +\\ 0.656+0.000j XZ +\\ 0.292+0.000j IY +\\ -0.648+0.000j ZZ +\\ -0.006+0.000j YZ\end{tabular}                                                                                                                                                                                                                                                                                                                                                                                                                                 \\ \cline{3-6} 
                                                                                                                                                                                                                                                                                                                                                                                                                                                                                                                                                                                                                                                            &                                                                                                                                                                                                                                                                                                                                                                                                                                                                                                                                                                                                                                                                                                                                & -IIIZ                                                                         & $ I \otimes I \otimes  I  \otimes  \ket{1}\bra{1}  $                            & \begin{tabular}[c]{@{}l@{}}-0.500+0.000j III +\\ 0.932+0.000j ZII +\\ -0.056+0.000j YII +\\ -0.025+0.000j ZXI +\\ -0.025+0.000j YXI +\\ 0.057+0.000j ZIX +\\ -0.197+0.000j YIX +\\ -0.051+0.000j ZXX +\\ 0.051+0.000j YXX +\\ 0.560+0.000j XZI +\\ 0.395+0.000j ZZI +\\ 0.397+0.000j YZI +\\ 0.141+0.000j ZZX +\\ 0.142+0.000j YZX +\\ 0.345+0.000j XIZ +\\ 0.093+0.000j XXZ +\\ 0.467+0.000j IIY +\\ -0.187+0.000j IXY +\\ -0.496+0.000j ZIZ +\\ 0.494+0.000j YIZ +\\ 0.215+0.000j XZZ +\\ -0.200+0.000j IYZ +\\ -0.152+0.000j ZZZ +\\ -0.153+0.000j YZZ +\\ 0.071+0.000j ZYY +\\ 0.071+0.000j YYY\end{tabular} & \begin{tabular}[c]{@{}l@{}}-0.500+0.000j III +\\ 0.261+0.000j XII +\\ 0.932+0.000j ZII +\\ -0.230+0.000j YII +\\ -0.025+0.000j ZXI +\\ -0.071+0.000j YXI +\\ 0.295+0.000j ZIX +\\ -0.197+0.000j YIX +\\ -0.142+0.000j ZXX +\\ 0.051+0.000j YXX +\\ 0.395+0.000j XZI +\\ 0.037+0.000j IYX +\\ 0.395+0.000j ZZI +\\ 0.223+0.000j YZI +\\ 0.120+0.000j ZZX +\\ 0.142+0.000j YZX +\\ 0.263+0.000j XIZ +\\ 0.066+0.000j XXZ +\\ 0.366+0.000j IIY +\\ -0.132+0.000j IXY +\\ -0.496+0.000j ZIZ +\\ 0.419+0.000j YIZ +\\ 0.393+0.000j XZZ +\\ -0.200+0.000j IYZ +\\ -0.074+0.000j IZY +\\ -0.152+0.000j ZZZ +\\ -0.425+0.000j YZZ +\\ 0.060+0.000j ZYY +\\ 0.071+0.000j YYY\end{tabular} \\ \hline
\end{tabular}
\end{adjustbox}
\caption{Different contextual subspace Hamiltonians defined from $H$ (Equation \ref{eq:H_full_Example}). Here $\mathcal{W}$ has been set to $\mathcal{W}_{all}$, which defines $U_{\mathcal{W}_{all}}^{\dagger}=e^{1i \frac{\pi}{4} XIYI} e^{1i \frac{\pi}{4} IYYI}  R_{S/LCU}$. $R_{S}$ and $R_{LCU}$ are defined in Equations \ref{eq:R_S_example_full} and \ref{eq:R_LCU_example_full}. The two left columns ($H_{SeqRot}$ and $H_{LCU}$) give $H$ rotated by $U_{\mathcal{W}}$. Each projected Hamiltonian is generated from these, where the eigenvalue of certain stabilizers are fixed according to the projector $Q_{\mathcal{W}}$. For the last two rows, the eigenvalue of $ \mathcal{A}(\vec{r}_{0}) $ has not been fixed, but the non-Clifford operator $R_{S/LCU}$ is still included within $U_{\mathcal{W}_{all}}^{\dagger}$. This leads to an unnecessary increase in the number of Pauli operators for these two cases, as these transformed operators are isospectral with associated Hamiltonians in Table \ref{tab:SeqRot_example_tab}.  \label{tab:old_approach_example_tab}}
\end{table*}

\clearpage
\newpage
\section{\label{sec:Molecular_Results_graphic}Graphical Results for CS-VQE simulation of each Molecular Hamiltonian}

\begin{figure}[H]
 \label{fig: all_plots1}
  \centering
  \subfloat[]{\label{fig:0}\includegraphics[width=0.5\columnwidth]{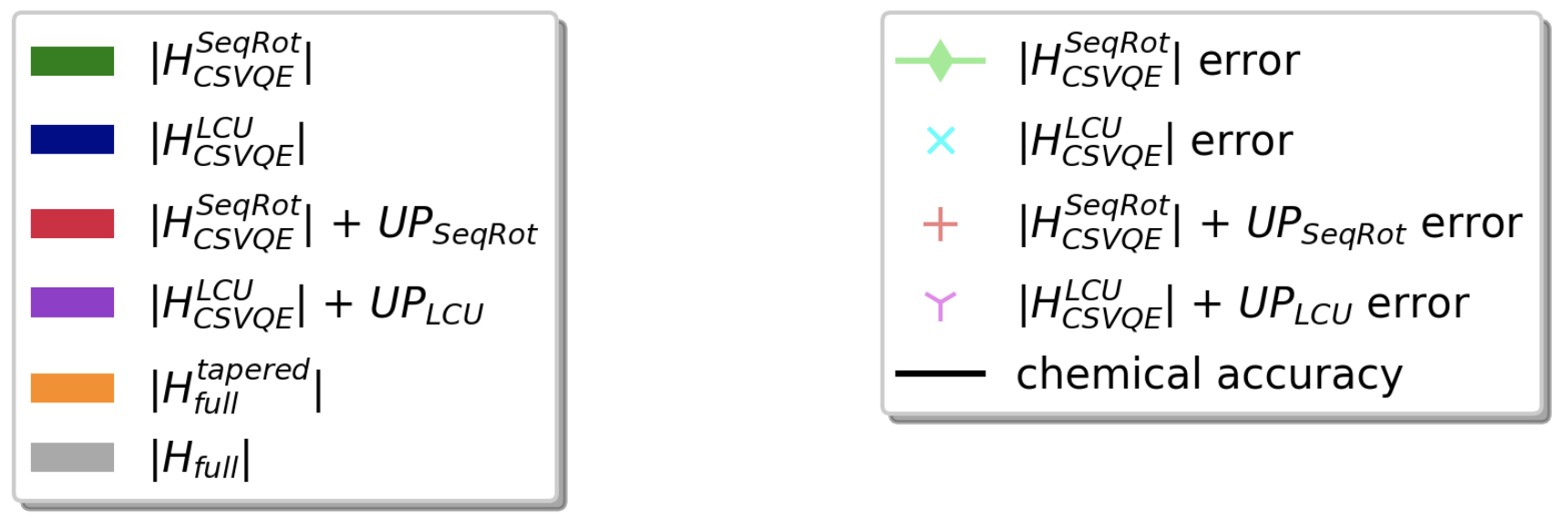}}
  \\
  \subfloat[]{\label{fig:1}\includegraphics[width=0.45\columnwidth]{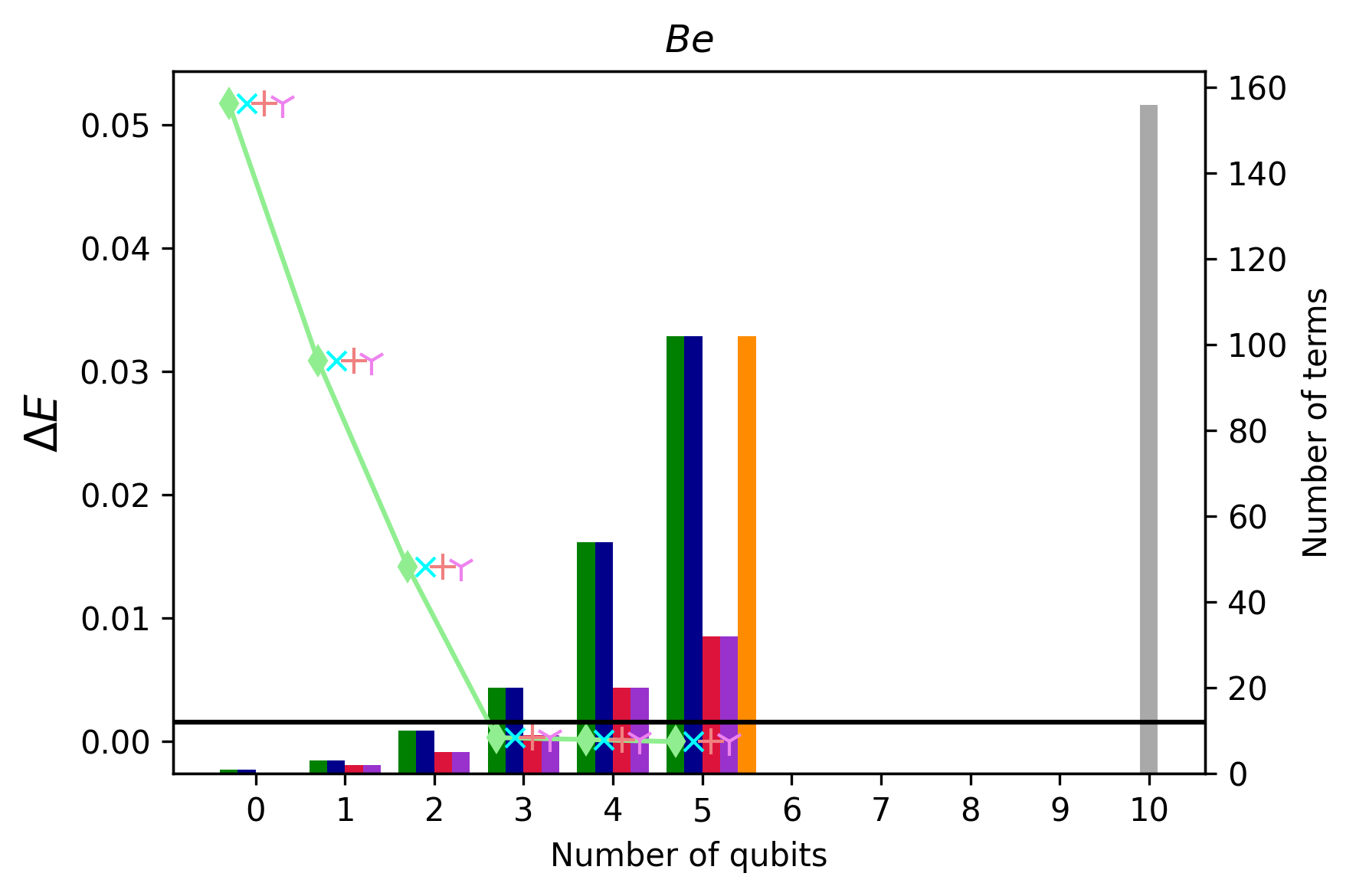}}
  \hfill
  \subfloat[]{\label{fig:2}\includegraphics[width=0.45\columnwidth]{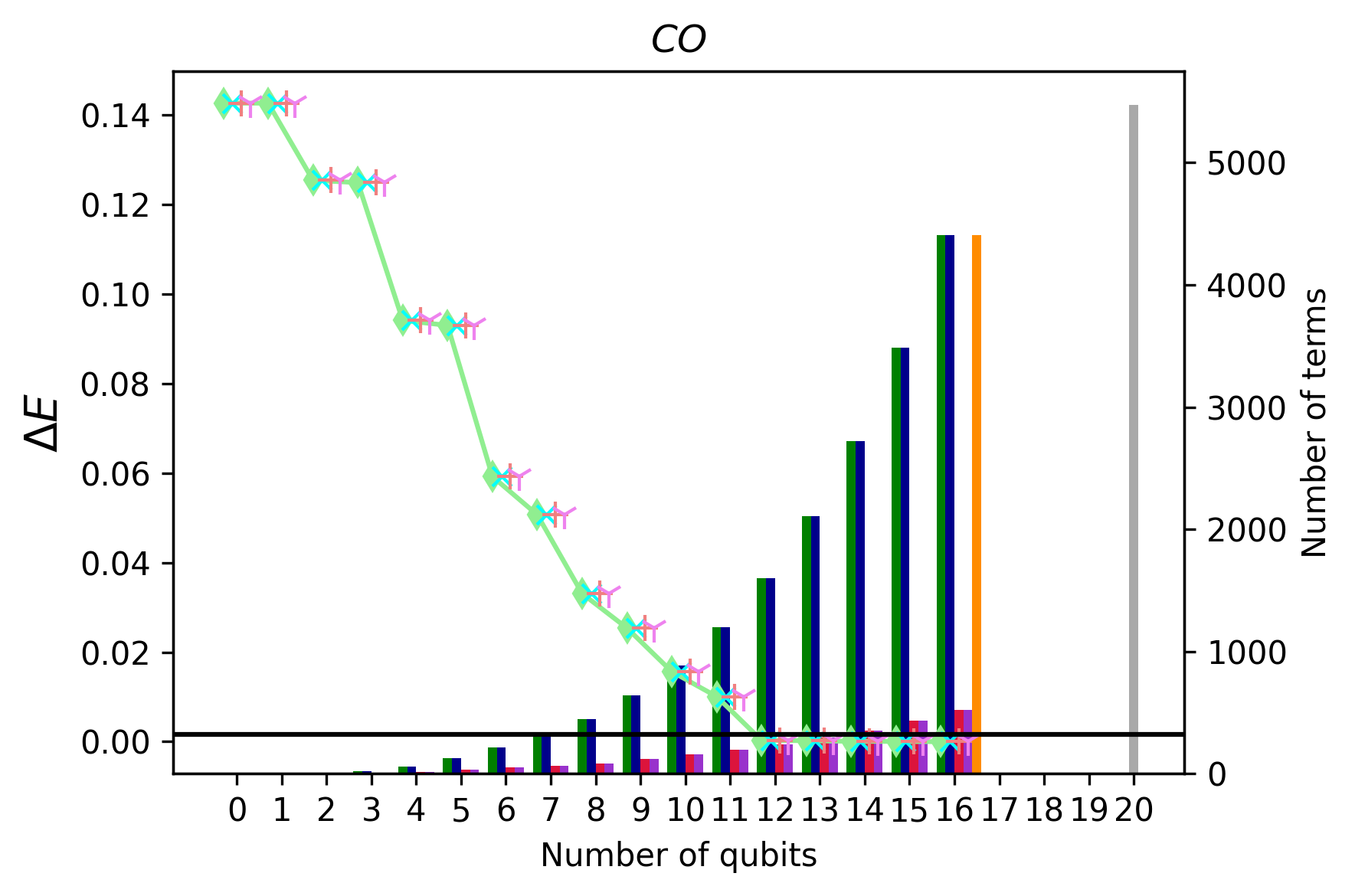}}
  \\
  \subfloat[]{\label{fig:3}\includegraphics[width=0.45\columnwidth]{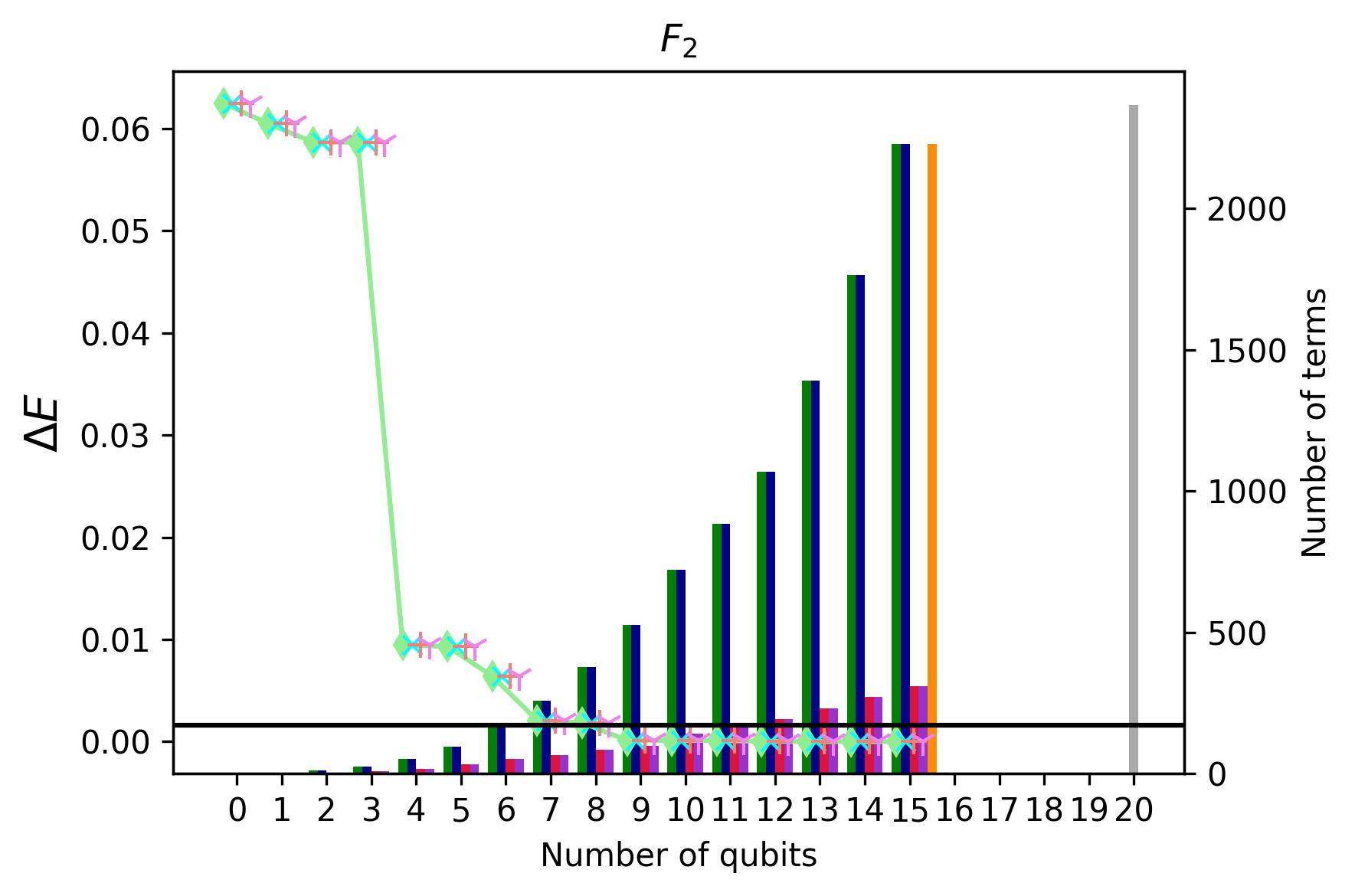}}
  \hfill
  \subfloat[]{\label{fig:4}\includegraphics[width=0.45\columnwidth]{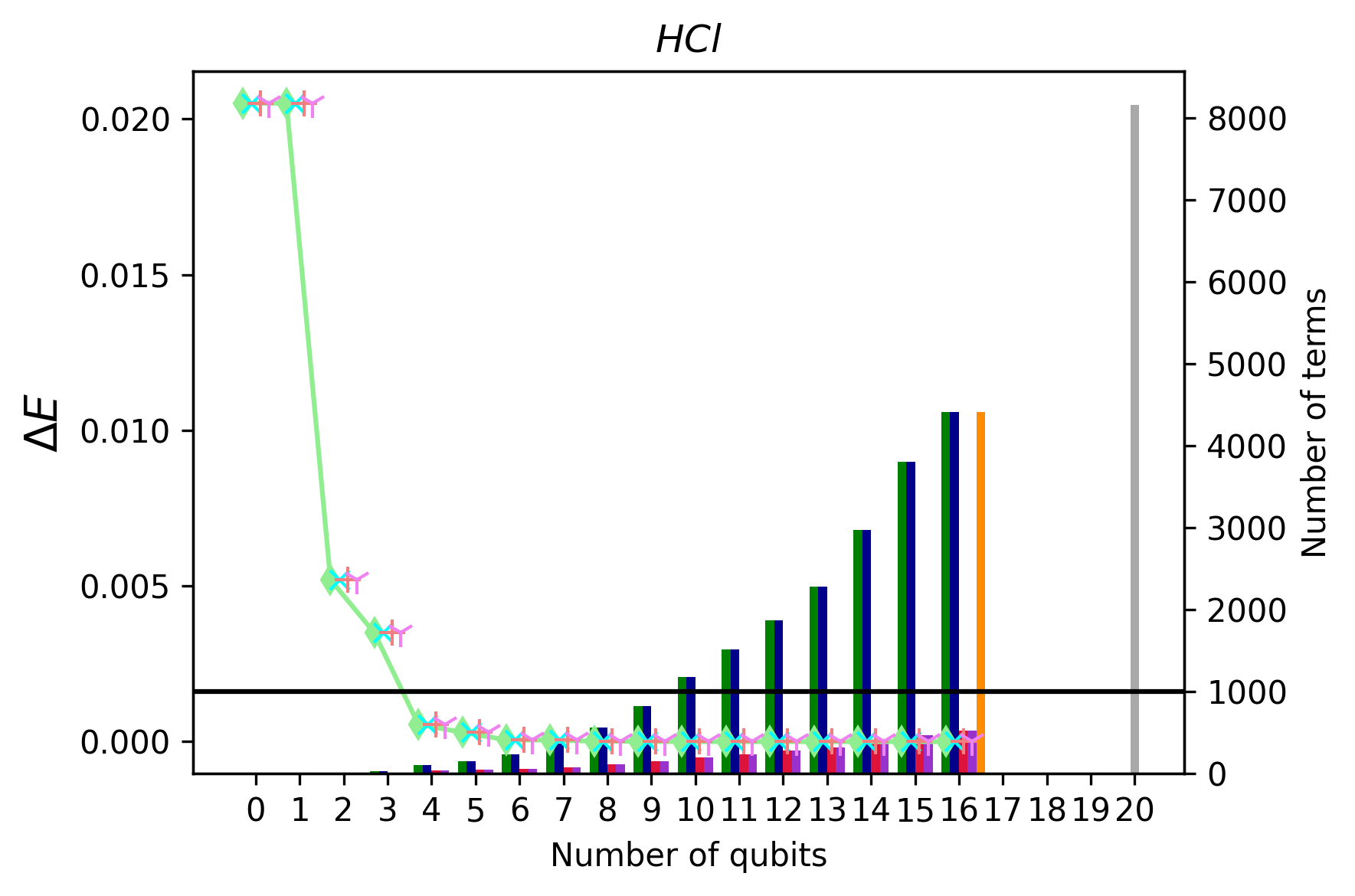}}
  \\
  \subfloat[]{\label{fig:5}\includegraphics[width=0.45\columnwidth]{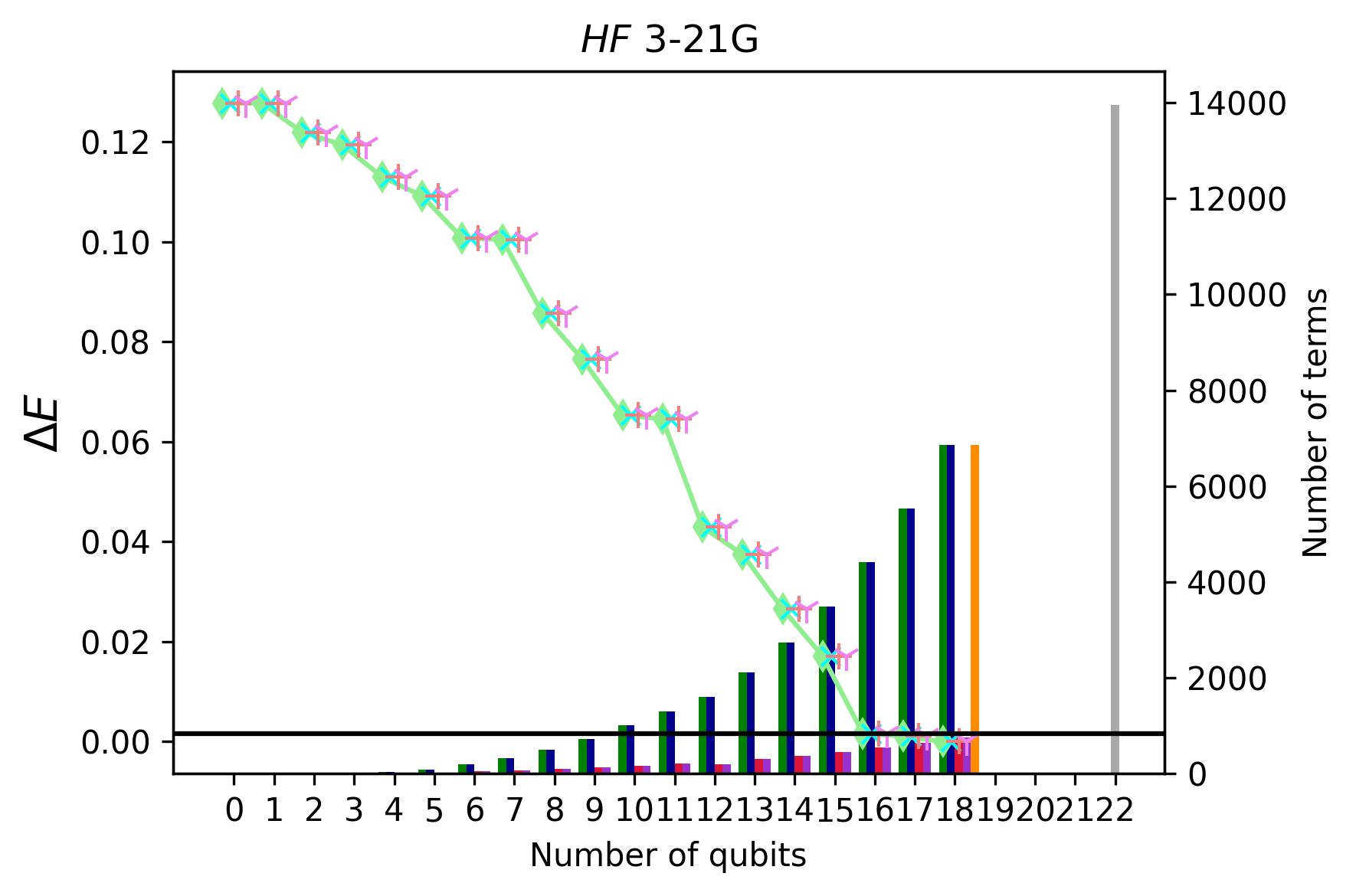}}
  \hfill
  \subfloat[]{\label{fig:6}\includegraphics[width=0.45\columnwidth]{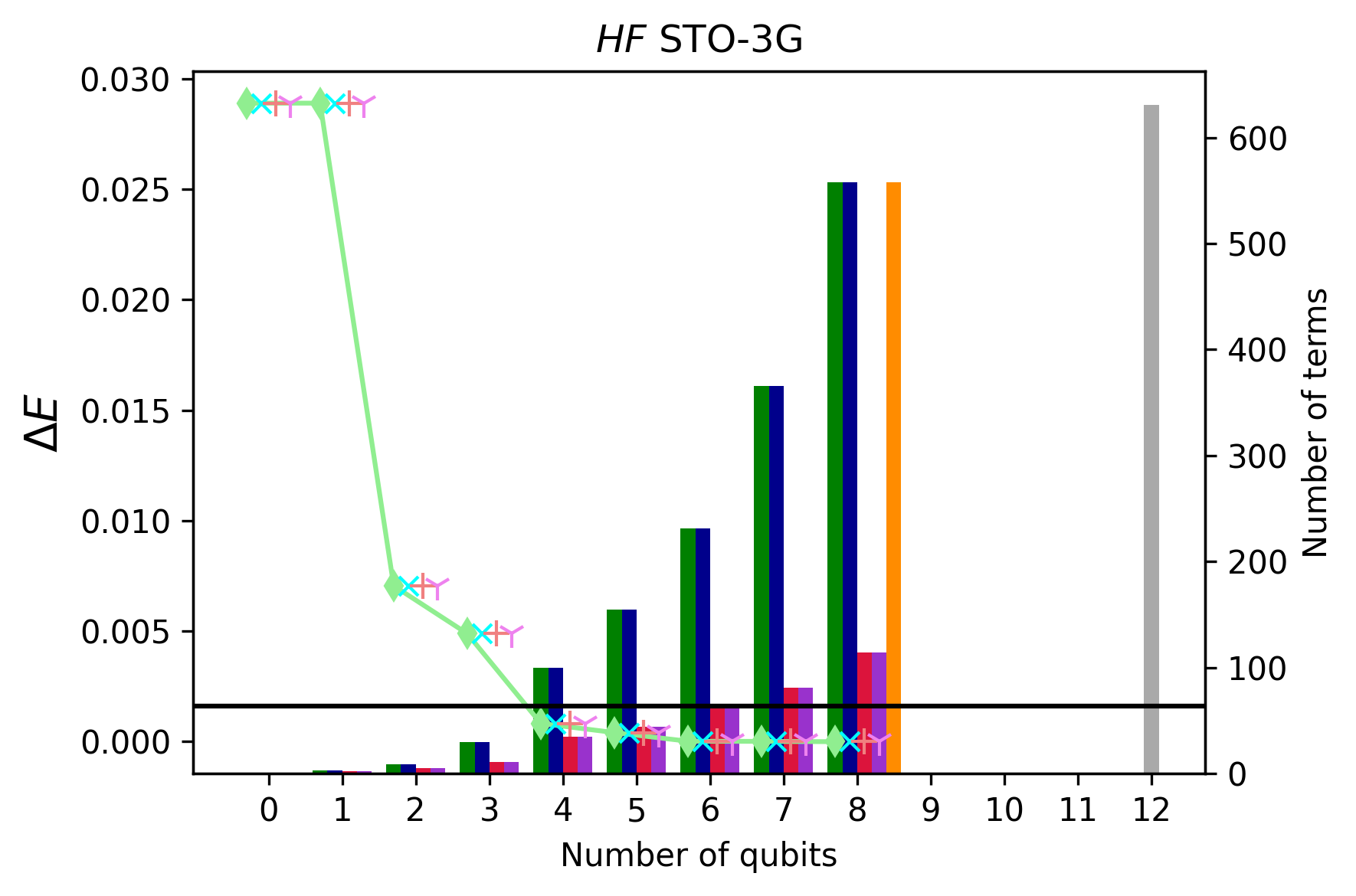}}
\end{figure}

\begin{figure}[H]
\ContinuedFloat 
 \label{fig: all_plots2}
  \centering
  \subfloat[]{\label{fig:7}\includegraphics[width=0.45\columnwidth]{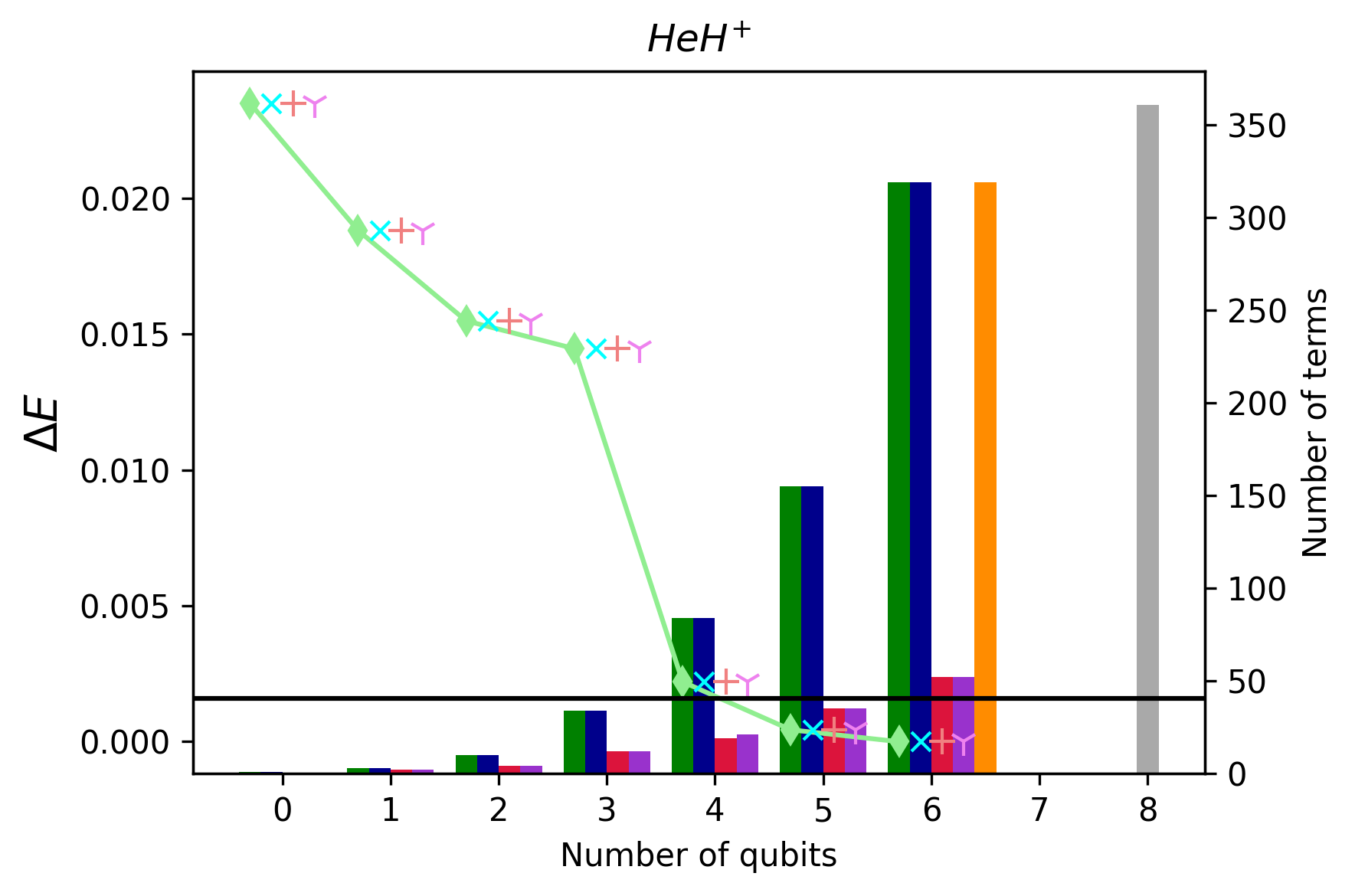}}
  \hfill
  \subfloat[]{\label{fig:8}\includegraphics[width=0.45\columnwidth]{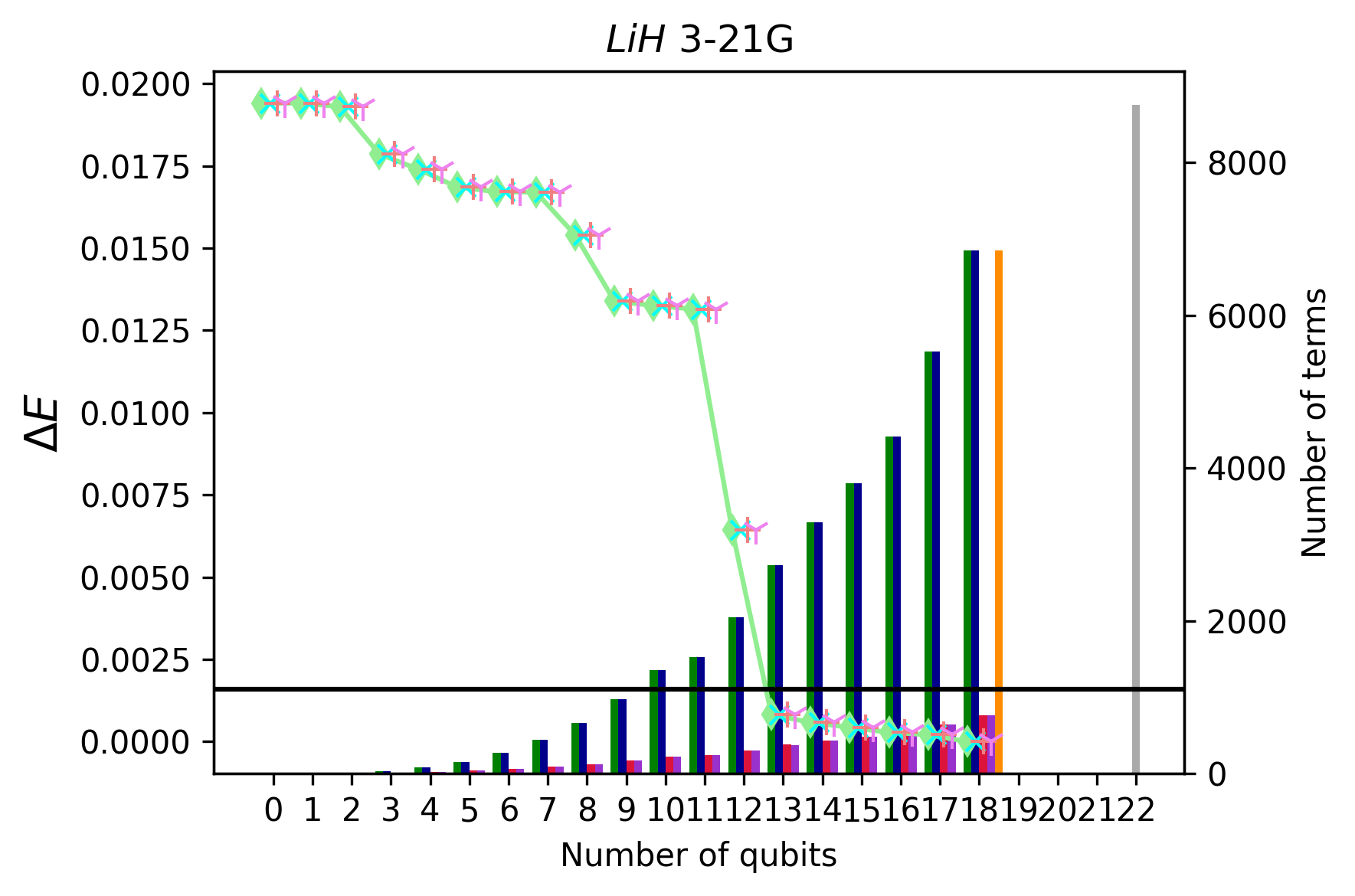}}
  \\
  \subfloat[]{\label{fig:9}\includegraphics[width=0.45\columnwidth]{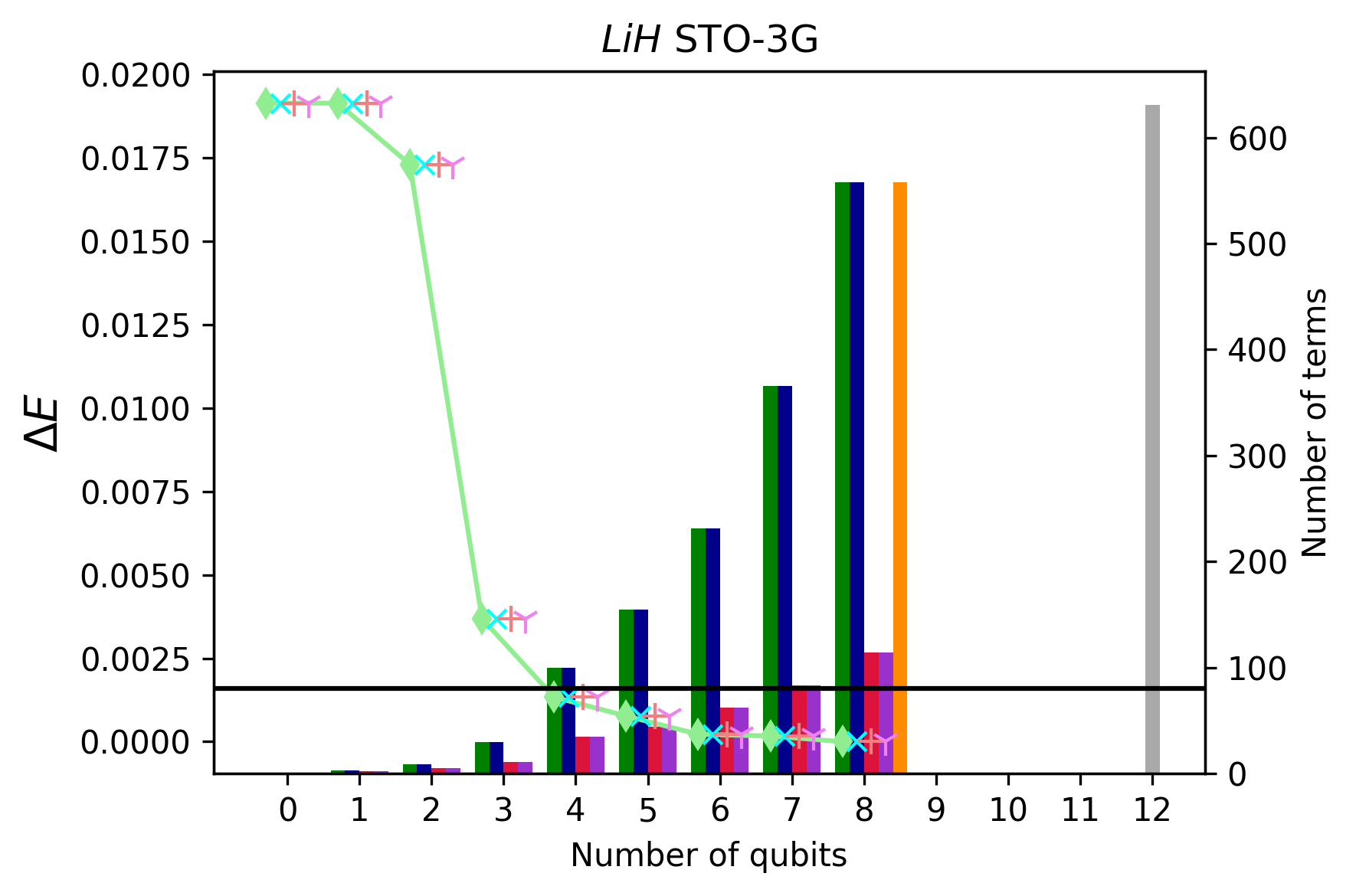}}
  \hfill
  \subfloat[]{\label{fig:10}\includegraphics[width=0.45\columnwidth]{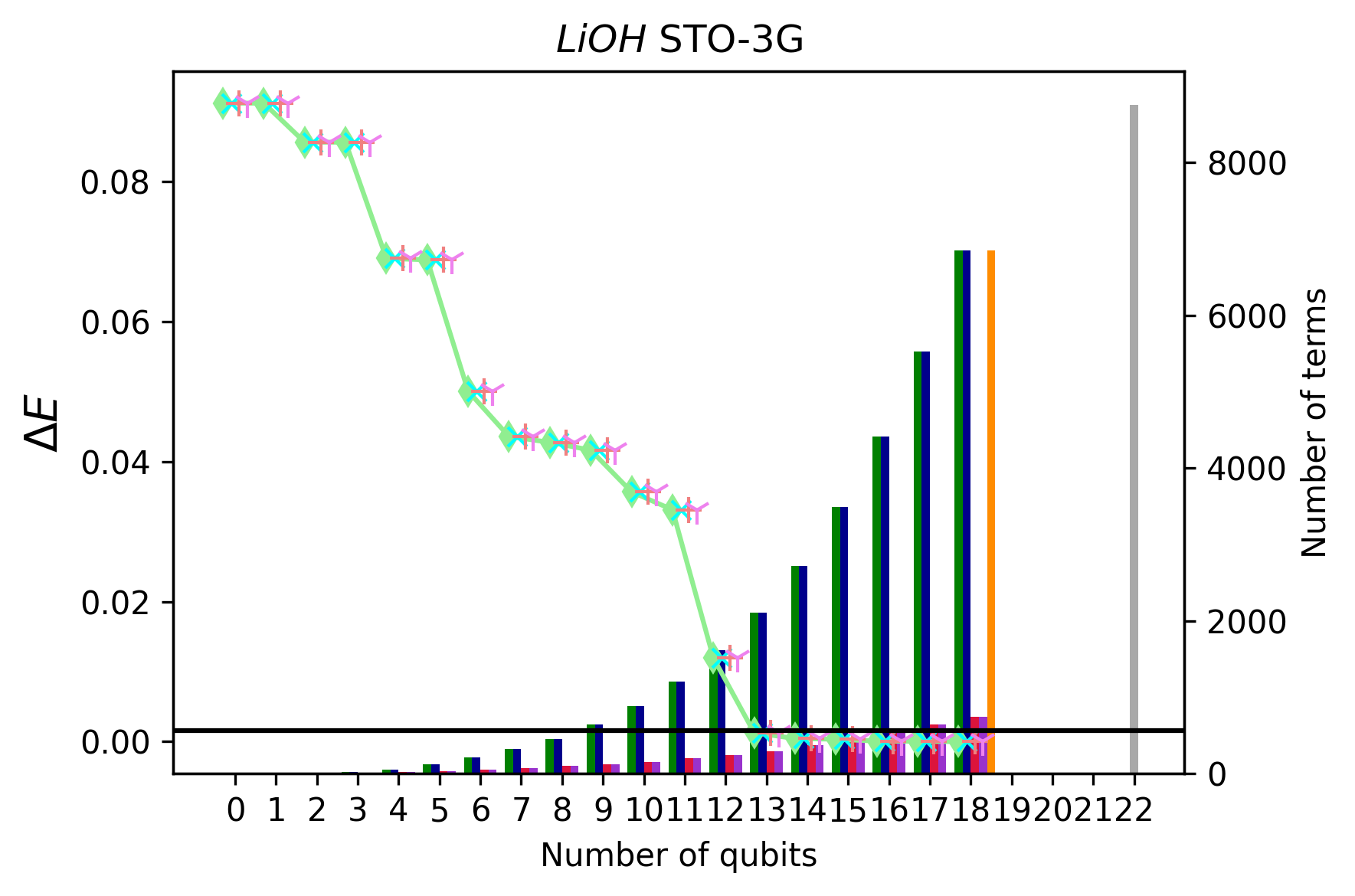}}
  \\
  \subfloat[]{\label{fig:11}\includegraphics[width=0.45\columnwidth]{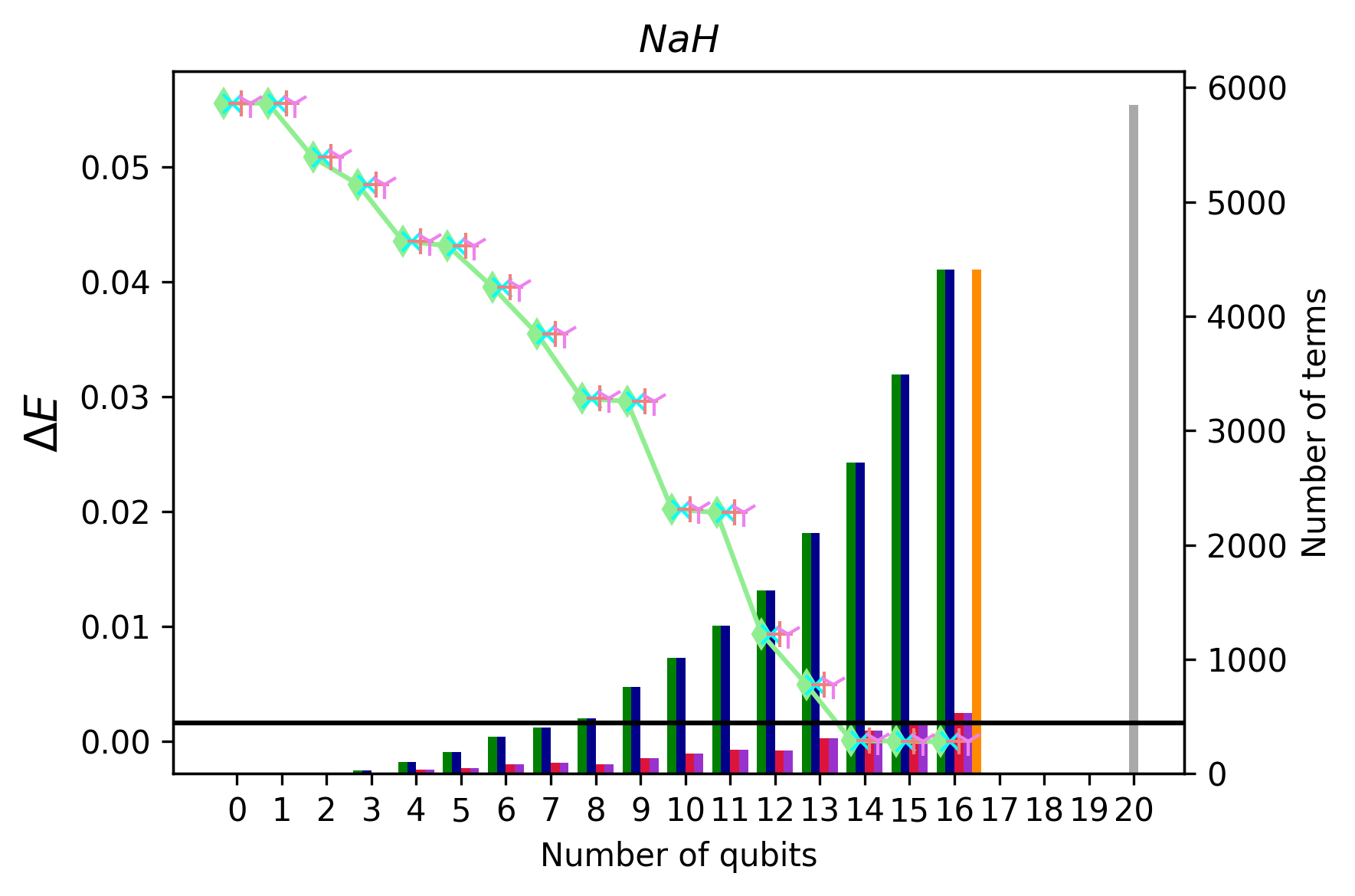}}
  \hfill
  \subfloat[]{\label{fig:12}\includegraphics[width=0.45\columnwidth]{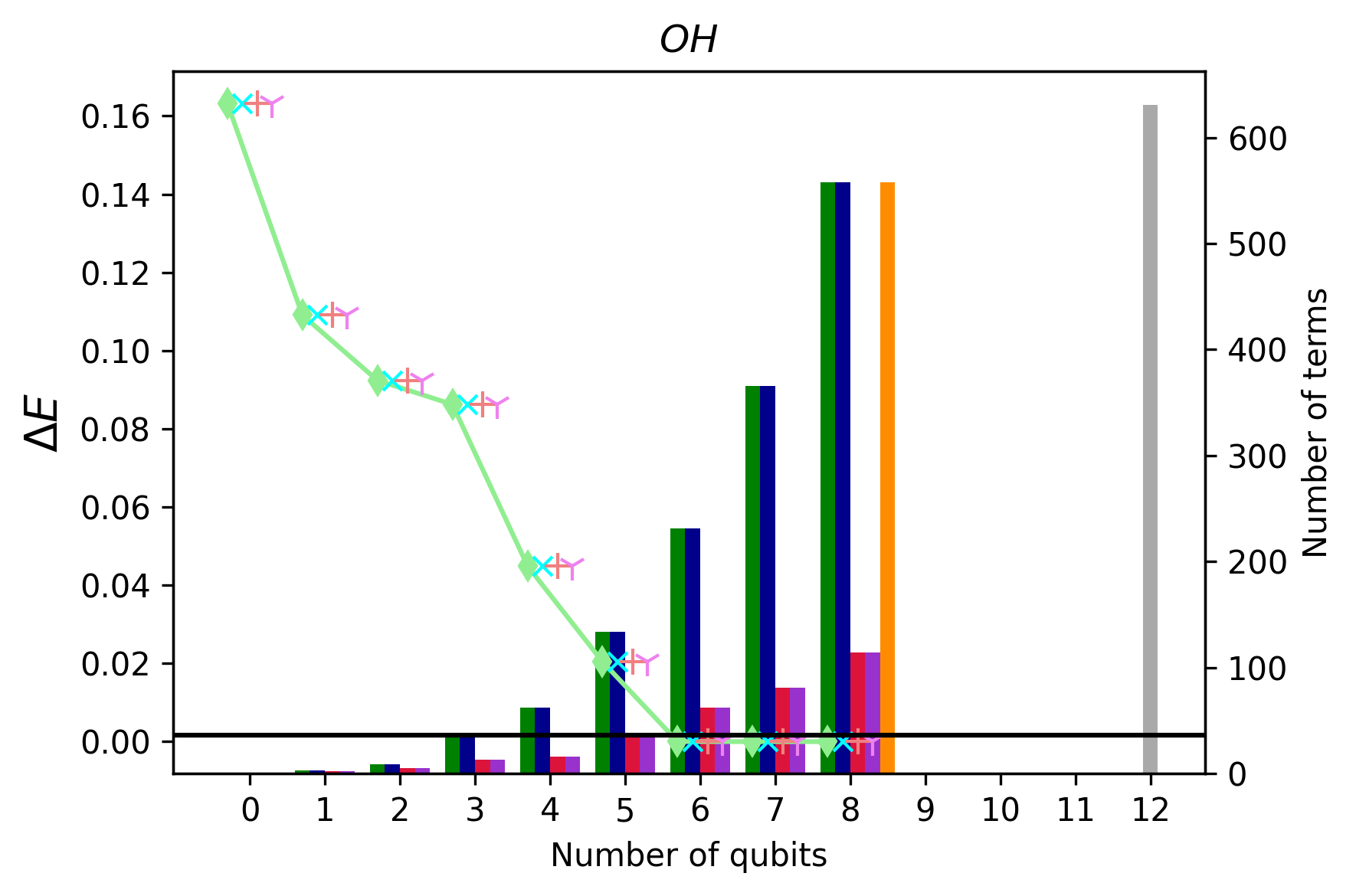}}
  \\
  \subfloat[]{\label{fig:13}\includegraphics[width=0.45\columnwidth]{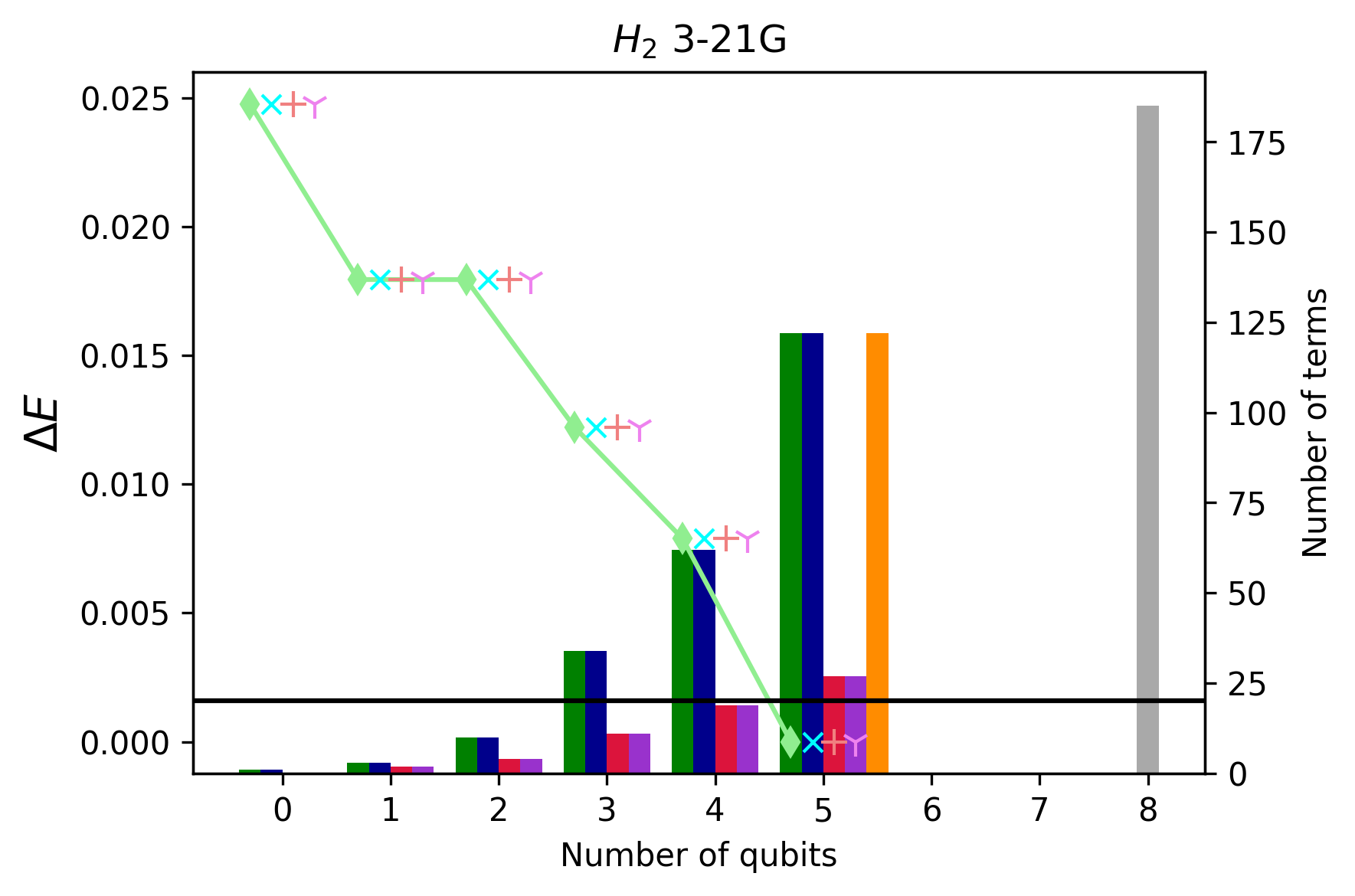}}
  \hfill
  \subfloat[]{\label{fig:14}\includegraphics[width=0.45\columnwidth]{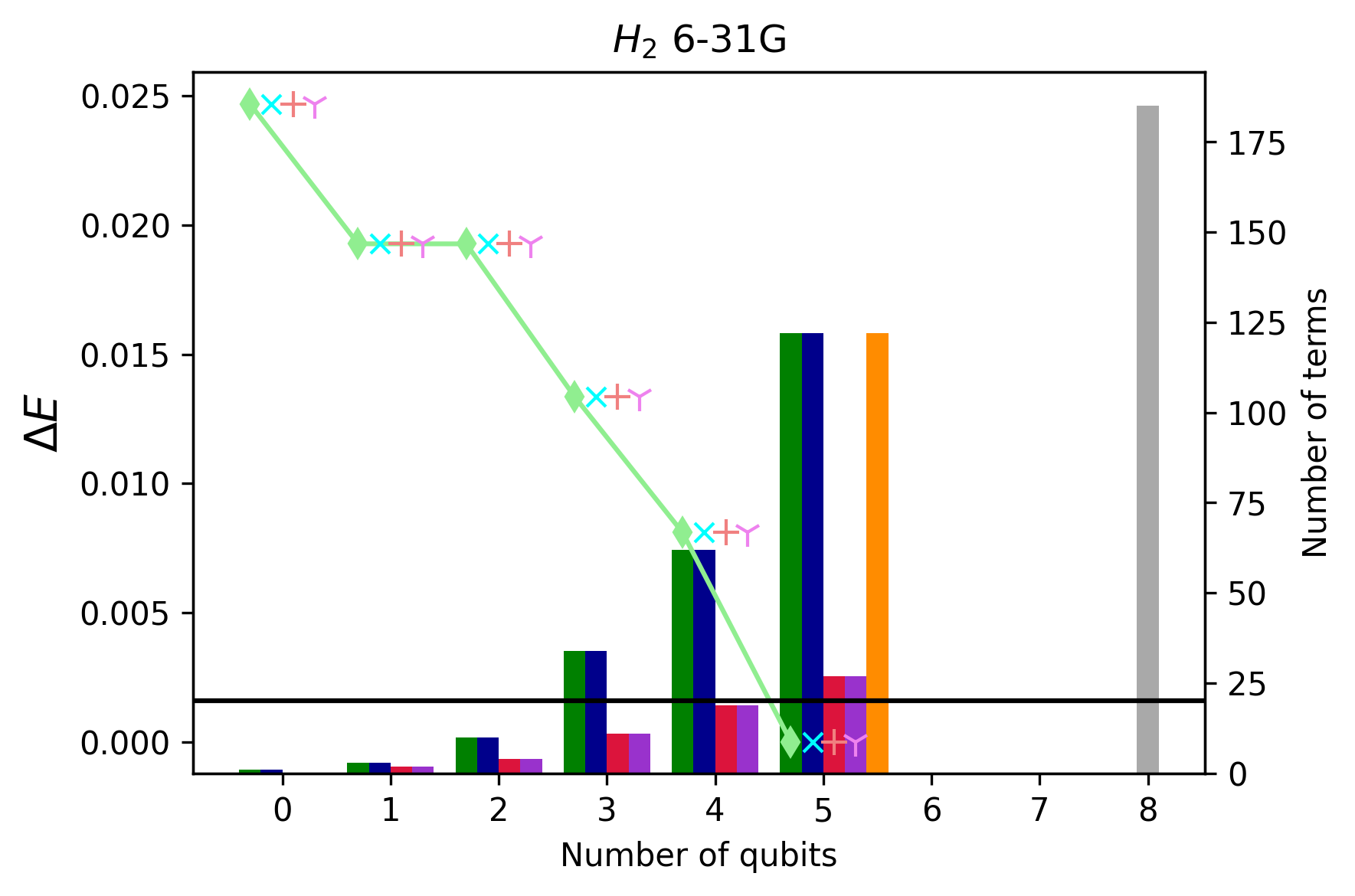}}
\end{figure}

\begin{figure}[H]
\ContinuedFloat 
 \label{fig: all_plots3}
  \centering
  \subfloat[]{\label{fig:15}\includegraphics[width=0.45\columnwidth]{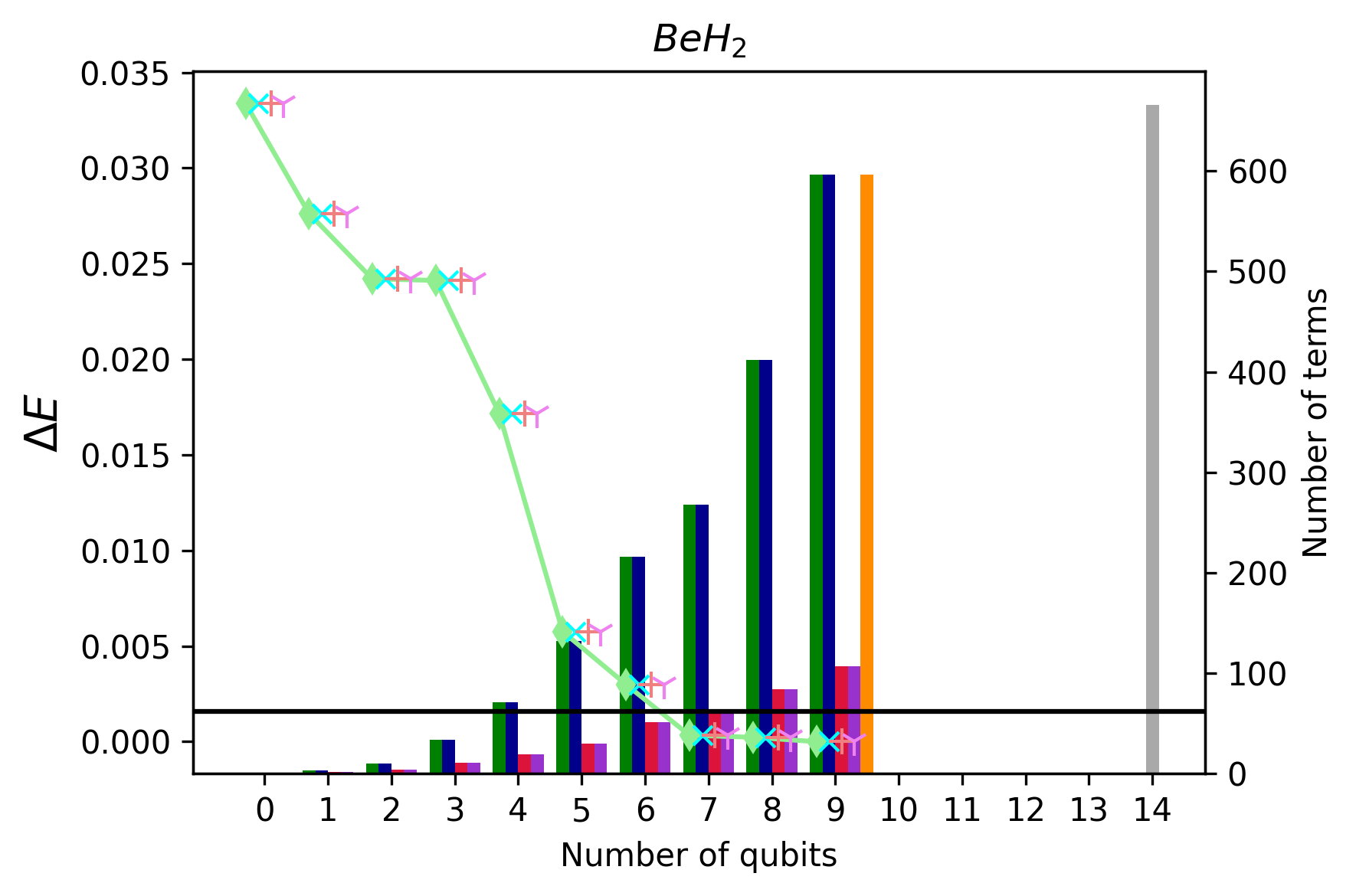}}
  \hfill
  \subfloat[]{\label{fig:16}\includegraphics[width=0.45\columnwidth]{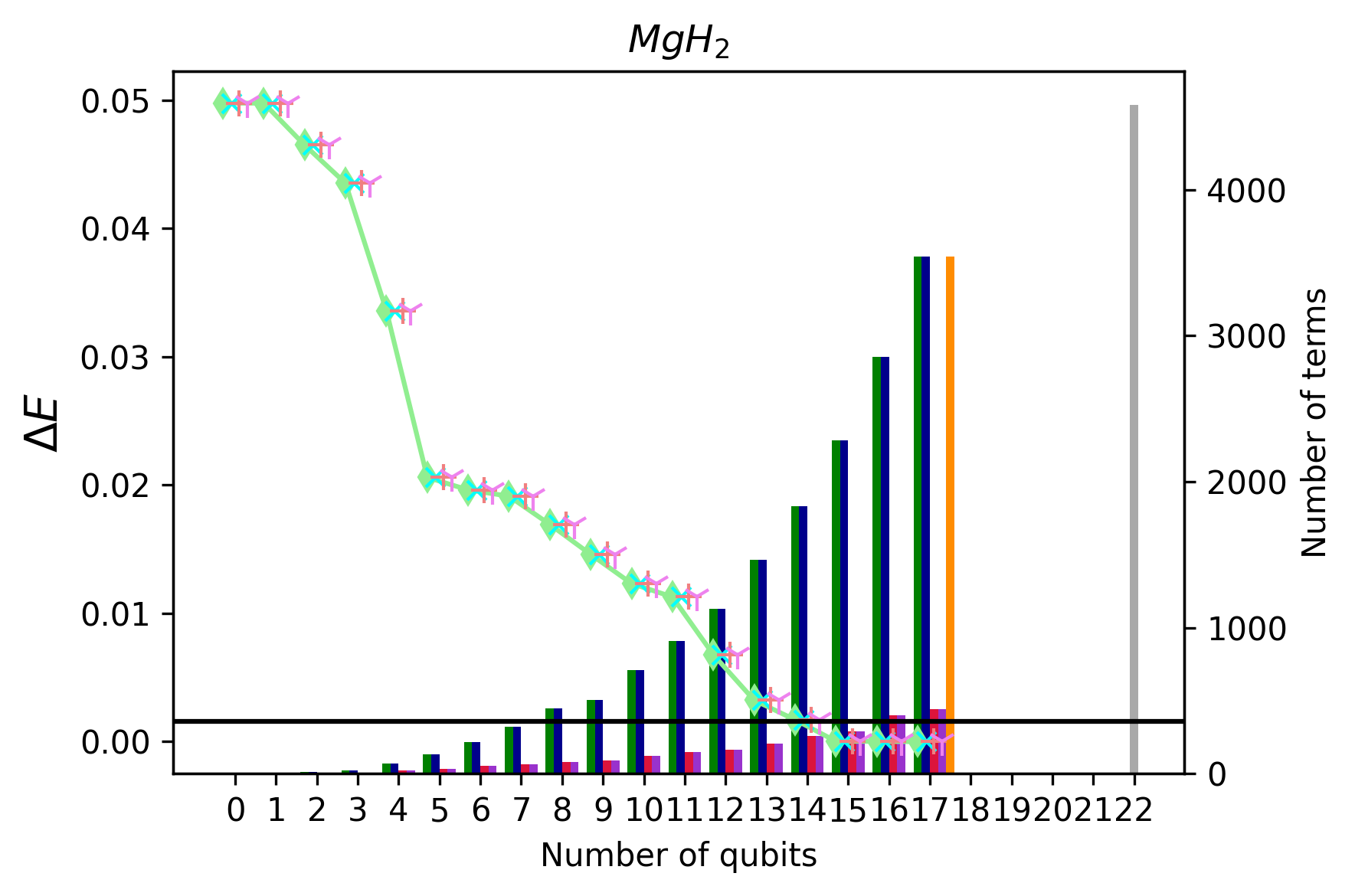}}
  \\
  \subfloat[]{\label{fig:17}\includegraphics[width=0.45\columnwidth]{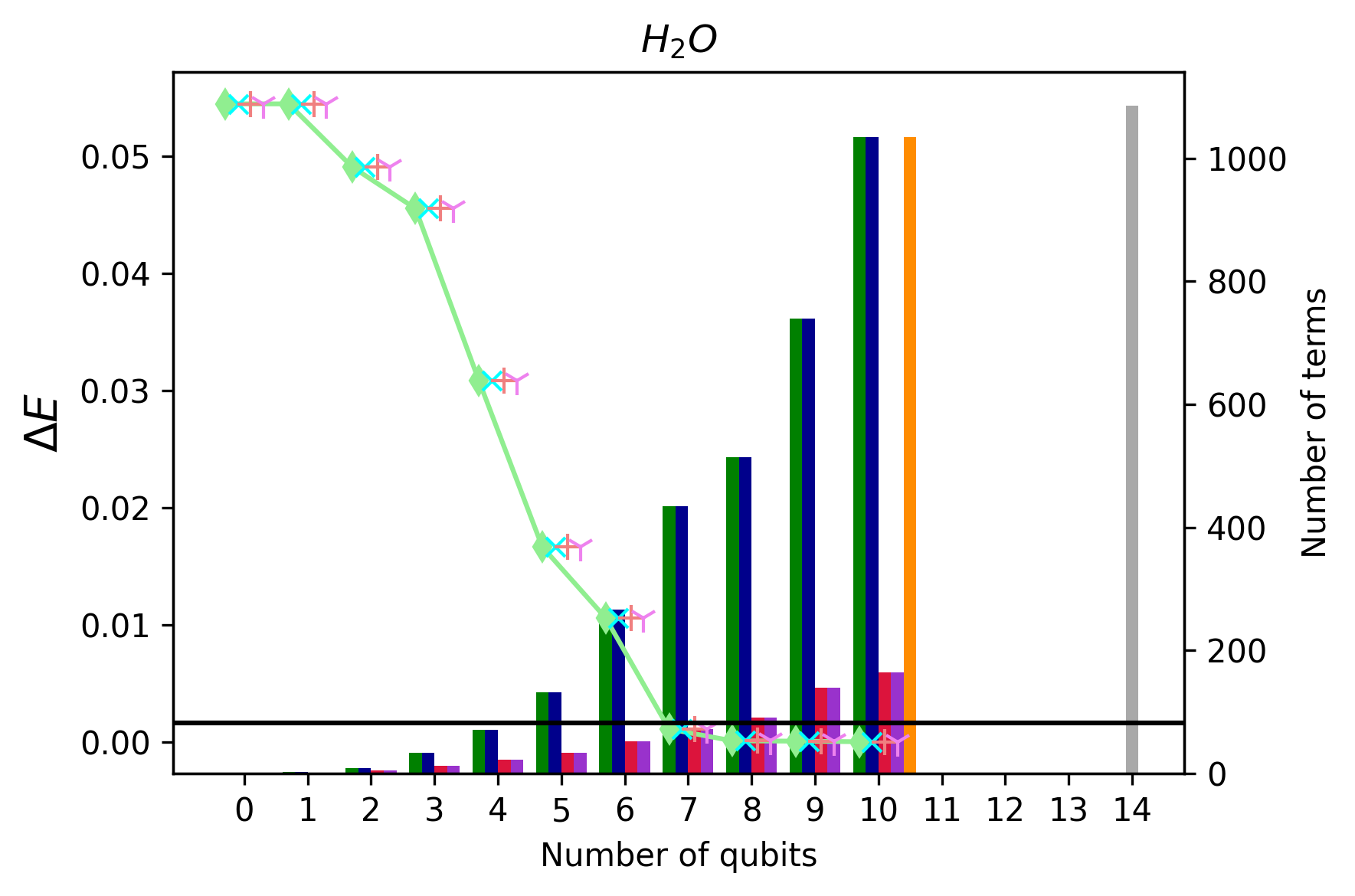}}
  \hfill
  \subfloat[]{\label{fig:18}\includegraphics[width=0.45\columnwidth]{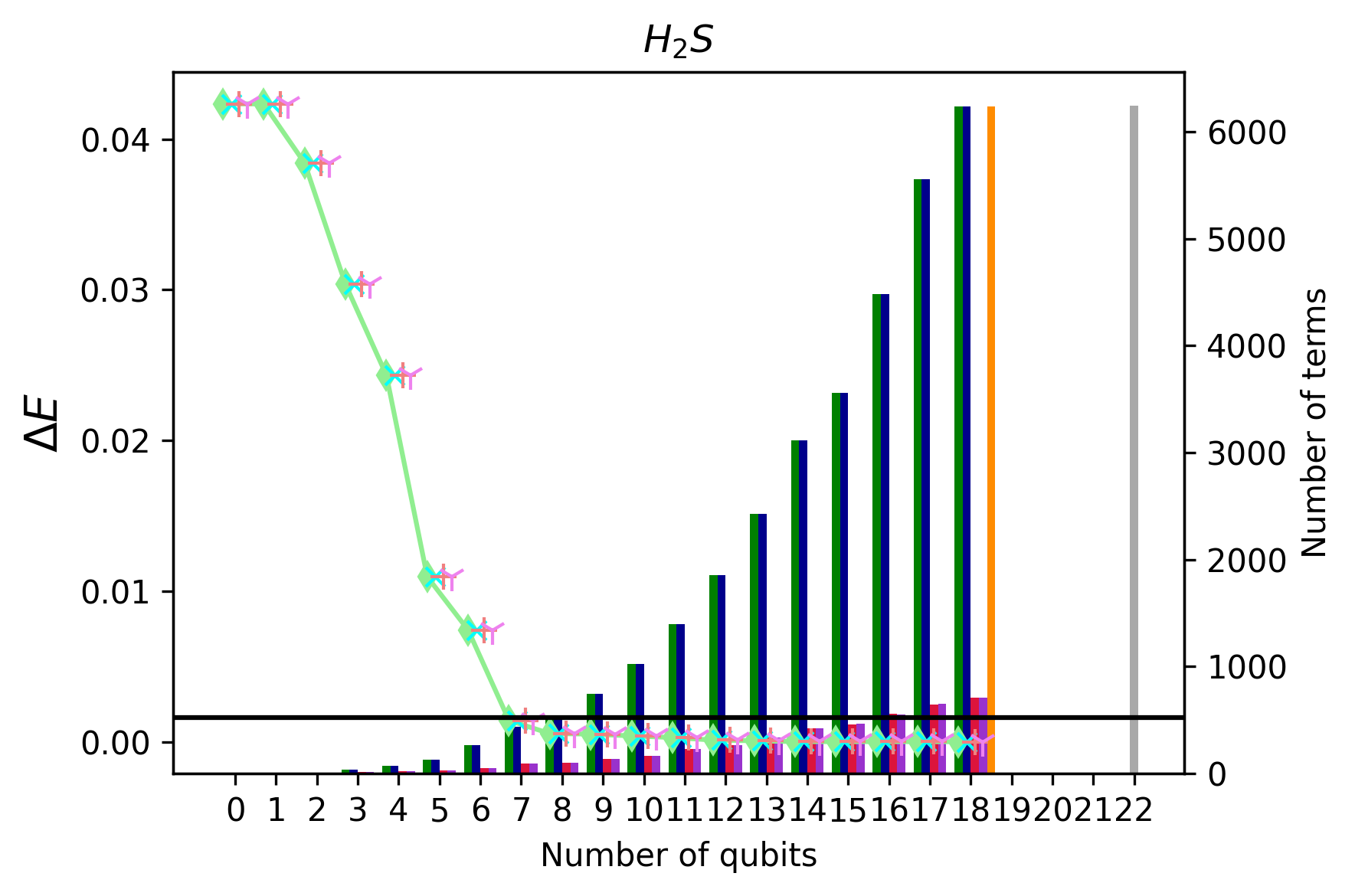}}
  \\
  \subfloat[]{\label{fig:19}\includegraphics[width=0.45\columnwidth]{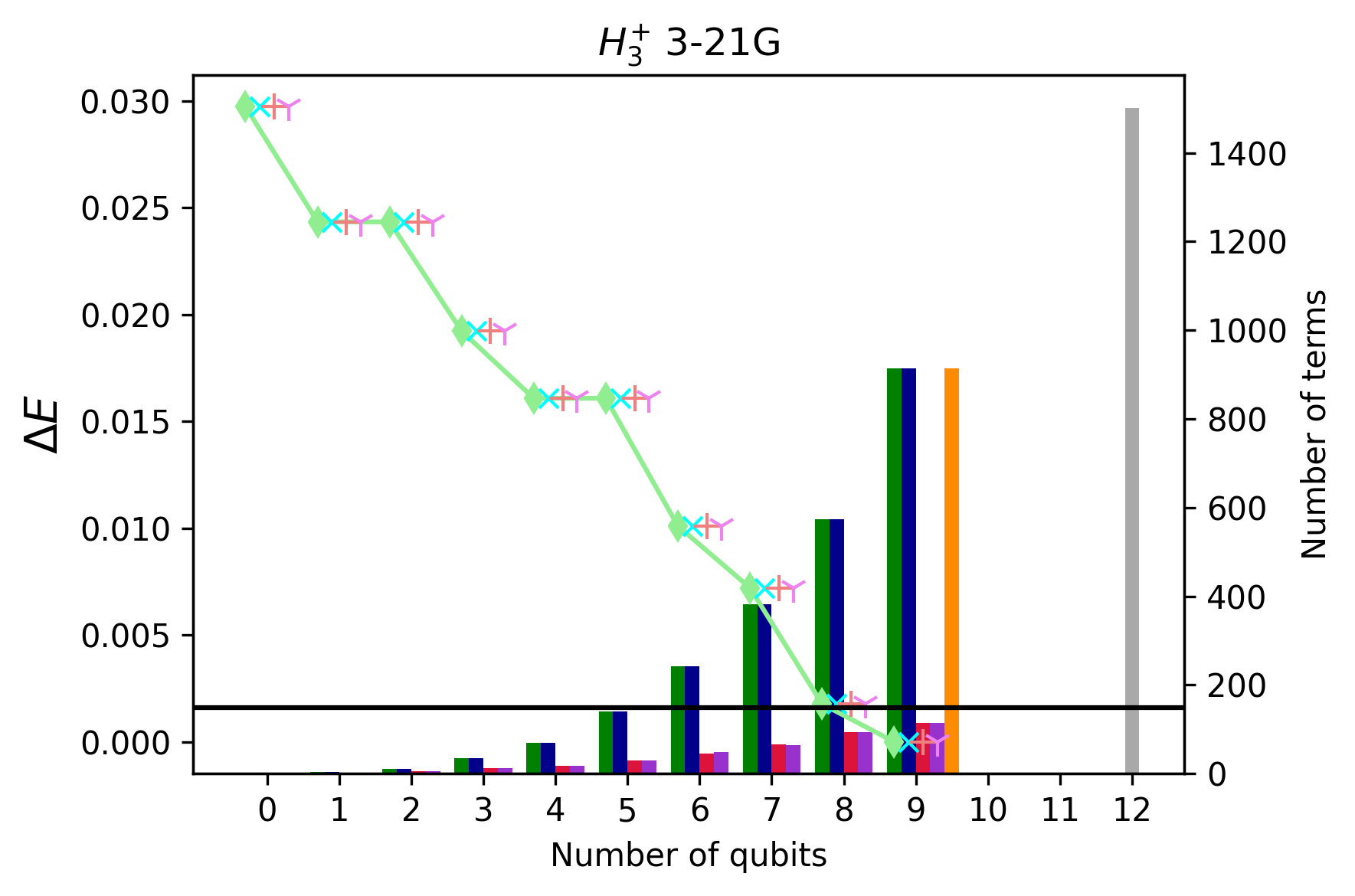}}
  \hfill
  \subfloat[]{\label{fig:20}\includegraphics[width=0.45\columnwidth]{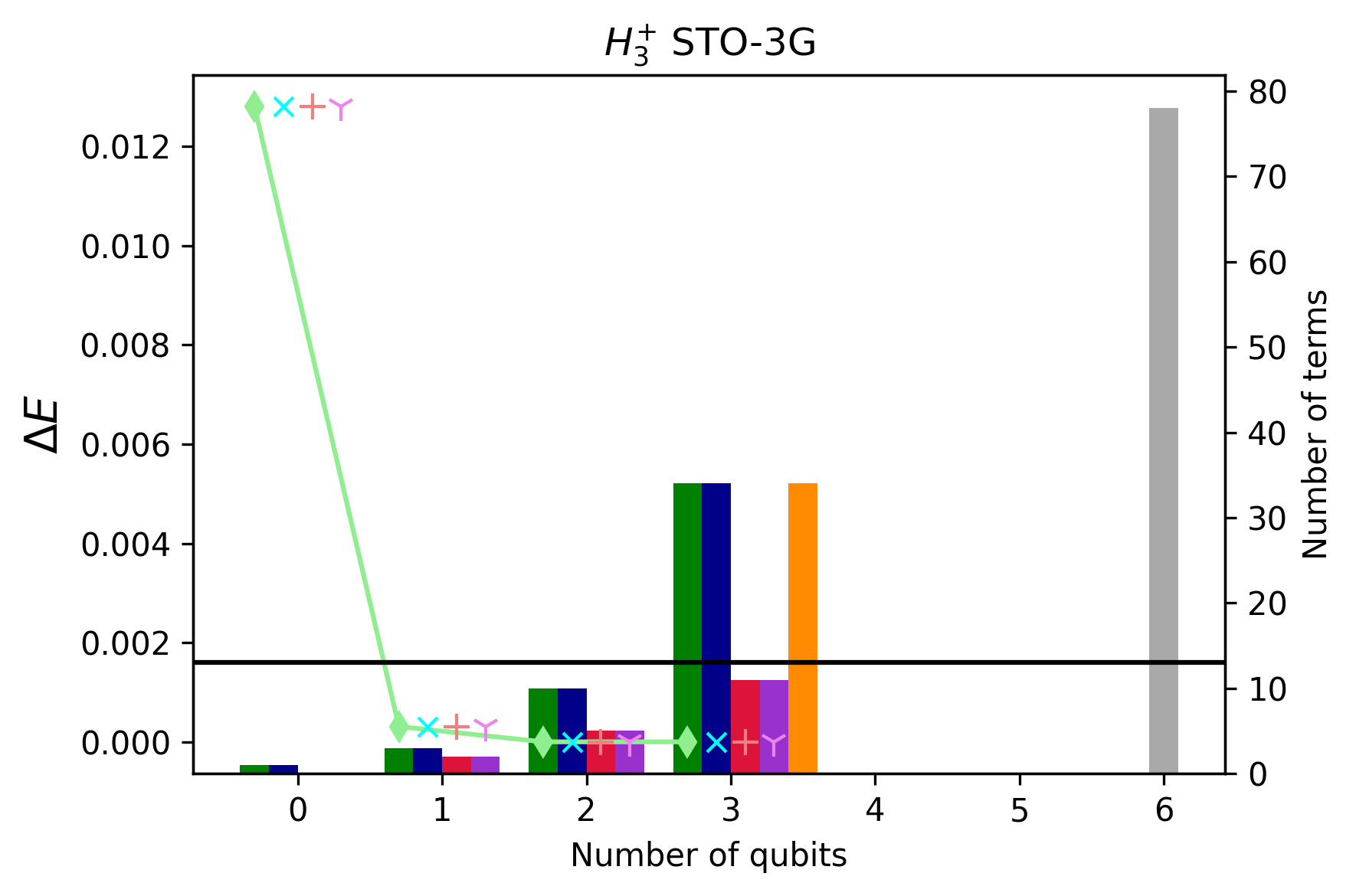}}
  \\
  \subfloat[]{\label{fig:21}\includegraphics[width=0.45\columnwidth]{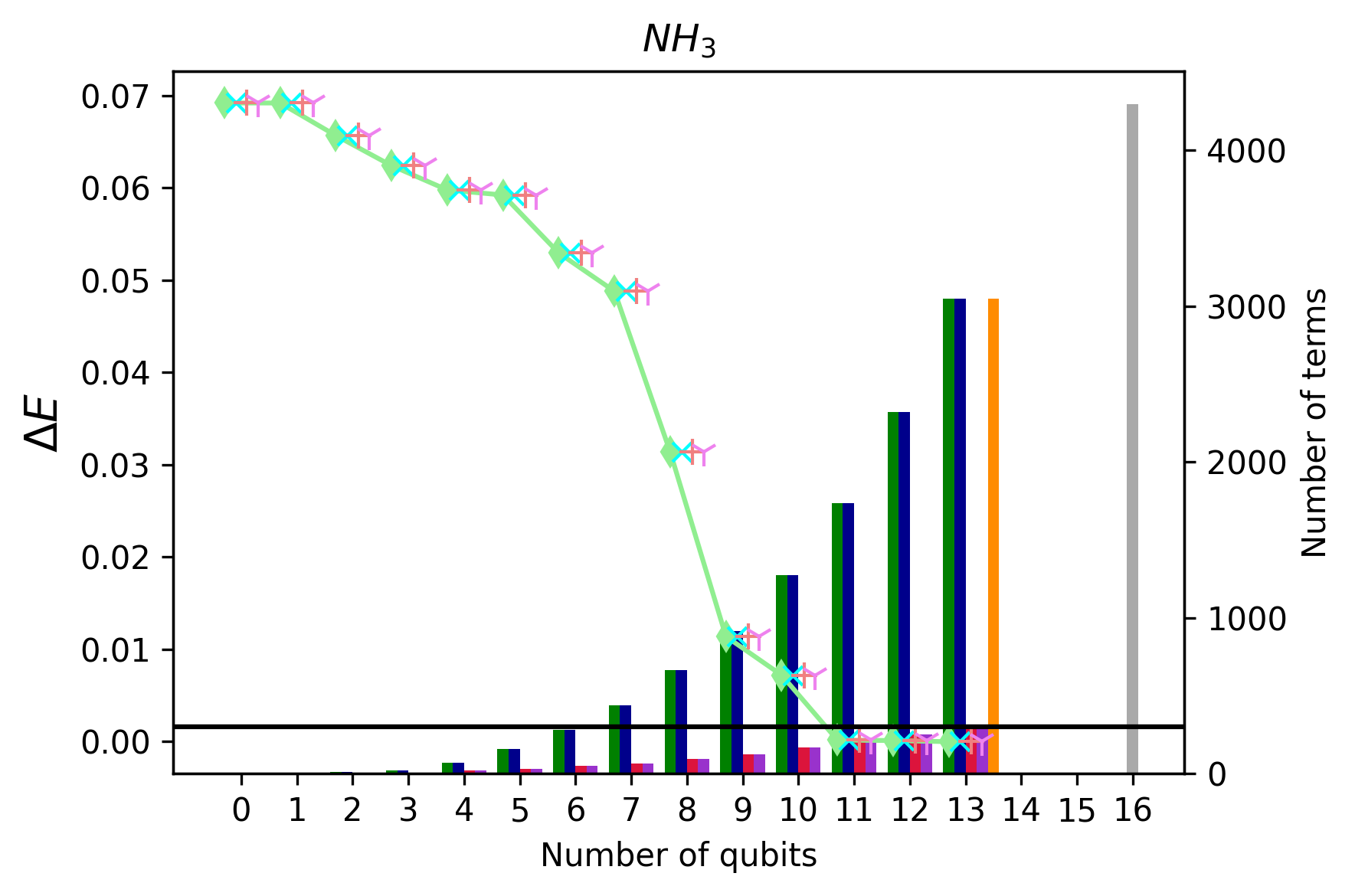}}
  \hfill
  \subfloat[]{\label{fig:22}\includegraphics[width=0.45\columnwidth]{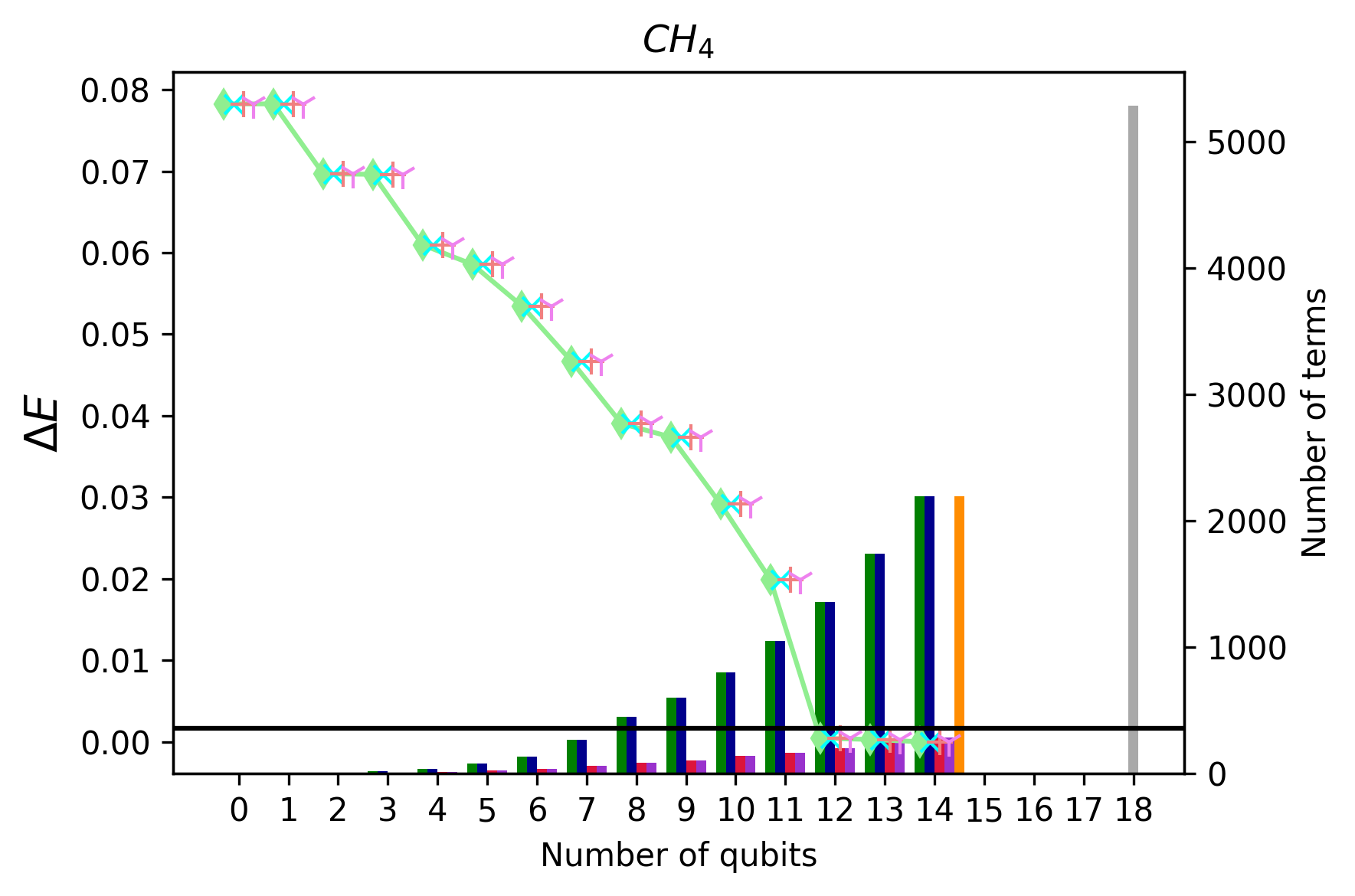}}
\end{figure}

\begin{figure}[H]
\ContinuedFloat 
 \label{fig: all_plots4}
  \centering
  \subfloat[]{\label{fig:23}\includegraphics[width=0.45\columnwidth]{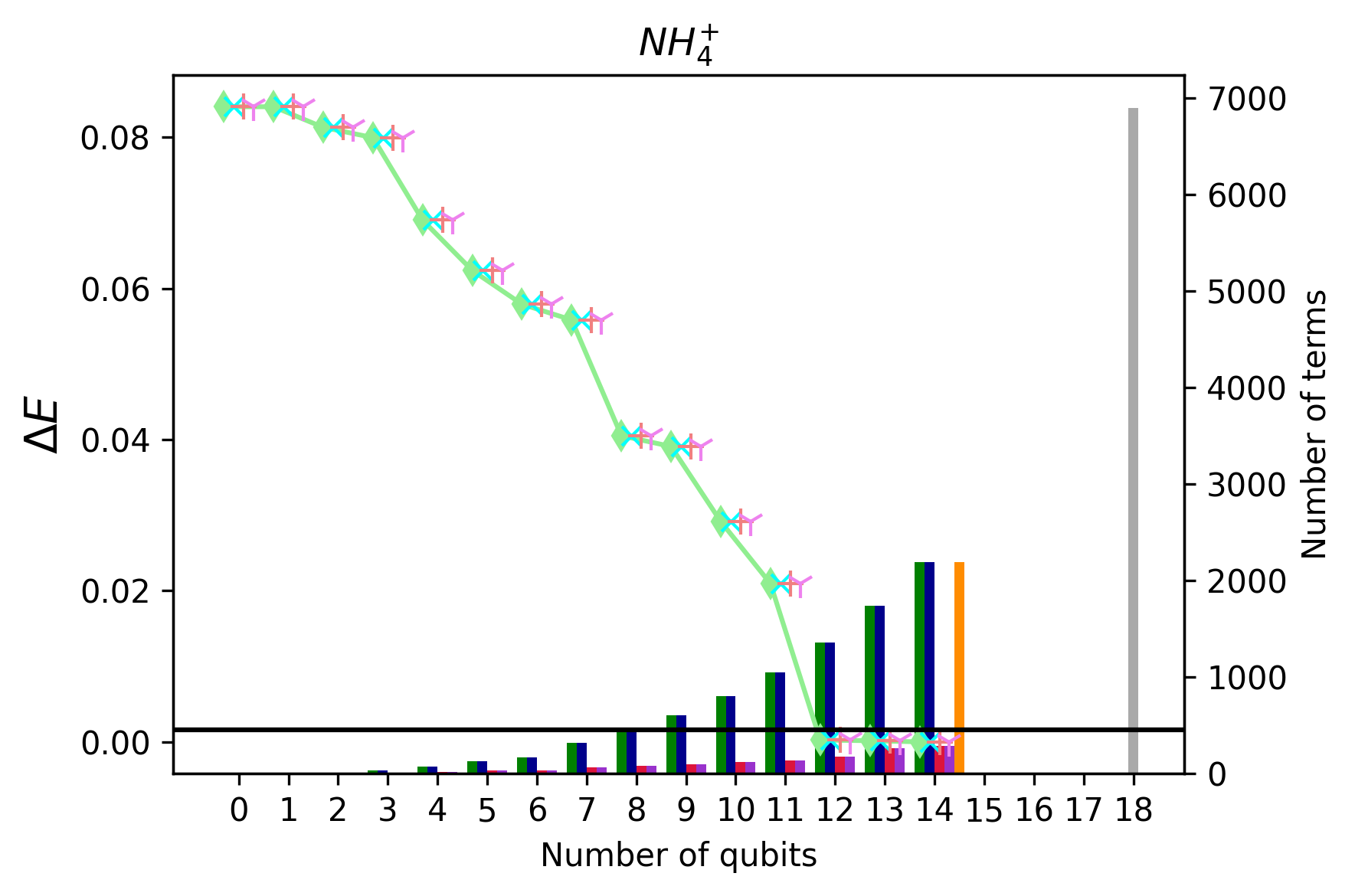}}
  \hfill
  \subfloat[]{\label{fig:24}\includegraphics[width=0.45\columnwidth]{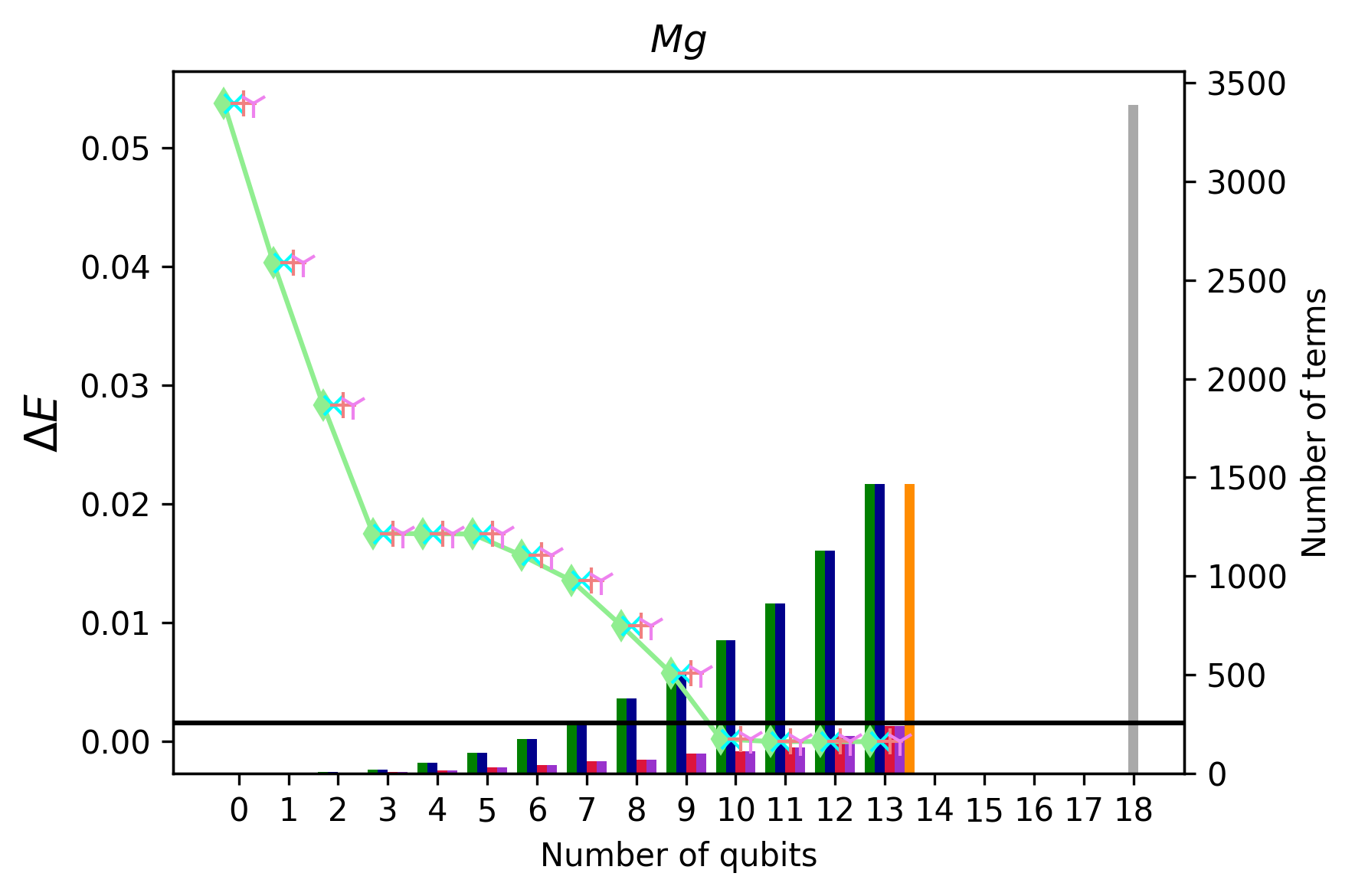}}
    \\
  \subfloat[]{\label{fig:25}\includegraphics[width=0.45\columnwidth]{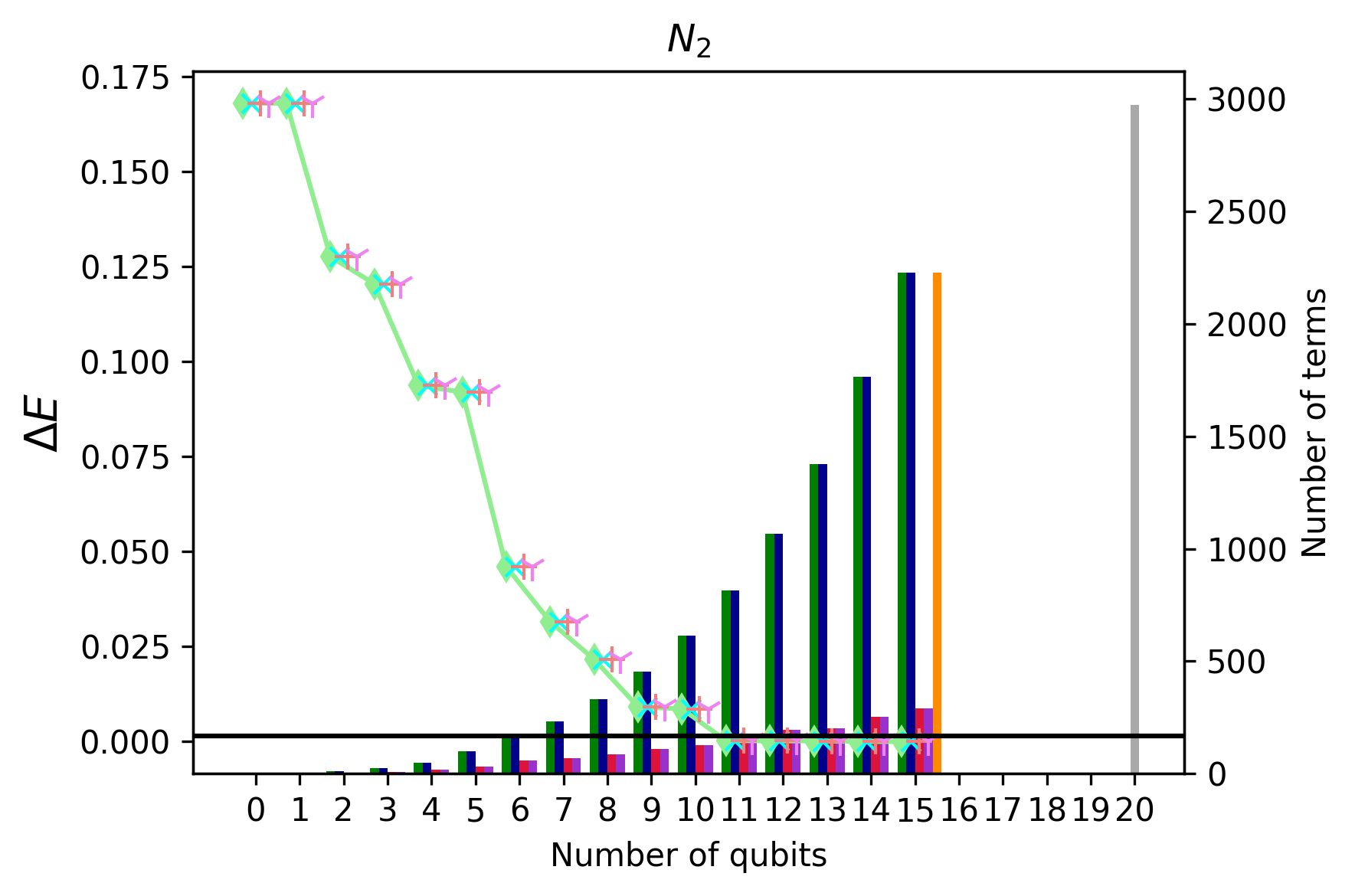}}
  \hfill
  \subfloat[]{\label{fig:26}\includegraphics[width=0.45\columnwidth]{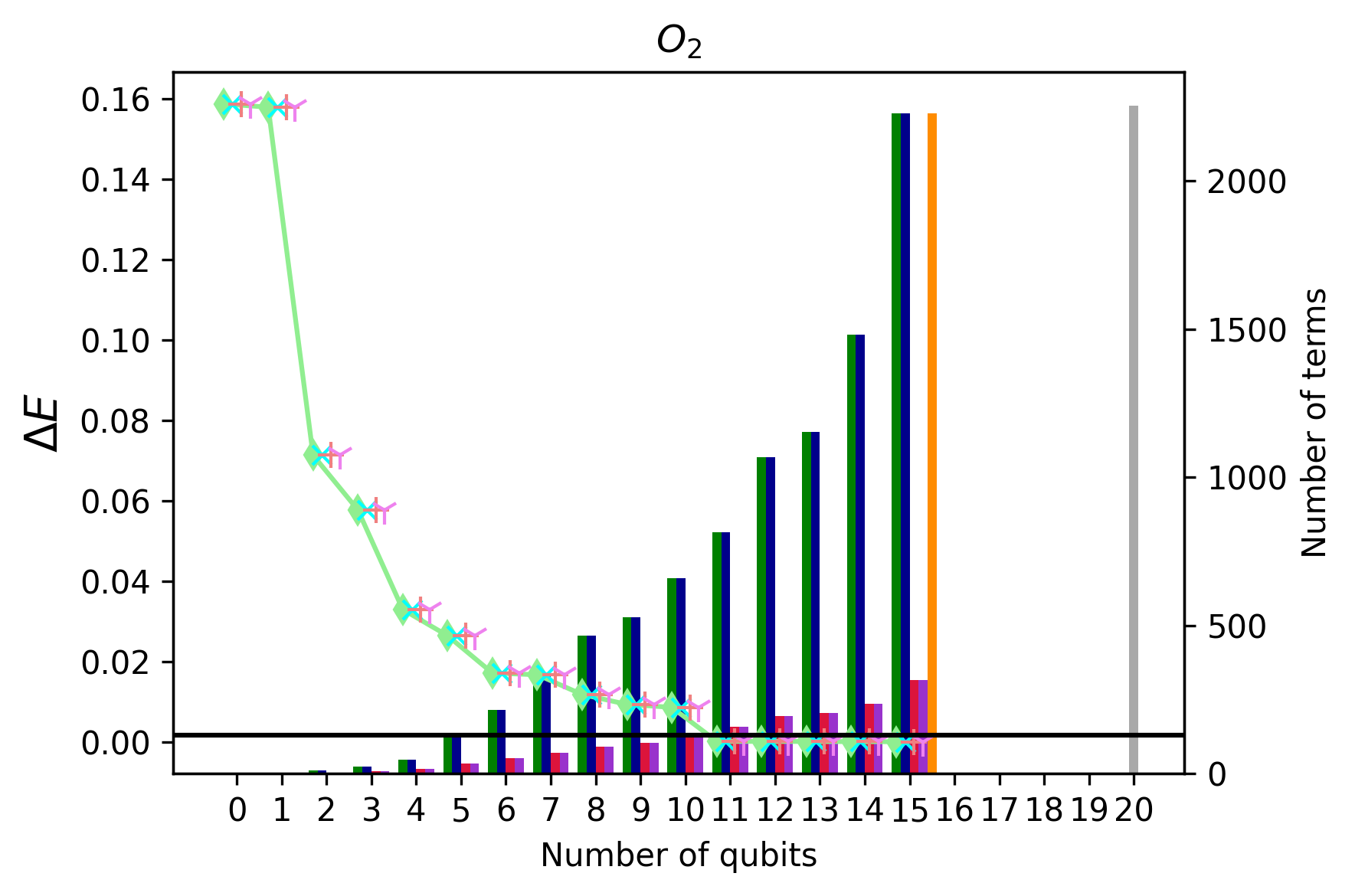}}
\caption{CS-VQE approximation errors $\Delta E$ versus number of qubits used on the quantum computer (scatter plot). The horizontal solid black lines indicate chemical accuracy. The number of terms in each Hamiltonian is given by the bar chart.   \label{fig: all_plots}}
\end{figure}

All the subplots in Figure \ref{fig: all_plots} give the simulation results of each molecular Hamiltonian at different levels of noncontextual approximations. This is equivalent to how many contextual stabilizers $\mathcal{W}$ eigenvalues are fixed. In each plot, the leftmost data represents the case when all the noncontextual stabilizer eigenvalues are fixed and is the case for the full noncontextual approximation to a given problem \cite{kirby2020classical}. Moving right, we remove a single stabilizer from $\mathcal{W}$ and thus don't fix the eigenvalue of that stabilizer. This reintroduces a qubits worth degree of freedom into the problem. At the limit that no stabilizer eigenvalues are fixed ($\mathcal{W}= \{ \}$) we return to standard VQE over the full problem and no noncontextual approximation is made. In each plot this scenario is represented by the far right data point (excluding the data for the full non tapered Hamiltonian that is supplied for reference only). The raw data for these results is supplied in the Supplemental Material (see the zipped file). We include data beyond Hamiltonians achieving chemical accuracy, to show the different possible approximations, rather than stopping once chemical accuracy was achieved.

\section{\label{sec:chemical_acc_table}Tabulated Results of simulation}

Table \ref{tab:Tab_results} summarises the numerical results of Figures \ref{fig:N_terms_plot} and \ref{fig:N_qubits_plot}.

\begin{table*}[!ht]
\centering
\begin{adjustbox}{width=1\textwidth}
\small
\begin{tabular}{cccccccc}
\hline
molecule &  basis & $H_{ \text{CS-VQE} }$ & $H_{\text{ CS-VQE} }$ $+\: UP^{(LCU)}$ & $H_{ \text{CS-VQE} }$ $+\: UP^{(SeqRot)}$ &    $H_{\text{tapered}}$  & $R H_{tapered}R^{\dagger}$ &        $H_{\text{full}}$  \\
\hline
    \ce{BeH2}&  STO-3G &   (7, 268) &               (7, 61) &                  (7, 61) &   (9, 596) &          (9, 614) &   (14, 666)  \\
    \ce{Mg}  &  STO-3G &  (10, 675) &             (10, 114) &                (10, 114) & (13, 1465) &        (13, 1465) &  (18, 3388)  \\
    \ce{H3+} &  3-21G  &   (9, 914) &              (9, 115) &                 (9, 115) &   (9, 914) &          (9, 786) &  (12, 1501)  \\
    \ce{O2}  &  STO-3G &  (11, 815) &             (11, 157) &                (11, 157) & (15, 2229) &        (15, 2374) &  (20, 2255)  \\
    \ce{OH}  &  STO-3G &   (6, 231) &               (6, 62) &                  (6, 62) &   (8, 558) &          (8, 558) &   (12, 631)  \\
    \ce{CH4} &  STO-3G & (12, 1359) &             (12, 203) &                (12, 203) & (14, 2194) &        (14, 2194) &  (18, 5288)  \\
    \ce{Be}  &  STO-3G &    (3, 20) &                (3, 9) &                   (3, 9) &   (5, 102) &          (5, 108) &   (10, 156)  \\
    \ce{NH3} &  STO-3G & (11, 1733) &             (11, 200) &                (11, 200) & (13, 3048) &        (13, 2738) &  (16, 4293)  \\
    \ce{H2S} &  STO-3G &   (7, 435) &               (7, 92) &                  (7, 92) & (18, 6237) &        (18, 6237) &  (22, 6246)  \\
    \ce{H2}  &  3-21G  &   (5, 122) &               (5, 27) &                  (5, 27) &   (5, 122) &          (5, 124) &    (8, 185)  \\
    \ce{HF}  &  3-21G  & (17, 5530) &             (17, 648) &                (17, 648) & (18, 6852) &        (18, 6852) & (22, 13958)  \\
    \ce{F2}  &  STO-3G &   (9, 527) &               (9, 99) &                  (9, 99) & (15, 2229) &        (15, 2229) &  (20, 2367)  \\
    \ce{HCl} &  STO-3G &   (4, 100) &               (4, 35) &                  (4, 35) & (16, 4409) &        (16, 4409) &  (20, 8159)  \\
    \ce{HeH+}&  3-21G  &   (5, 155) &               (5, 35) &                  (5, 35) &   (6, 319) &          (6, 319) &    (8, 361)  \\
    \ce{MgH2}&  STO-3G & (15, 2285) &             (15, 289) &                (15, 289) & (17, 3540) &        (17, 3540) &  (22, 4582)  \\
    \ce{CO}  &  STO-3G & (12, 1599) &             (12, 241) &                (12, 241) & (16, 4409) &        (16, 4409) &  (20, 5475)  \\
    \ce{LiH} &  STO-3G &   (4, 100) &               (4, 35) &                  (4, 35) &   (8, 558) &          (8, 586) &   (12, 631)  \\
    \ce{N2}  &  STO-3G &  (11, 815) &             (11, 153) &                (11, 153) & (15, 2229) &        (15, 2229) &  (20, 2975)  \\
    \ce{NaH} &  STO-3G & (14, 2722) &             (14, 375) &                (14, 375) & (16, 4409) &        (16, 4409) &  (20, 5851)  \\
    \ce{H2O} &  STO-3G &   (7, 435) &               (7, 73) &                  (7, 73) & (10, 1035) &        (10, 1035) &  (14, 1086)  \\
    \ce{H3+} &  STO-3G &     (1, 3) &                (1, 2) &                   (1, 2) &    (3, 34) &           (3, 35) &     (6, 78)  \\
    \ce{LiOH}&  STO-3G & (13, 2104) &             (13, 296) &                (13, 296) & (18, 6852) &        (18, 6852) &  (22, 8758)  \\
    \ce{LiH} &  3-21G  & (13, 2732) &             (13, 375) &                (13, 383) & (18, 6852) &        (18, 6852) &  (22, 8758)  \\
    \ce{H2}  &  6-31G  &   (5, 122) &               (5, 27) &                  (5, 27) &   (5, 122) &          (5, 124) &    (8, 185)  \\
    \ce{NH4+}&  STO-3G & (12, 1359) &             (12, 176) &                (12, 176) & (14, 2194) &        (14, 2194) &  (18, 6892)  \\
    \ce{HF}  &  STO-3G &   (4, 100) &               (4, 35) &                  (4, 35) &   (8, 558) &          (8, 558) &   (12, 631)  \\
    \hline
      \end{tabular}
\end{adjustbox}
\caption{Different resource requirements to study different electronic structure Hamiltonians required to achieve chemical accuracy. Each round bracket tuple reports $(n, |H|)$ and gives the number of qubits and terms for each Hamiltonian considered. $R H_{tapered}R^{\dagger}$ describes the effect of the CS-VQE unitary partitioning rotation on the problem Hamiltonian and $H_{\text{CS-VQE}} = Q_{\mathcal{W}}U_{\mathcal{W}}^{\dagger} H_{\text{full}}U_{\mathcal{W}} Q_{\mathcal{W}}^{\dagger}$. The square tuple gives the upper bound of single qubit gates (SQG) and CNOT gates $[SQG, CNOT]$ required to perform $R$ as a sequence of rotations in the unitary partitioning measurement reduction step, based on the largest anticommuting clique - representing the largest possible circuit for $R_{S}$. The size of the Hamiltonian for \ce{LiH} (3-21G singlet) with measurement reduction applied is different for the sequence of rotations and LCU  unitary partitioning methods. This is an artifact of the graph colour heuristic finding different anticommuting cliques in the CS-VQE Hamiltonian. \label{tab:Tab_results}} 
\end{table*}

\begin{table*}[!ht]
\end{table*}

\bibliography{references}